\documentclass[twocolumn,showpacs,nofootinbib,preprintnumbers,amsmath,amssymb,showkeys,superscriptaddress]{revtex4}
\usepackage{epsfig}
\usepackage{amssymb}
\usepackage{amsmath}
\usepackage{graphicx}
\usepackage{bm}
\usepackage{color}

\newcommand{\beq}{\begin{equation}}
\newcommand{\eeq}{\end{equation}}
\newcommand{\bea}{\begin{eqnarray}}
\newcommand{\eea}{\end{eqnarray}}
\newcommand{\nn}{\nonumber \\}

\newcommand\eqn[1]{(\ref{#1})}      
\newcommand\Eqn[1]{Eq.~(\ref{#1})}  
\newcommand\Fig[1]{Fig.~\ref{#1}}  

\newcommand{\tr}{\hbox{tr}}
\newcommand{\bartheta}{\,\bar{\!\theta}}
\newcommand{\thetab}{{\bartheta}}
\newcommand{\tb}{{\bartheta}}
\newcommand{\ts}{{\theta}}
\newcommand{\cb}{{\bar c}}

\begin{document}

\title{Influence of Gribov ambiguities in a class of nonlinear covariant gauges}

\author{J. Serreau}
\affiliation{(APC), Astro-Particule et Cosmologie, Universit\'e Paris Diderot CNRS/IN2P3, CEA/Irfu, Observatoire de Paris Sorbonne Paris Cit\'e \\ 10, rue Alice Domon et L\'eonie Duquet, 75205 Paris Cedex 13, France.}
\author{M. Tissier}
\affiliation{LPTMC, Laboratoire de Physique Th\'eorique de la Mati\`ere Condens\'ee, CNRS UMR 7600, Universit\'e Pierre et Marie Curie, \\ boite 121, 4 place Jussieu, 75252 Paris Cedex 05, France}
\author{A. Tresmontant}
\affiliation{Astro-Particule et Cosmologie (APC), CNRS UMR 7164, Universit\'e Paris Diderot\\ 10, rue Alice Domon et L\'eonie Duquet, 75205 Paris Cedex 13, France.}
\affiliation{LPTMC, Laboratoire de Physique Th\'eorique de la Mati\`ere Condens\'ee, CNRS UMR 7600, Universit\'e Pierre et Marie Curie, \\ boite 121, 4 place Jussieu, 75252 Paris Cedex 05, France}

\date{\today}

\begin{abstract}
 {We consider Yang-Mills theories in a recently proposed family of nonlinear covariant gauges that consistently deals with the issue of Gribov ambiguities. Such gauges  provide a generalization of the Curci-Ferrari-Delbourgo-Jarvis gauges which can be formulated as an extremization procedure and  might be implemented in numerical calculations. This would allow for nonperturbative studies of Yang-Mills correlators in a broad class of covariant gauges continuously connected to the well-studied Landau gauge.
We compute the ghost and gluon propagators in the continuum formulation at one-loop order in perturbation theory and we study their momentum dependence down to the deep infrared regime, with and without renormalization-group improvement. In particular, we show that the theory admits infrared-safe renormalization-group trajectories with no Landau pole. Both the gluon and the ghost behave as massive fields at low energy, and the gluon propagator is transverse even away from the Landau gauge limit. We compare our results to those obtained in the usual Curci-Ferrari model, which allows us to pinpoint the specific effects arising from our treatment of Gribov copies. }
\end{abstract}


\maketitle

\section{Introduction}\label{sec:introduction}

Describing the long-distance dynamics of non-Abelian gauge fields is a problem of notorious difficulty, mainly due to the breakdown of standard perturbative tools at low energies.
Various complementary approaches have been developed to tackle the problem, either through direct Monte Carlo simulations, or by means of approximate continuum methods. The former, which use a Euclidean lattice discretization, can directly access physical observables but are essentially limited to static quantities. In contrast, continuum approaches, such as truncations of Dyson-Schwinger equations \cite{vonSmekal97,Alkofer00}, or  the nonperturbative renormalization group \cite{Ellwanger96}, can access real-time quantities \cite{Strauss:2012dg,Haas:2013hpa}. However, they essentially rely on computing the basic correlation functions of Yang-Mills fields and require a gauge-fixing procedure. It is thus important to have a quantitative understanding of such correlators in order to assess the (necessary) approximations underlying such approaches. This can be achieved through gauge-fixed lattice calculations. 

An important issue concerns the algorithmic complexity of fixing a gauge numerically, in particular, in the case of covariant gauges, the most convenient ones for continuum calculations. Generically, this requires one to find the roots of a large set of coupled nonlinear equations, which is numerically demanding. However in some cases, the  gauge condition can be formulated as an extremization problem and efficient minimization algorithms can be employed.
This strategy was successfully applied for  the Landau gauge \cite{Boucaud08,Cucchieri_08b,Cucchieri_08,Bornyakov2008,Bogolubsky09,Dudal10,Maas:2011se}. Attempts to formulate general linear covariant gauges on the lattice have been made in Refs.~\cite{Giusti:1996kf,Giusti:1999im,Cucchieri:2008zx,Cucchieri:2009kk}. However, the extremization functional proposed in Refs.~\cite{Giusti:1996kf,Giusti:1999im} has spurious solutions, while the procedure of Ref.~\cite{Cucchieri:2008zx} is limited to infinitesimal deviations from the Landau gauge. The numerical procedure put forward in Ref.~\cite{Cucchieri:2009kk} and further developed in Refs.~\cite{Cucchieri:2010ku,Cucchieri:2011pp,Cucchieri:2011aa,Bicudo:2015rma} seems to correctly produce linear gauge-fixed configurations although its practical implementation does not really correspond to solving an extremization problem. Incidentally, it is problematic in what concerns the ghost sector. Linear gauges have been recently investigated in the context of Dyson-Schwinger equations in Refs.~\cite{Aguilar:2007nf,Aguilar:2015nqa,Huber:2015ria}.

In a recent work \cite{Serreau:2013ila,Serreau:2014xua}, we have proposed a simple generalization of the Landau gauge extremization functional, which generates a one-parameter family of nonlinear covariant gauges with a well-defined continuum formulation. Neglecting the issue of Gribov copies and applying the standard Faddeev-Popov (FP) procedure, this reduces to the Curci-Ferrari-Delbourgo-Jarvis (CFDJ) Lagrangian \cite{Curci:1976bt,Delbourgo:1981cm}. The latter is unitary and renormalizable in four dimensions. The proposed extremization functional satisfies the required properties for powerful numerical minimization algorithms, such as, e.g., the Los Alamos algorithm \cite{gupta87}. A numerical implementation of this proposal would open the way for nonperturbative calculations of Yang-Mills correlators in a broad class of covariant gauges continuously connected to the Landau gauge. 

The purpose of the present work is to study the ghost and gluon propagators in the continuum Euclidean formulation of this class of gauges.
Continuum approaches have to deal with the issue of Gribov copies \cite{Gribov77}. In the present context, this originates from the fact that the extremization functional possesses many extrema. Numerical minimization algorithms allow one to isolate a single extremum (minimum), but such a procedure is difficult to formulate in terms of a local action, the necessary starting point for continuum techniques. The standard FP procedure completely ignores the Gribov issue and is valid, at best, in the high-energy regime, where one expects Gribov copies to be irrelevant. It is thus of key importance to devise gauge-fixing procedures which deal with the issue of Gribov ambiguities. A famous example is the (refined) Gribov-Zwanziger proposal \cite{Gribov77,Zwanziger89,Dell'Antonio:1991xt,Dudal08}, where one restricts the path integral to the first Gribov region, in which the FP operator is positive definite. This has been mostly worked out in the case of the Landau gauge but some studies exist for linear covariant gauges \cite{Sobreiro:2005vn,Lavrov:2012ta,Lavrov:2013boa,Reshetnyak:2013bga,Moshin:2014xka}. This procedure is, however, not completely satisfactory since the first Gribov region itself is not free of Gribov ambiguities~\cite{vanBaal:1991zw,Vandersickel:2012tz}.

An alternative strategy, put forward in Ref.~\cite{Serreau:2012cg} in the case of the Landau gauge, is to average over the Gribov copies in such a way as to lift their degeneracy. For an appropriate choice of the averaging procedure, this can be formulated in terms of a local renormalizable field theory in four dimensions, which turns out to be perturbatively equivalent to the Landau limit of the Curci-Ferrari (CF) model \cite{Curci:1976bt}---a simple massive extension of the FP Lagrangian---for the calculation of ghost and gluon correlators. Here, the bare gluon mass is related to the gauge-fixing parameter which lifts the degeneracy between Gribov copies. Remarkably, this theory admits infrared-safe renormalization-group (RG) trajectories, allowing the use of perturbation theory down to arbitrarily low momenta \cite{Tissier_10}. One-loop calculations of ghost and gluon correlators in the vacuum and at finite temperature are in good agreement with lattice results  \cite{Tissier_10,Pelaez:2013cpa,Reinosa:2013twa}. This has been extended to QCD in Ref.~\cite{Pelaez:2014mxa} and to background field methods in Ref.~\cite{Reinosa:2014ooa,Reinosa:2014zta,Reinosa:2015oua}, allowing for a simple perturbative description of the confinement-deconfinement transition in pure SU($N$) theories. 

We have generalized such an averaging procedure over Gribov copies to the family of nonlinear covariant gauges mentioned above in Ref.~\cite{Serreau:2013ila}; see also Ref.~\cite{Serreau:2014xua} for a short overview. Again, for a suitable choice of the averaging procedure, this can be formulated as a local field theory, which is perturbatively renormalizable in four dimensions and, thus, well suited for continuum approaches. As in the case of the Landau gauge, the resulting gauge-fixed action consists of a massive extension of the CFDJ action---the general CF action \cite{Curci:1976bt}---coupled to a set of replicated supersymmetric nonlinear sigma models. The superfield sector is directly related to the treatment of Gribov ambiguities. However, unlike in the special case of the Landau gauge, this superfield sector does not decouple in the calculation of ghost and gluon correlators and the resulting effective gauge-fixed theory is very different from the general CF model. In this article, we present a complete calculation of the ghost and gluon propagators at one-loop order using this gauge-fixed action, with and without RG improvement, and we investigate possible infrared-safe RG trajectories. We compare our results to those in the CF model, which gives direct information on the role of Gribov ambiguities.

 One aim of the present work is to motivate lattice calculations of ghost and gluon correlators in the proposed family of gauges. Such calculations would typically be performed by selecting one minimum (maximum) of the extremization functional. It is by no means clear how this would be related with the averaging over Gribov copies studied here. However, we may hope that the situation is similar to the case of the Landau gauge, or at least close to the latter. As mentioned above, in this case, the weighted average procedure put forward in Ref. \cite{Serreau:2012cg} produces perturbative results which agree well with lattice calculations in the minimal Landau gauge---where one selects a unique (random) Gribov copy in the first Gribov region, that is, among the ensemble of minima (maxima) of the extremization functional. A possible interpretation of this observation is that, for some range of the averaging parameter, all Gribov copies in the first region are essentially equiprobable while other Gribov regions are suppressed, such that averaging over copies is equivalent to randomly choosing a particular one in the first region \cite{Serreau:2012cg}. Understanding this issue further would require a detailed investigation of the landscape of the extremization functional. This can be done on small lattices \cite{Hughes:2012hg} but it is a formidable task in general both because the number of copies typically grows exponentially with increasing volume and because it is virtually impossible to numerically find copies that are not in the first Gribov region (i.e., saddle points of the extremization functional). 

The paper is organized as follows. In Sec.~\ref{sec_gaugefixing}, we recall the main steps of the gauge-fixing procedure proposed in Ref.~\cite{Serreau:2013ila}. We fix the notations and we introduce a crucial technical aspect of the whole procedure, namely the replica trick. In Sec.~\ref{Perturbative_computations}, we review the Feynman rules relevant for one-loop computations and calculate the self-energies of the theory. We show in an explicit example how the calculation of loop diagrams involving the superfield sector are performed in practice. In Sec. \ref{sec_renormalization} we present two renormalization schemes and we discuss, in particular, the role of the replica. In Sec.~\ref{sec_perturbativ_results}, we present our results for the ghost and gluon propagators for different choices of the gauge-fixing parameter. Finally, we investigate RG improvement in Sec.~\ref{sec_RG} and we find infrared-safe RG trajectories  where the coupling constant remains under control all along the flow. This justifies the use of perturbation theory down to deep infrared momenta. In this regime, we find that both the gluon and the ghost are massive, and that the gluon propagator is always strictly transverse. The superfield sector is crucial in that it ensures that the present approach is indeed a bona fide gauge fixing, in contrast, e.g., to the CF model. In order to emphasize the role of the superfield sector, we systematically compare our results to those of the CF model at the same order of approximation. Finally, we give some technical details in Appendices~\ref{sec_complete_propagator}--\ref{appsec:UV}. 

\section{The gauge-fixing procedure}
\label{sec_gaugefixing}
The classical action of the SU($N$) Yang-Mills theory reads, in $d$-dimensional Euclidean space,
\beq 
\label{eq:SYM}
 S_{\rm YM}[A]=\frac{1}{4}\int_x \left(F_{\mu\nu}^a\right)^2\,, 
\eeq 
where $\int_x\equiv\int d^dx$ and 
\beq
 F_{\mu\nu}^a=\partial_\mu A_\nu^a-\partial_\nu A_\mu^a +g_0f^{abc}A_\mu^bA_\nu^c,
\eeq
where $g_0$ is the (bare) coupling constant and a summation over repeated spacetime and color indices is understood. In the following, we use the convention that fields written without an explicit color index are contracted with the generators $t^a$ of SU($N$) in the fundamental representation and are thus $N\times N$ matrix fields, e.g., $A_\mu=A_\mu^a t^a$.  Our normalization for the generators is such that 
\begin{equation}
  \label{eq_prod_gene}
  t^a t^b=\frac{\delta ^{ab}}{2N}\openone +\frac{if^{abc}+d^{abc}}2 t^c,
\end{equation}
where $f^{abc}$ and $d^{abc}$ are the usual totally antisymmetric and totally symmetric tensors of SU($N$). In particular, we have
\beq
  \tr \left(t^a t^b\right)=\frac{\delta^{ab}}{2}.
\eeq

In order to fix the gauge, we consider the functional
\begin{equation}
  \label{eq_func}
  {\cal H}[A,\eta,U]=\int_x\,\tr\left[\left(A^U_\mu\right)^2+ \frac{U^\dagger \eta+\eta^\dagger U}2\right],
\end{equation}
where $\eta$ is an arbitrary $N\times N$ matrix field, and 
\begin{equation}
  \label{eq_gauge_transfo}
  A_\mu^U=UA_\mu U^\dagger+\frac i {g_0}U\partial_\mu U^\dagger
\end{equation}
is the gauge transform of $A_\mu$ with $U\in$ SU($N$). We define our gauge condition as (one of) the extrema of ${\cal H}$ with respect to $U$. This leads to the following covariant gauge condition:
\begin{equation}
  \label{eq_eq_mot}
  \left(\partial_\mu A_\mu^{U}\right)^a=\frac{ig_0}{2}\tr\left[t^a\left(U\eta^\dagger-\eta U^\dagger\right)\right].
\end{equation}

The functional ${\cal H}$ admits many extrema $U_i\equiv U_i[A,\eta]$ which correspond to Gribov copies. The possible numerical implementation of this gauge condition has been discussed in Ref.~\cite{Serreau:2013ila}. Once the gauge condition \eqn{eq_eq_mot} has been implemented, we average over the random field $\eta$, with a Gaussian weight\footnote{The weight \eqn{p_eta} is invariant under the change $\eta\to\eta U$, which ensures that one can eventually factor out the volume of the gauge group in the FP construction.} ($\mathcal{N}$ is a normalization factor) 
\begin{equation}\label{p_eta}
\mathcal{P}\left[\eta\right]= \mathcal{N} \exp \left(-\frac{g_0^2}{4 \xi_0} \int_x\,\tr ~\eta^\dagger \eta \right),
\end{equation}
where $\xi_0$ is a (bare) gauge-fixing parameter. The case $\xi_0=0$ corresponds to $\eta=0$, that is to the Landau gauge $\partial_\mu A_\mu^{U}=0$.

Let us briefly recall the main lines of the formalism developed in Ref.~\cite{Serreau:2013ila}. We define the vacuum expectation values of an operator $\mathcal O[A]$ by a two-step averaging procedure. The first step, hereafter denoted with brackets, consists in an average over the Gribov copies of any gauge field configuration $A$:
\begin{equation}
  \label{eq_average_G}
  \langle\mathcal O[A]\rangle=\frac{\int \mathcal D\eta \mathcal
    P[\eta]\sum_i \mathcal O [A^{U_i}]s(i)e^{-\beta_0 {\cal H}[A,\eta,U_i]}}{\int \mathcal D\eta \mathcal
    P[\eta]\sum_i s(i)e^{-\beta_0 {\cal H}[A,\eta,U_i]}},
\end{equation}
where the discrete sums run over all Gribov copies, $s(i)$ is the sign of the functional determinant of the Faddeev-Popov operator---the Hessian of the functional \eqn{eq_func}---evaluated at $U={U_i}$, and $\beta_0$ is a free gauge-fixing parameter which controls the lifting of degeneracy between Gribov copies. The case $\beta_0 =0$ would correspond to a flat weight over the Gribov copies and would reduce to the standard FP construction. The second step, denoted by an overall bar, is a standard average over gauge field configurations with the Yang-Mills weight. The expectation value of the operator ${\mathcal O[A]}$ is thus obtained as
\begin{equation}
  \label{eq_av_A}
\overline{\langle {\mathcal O[A]}\rangle}=\frac{\int\mathcal
  DA\, \mathcal \langle {\cal O}[A]\rangle \,e^{-S_{\rm YM}[A]}}{\int\mathcal
  DA\, e^{-S_{\rm YM}[A]}}.
\end{equation}

Using standard techniques, the sum over Gribov copies in \Eqn{eq_average_G} can be written in terms of a path integral over an SU($N$) matrix field $U$ as well as ghost and antighost fields $c$ and $\cb$ and a Nakanishi-Lautrup field $h$ which ensure the gauge condition \eqn{eq_eq_mot}. One can then explicitly perform the Gaussian integration over the field $\eta$.
A crucial point here is the presence of the denominator in \Eqn{eq_average_G}, which guarantees that gauge-invariant observables ${\cal O}_{inv}[A]$ are blind to the gauge-fixing procedure: $\langle\mathcal O_{inv}[A]\rangle = {\cal O}_{inv}[A]$. Such a denominator produces a nonlocal functional of the gauge field which can, however, be treated by means of a local action using the replica technique \cite{young}. The latter amounts to making $n$ replicas of the set of fields $(U,c,\cb,h)$ and taking the limit $n\to0$ at the end of (loop) calculations. One can then explicitly factor out the volume of the gauge group $\int {\cal D}U$, which effectively singles out one replica.  The $n-1$ remaining replicated fields  $(U_k,c_k,\cb_k,h_k)$, with $2\le k \le n$, are conveniently grabbed together by introducing SU($N$) matrix superfields ${\cal V}_k$ that depend on the Euclidean coordinate $x$ and a pair of Grassmannian coordinates $(\ts_k,\tb_k)\equiv\underline\theta_k$ as
\beq
\label{eq:super}
 \mathcal{V}_k(x,\underline\theta_k) = \exp\left\{ ig_0\Big(\bar{\theta}_kc_k+ \bar{c}_k\theta_k+\bar{\theta}_k\theta_k \hat h_k\Big)\right\}U_k,
\eeq
where $\hat h_k=ih_k+\frac{g_0}{2}\{\cb_k,c_k\}$. Here, the $x$ dependence on the right-hand side is only through the fields. Note that there is an independent pair of Grassmannian coordinates for each replica. 

After some manipulations, one finally gets \cite{Serreau:2013ila}
\begin{equation}
  \label{eq_average2ter}
  \overline{\langle{\cal O}[A]\rangle}=\lim_{n\to 0}\frac{\int\mathcal D (A,c,\bar c,h,\{{\cal V}\})\,{\cal O}[A]\, e^{-S[A,c,\bar c,h,\{{\cal V}\}]}}{\int\mathcal D (A,c,\bar c,h,\{{\cal V}\})\, e^{-S[A,c,\bar c,h,\{{\cal V}\}]}}\,,
\end{equation}
where $\mathcal D (A,c,\bar c,h,\{{\cal V}\})\equiv\mathcal D (A,c,\bar c,h)\times\prod_{k=2}^n \mathcal D {\cal V}_k$ and the gauge-fixed action for arbitrary $n$ is given by 
\begin{align}
  \label{eq_action2}
  S[A,c,\bar c,h,\{{\cal V}\}]&=S_{{\rm YM}}[A]+S_{{\rm CF}}[A,c,\cb,h]\nn
  &+\sum_{k=2}^n S_{\rm SUSY}[A,\mathcal V_k],
\end{align}
where $S_{{\rm YM}}[A]$ is the Yang-Mills action,
\begin{align}
  \label{eq_action_CF}
 S_{{\rm CF}}&[A,c,\cb,h]=\int_x\bigg\{ \partial_\mu \cb^aD_\mu c^a+ih^a \partial_\mu A_\mu^a\nn
  &+\xi_0 \bigg[\frac {(h^a)^2}2\!-\!\frac {g_0}2 f^{abc}ih^a\cb^b c^c\!-\!\frac {g_0^2}4 \left( f^{abc}\cb^b c^c\right)^2\bigg] \nn
  &+ {\beta_0}\left[ \frac{1}{2}(A_\mu^a)^2+\xi_0\cb^a c^a \right] \bigg\} ,
\end{align}
and 
\beq
  \label{eq_action_CF_susy}
 S_{\rm SUSY}[A,\mathcal V_k]=S_{{\rm CF}}[A^{U_k},c_k,\cb_k,h_k].
\eeq
It is remarkable that the integration over the random matrix field $\eta$ with the weight \eqn{p_eta} produces the CFDJ action, the first two lines on the right-hand side of \Eqn{eq_action_CF}. The third line, which corresponds to a bare mass term for both the gluon and the ghost fields proportional to the factor $\beta_0$, arises from the average over Gribov copies. Altogether, the first line of \Eqn{eq_action2} corresponds to the general CF action, whereas the second line is the contribution from the replicated superfields.

Finally, for actual calculations, it is useful to exploit the superfield formulation \Eqn{eq:super} and to rewrite \eqn{eq_action_CF_susy} as
\begin{align}
  \label{eq_av_superfield_bis}
 S_{\rm SUSY}[A,{\cal V}]\!=\!\frac{1}{g_0^2}\int_{x,\underline{ \theta}}&\!\tr \Big\{D_\mu{\cal V}^\dagger D_\mu{\cal V}\!+\!\frac{\xi_0}2 g^{MN}\partial_N{\mathcal V}^\dagger \partial_M {\mathcal V}\Big\},
\end{align}
with $D_\mu{\cal V}=\partial_\mu{\cal V}+ig_0{\cal V}A_\mu$ and where the uppercase latin letters $M,N$ stand for Grassmann variables $\ts,\tb$ (a sum over repeated indices is understood). Here, we have introduced a curved Grassmann manifold with line element $ds^2=g_{MN}dNdM=2g_{\ts\tb}d\tb d\ts$, where 
\beq
\begin{split}
  \label{eq_metric_grass}
  &g_{\tb\ts}=-g_{\ts\tb}=\beta_0\tb\ts+1,\\
  &g^{\tb\ts}=-g^{\ts\tb}=\beta_0\tb\ts-1.
\end{split}
\eeq
The curvature of the Grassmannian manifold is controlled by the lifting parameter $\beta_0$. As already noticed, there is such a curved Grassmannian manifold associated to each replica superfield ${\cal V}_k$. The integration measure is defined accordingly as  \cite{Tissier_08}
\begin{equation}
  \label{eq_note_int_grass}
\int_{\underline{\theta}}=\int d\theta d\thetab\,g^{1/2}(\theta,\thetab)  \,,
\end{equation}
where 
\begin{equation}
\label{eq_measure}
g^{1/2}(\theta,\thetab)\equiv g^{1/2}(\underline{\theta})=\beta_0\thetab\theta-1. 
\end{equation}
The formulation \eqn{eq_av_superfield_bis} makes transparent a large class of supersymmetries which mix the original bosonic $(U_k,h_k)$ and Grassmannian $(c_k,\cb_k)$ fields. These simply correspond to the isometries of the Grassmannian manifold \cite{Tissier_08,Serreau:2013ila}. 

We stress again that the limit $n \to 0$ in \Eqn{eq_average2ter} results from the presence of the denominator in \Eqn{eq_average_G} and is thus crucial in order for the present theory to correspond to a gauge-fixing procedure, as discussed above. Instead, ignoring the denominator in \Eqn{eq_average_G} is equivalent to setting $n=1$ which does not correspond to a gauge-fixed theory. In that case, there is no superfield sector and the model reduces to the CF action, given by the first line of \Eqn{eq_action2}. It is, therefore, interesting to compare the results obtained in the limit $n\to0$ to the corresponding ones in the CF model, obtained by setting $n=1$ in order to highlight the importance of the superfield sector and of the limit $n\to0$. This allows us to study the influence of the Gribov ambiguities. It is worth emphasizing that the action \eqn{eq_action2} is perturbatively renormalizable and asymptotically free in four dimensions for an arbitrary value of $n$ despite the presence of the replicated matrix superfield \cite{Serreau:2013ila}. We can thus use the same renormalization schemes in the cases $n\to0$ and $n=1$.

To summarize,  we have presented a genuine gauge-fixing procedure of the Yang-Mills action in a certain class of nonlinear covariant gauges, which consistently deals with the issue of Gribov ambiguities and which can be written as a local, renormalizable gauge-fixed action. These gauges are indexed by a parameter $\xi_0$, and the lifting over the Gribov copies is controlled by an extra gauge-fixing parameter $\beta_0>0$ homogeneous to a square mass. The resulting gauge-fixed action \eqn{eq_action2} depends on the gauge field $A$, the usual ghost and antighost fields $c,~ \cb$ and the Nakanishi-Lautrup field $h$, and a set of $n-1$ replicated SU($N$) matrix superfields. In the following, we compute the two-point vertex functions of the theory in the vacuum with the action \eqn{eq_action2} for fixed $n$, from which we obtain the various propagators after inversion. We then either perform the limit\footnote{This has to be done after inversion of the matrix of two-point vertex functions because the latter involves summations over the replica.} $n\to0$  as required by our gauge-fixing procedure, or set $n=1$ in order to compare with the CF model.

\section{Perturbation theory}\label{Perturbative_computations}

We compute the ghost and gluon propagators at one-loop order in perturbation theory. The calculations are lengthy but straightforward. Here, we solely display the relevant Feynman diagrams and detail one explicit example involving the superfield sector. The complete expressions of the diagrams are given in the supplemental material \cite{supp_mat}.

\subsection{Feynman rules}\label{feynman_rules}

For the purpose of the perturbative expansion, we introduce the replicated superfields $\Lambda^a_k(x,\underline\theta_k)$ as
\begin{equation}\label{super_field}
{\cal V}_k(x,\underline\theta_k)=\exp\left[ig_0 t^a\Lambda^a_k(x,\underline\theta_k) \right].
\end{equation}
Expanding the action \eqn{eq_action2} in powers of $\Lambda_k$ up to order $g_0^2$, we get the free propagators and the vertices relevant for one-loop computations. We work in momentum Euclidean space with the Fourier convention $\partial_\mu\to-ip_\mu$. Since the Grassmann spaces are curved, it is not useful to introduce associated Fourier variables.

The quadratic part of the action \eqn{eq_action2} couples the different fields, which leads to mixed propagators. In the $(A,c,\cb,h)$ sector, this yields the following expressions for the bare tree-level propagators \cite{Serreau:2013ila} (see also Appendix \ref{sec_complete_propagator}):
\begin{equation}
  \label{eq_propagAA}
   \left[ A^a_\mu(-p)\,A^b_\nu(p)\right]_{\!0}=\delta^{ab}\left(\frac{P_{\mu\nu}^T(p)}{p^2+n\beta_0}+\frac{\xi_0P_{\mu\nu}^L(p)}{p^2+\beta_0\xi_0}\right),
\end{equation}
\beq
   \label{eq_propagcc}
\left[ c^a(-p)\,\cb^b(p)\right]_{\!0}=\frac{\delta^{ab}}{p^2+\beta_0\xi_0},
\eeq
\beq
   \label{eq_propaghh}
 \left[ih^a(-p)ih^b(p)\right]_{\!0}=\frac{-\beta_0\delta^{ab}}{p^2+\beta_0\xi_0},
\eeq
and
\beq
   \label{eq_propagAh}
 \left[ih^a(-p)A_\mu^b(p)\right]_{\!0}=\frac{i\delta^{ab}p_\mu}{p^2+\beta_0\xi_0}.
\eeq
Here, $P_{\mu\nu}^L(p)=p_\mu p_\nu/p^2$,  $P_{\mu\nu}^T(p)=\delta_{\mu\nu}-p_\mu p_\nu/p^2$, and the square brackets with subscript $0$ denote an average with the quadratic part of the action \eqn{eq_action2}, with $n$ finite. The only reminiscence of the replica sector of the theory in the correlators \eqn{eq_propagAA}--\eqn{eq_propagAh} is through the square mass $n\beta_0$ appearing in the transverse part of the gluon propagator. For $n=1$ we recover the tree-level propagators of the CF model.

The correlator of the superfields $\Lambda_k$ reads 
\begin{equation}
  \label{eq_propagLL}
  \left[\Lambda^a_k(-p,\underline{\theta})\,\Lambda^b_l(p,\underline{\theta}')\right]_{\!0}=\delta^{ab}\!\left[\frac{\delta_{kl}\delta(\underline\theta,\underline\theta')}{p^2+\beta_0\xi_0}+\frac{\xi_0(1+\delta_{kl})}{p^2(p^2+\beta_0\xi_0)}\!\right]\!\!,
\end{equation}
where $\delta(\underline\theta,\underline\theta')=g^{-1/2}(\underline{\theta})\,(\thetab-\bar\theta ')(\theta-\theta')$
is the covariant Dirac delta function on the curved Grassmann space: 
$\int_{\underline{\theta}}\delta(\underline\theta,\underline\theta')f(\underline
{\theta})=f(\underline{\theta}')$. Notice that, for $\xi_0\neq 0$, there is a 
nontrivial correlation between different replica. Finally, there are nontrivial 
mixed correlators
\beq
  \label{eq_propaghL}
 \left[ih^a(-p)\Lambda_k^b(p,\underline{\theta})\right]_{\!0}=\frac{\delta^{ab}}{p^2+\beta_0\xi_0}
\eeq
and
\beq
  \label{eq_propagLA}
 \left[\Lambda_k^a(-p,\underline\theta)A_\mu^b(p)\right]_{\!0}=\frac{i\xi_0\delta^{ab}p_\mu}{p^2(p^2+\beta_0\xi_0)}.
\eeq

The interaction vertices are obtained from terms higher than quadratic in the fields. From \Eqn{eq_action2}, it appears clearly that the vertices of the sector ($A,c,\cb,h$) are identical to those of the CF model or, equivalently, to those of the CFDJ action. These include the Yang-Mills vertices with three and four gluons as well as the standard gluon-ghost-antighost vertex. In addition, there is a four-ghost vertex as well as a $hc\cb$ vertex, both proportional to $\xi_0$.  These vertices are well known \cite{deBoer:1995dh} and we do not recall their expressions here. The vertices of the replicated nonlinear sigma model sector are obtained by expanding the exponential \eqn{super_field} in powers of $\Lambda_k$. We refer the reader to Ref.~\cite{Serreau:2013ila} for details. For illustration, we simply recall the expression of one cubic vertex involving two superfields $\Lambda_k^a(p_1,\underline\theta)$, $\Lambda_l^b(p_2,\underline\theta')$ and one gauge field $A_\mu^c(p_3)$:
\beq
\label{eq:vertexLLA}
 i\frac {g_0}4f^{abc}\delta_{kl}(2\pi)^d\delta^{(d)}(p_1+p_2+p_3)\delta(\underline\ts,\underline\ts')(p_1-p_2)_\mu.
\eeq

\subsection{Two-point vertex functions at one-loop order}
\label{sec:twopt}

We are primarily interested in the ghost and gluon propagators. The latter are obtained from the two-point vertex functions, defined as the second derivatives of the effective action $\Gamma$ at fixed $n$:
\beq
\label{eq:defvertex}
 \Gamma_{XY}(p,\underline{\theta},\underline{\theta}')=\left.\frac{\delta_\theta^{(2)}\Gamma}{\delta X(p,\underline{\theta})\delta Y(-p,\underline{\theta}')}\right|_0,
\eeq
where $X$ and $Y$ denote any of the (super)fields, the subscript $0$ means that the derivative is evaluated at vanishing fields, and $\delta_\theta/\delta X$ is a covariant functional derivative \cite{Serreau:2012cg,Serreau:2013ila} which takes into account the curvature of the superspaces associated with each superfield. It is defined as  $\delta_\theta/\delta X(p)=\delta/\delta X(p)$ for normal fields and $\delta_\theta/\delta X(p,\underline{\theta})=g^{-1/2}(\underline{\theta})\delta/\delta X(p,\underline{\theta})$ for superfields. In particular, one has $\delta_\theta X(\underline{\theta})/\delta X(\underline{\theta'})=\delta(\underline{\theta},\underline{\theta}')$ for a given superfield $X$. 

All two-point vertex and correlation functions are diagonal in color space and we extract a trivial unit matrix from their expressions below. Spacetime symmetries imply the following general decompositions:
\beq
\label{eq:decompmunu}
\Gamma_{A_{\mu}A_{\nu}}(p)= P^T_{\mu \nu}(p)\Gamma_T(p)+P^L_{\mu \nu}(p)\Gamma_L (p)
\eeq
and
\beq
 \Gamma_{ihA_\mu}(p)=-\Gamma_{A_\mu ih}(p)= ip_\mu\Gamma_{ih A}(p),
\eeq
where the scalar functions $\Gamma_T(p)$, $\Gamma_L(p)$, and $\Gamma_{ihA}(p)$ only depend on $p^2$. 
For the superfield sector, the replica permutation symmetry and the isometries of the Grassmann subspaces associated to each replica imply the following general decompositions:
\begin{align}\label{grass_decomp}
\Gamma_{\Lambda_{k}\Lambda_{l}}(p,\underline{\theta},\underline{\theta}') &= \delta_{kl}\!\left[\Gamma_{1}(p) \delta(\underline{\theta},\underline{\theta}')+ \Gamma_2(p)\square_{\underline{\theta}}\delta(\underline{\theta},\underline{\theta}') \right]\nn
&+(\delta_{kl}-1)\Gamma_3(p)
\end{align}
and
\beq
\label{eq:decomplast}
 \Gamma_{\Lambda_{k}A_\mu}(p,\underline{\theta})=- ip_\mu \Gamma_4(p),
\eeq
where the scalar functions $\Gamma_{1,\ldots,4}(p)$ only depend on $p^2$ and where $\square_{\underline{\theta}}$ is the Laplace operator on the curved Grassmann space; see Appendix~\ref{sec_complete_propagator}. One has the identity $\square_{\underline{\theta}}\delta(\underline{\theta},\underline{\theta}')=-2+2\beta_0\delta(\underline{\theta},\underline{\theta}')$. The various components of the two-point vertex functions at tree level are given in Appendix~\ref{appsec:tree}.

The calculation of propagators from the two-point vertex functions is complicated by the fact that the fields $A_\mu$, $ih$ and $\Lambda_k$ are mixed at the quadratic level in the effective action. It is, in particular, the reason why we must take the limit $n\to0$ only after having inverted the matrix of vertex functions in field space. This is discussed in Appendix~\ref{sec_complete_propagator}. Because this mixing only involves the longitudinal component (with respect to momentum) of the gluon field, the ghost and transverse gluon propagator are given by the simple relations
 \begin{equation}\label{eq_ghost_propa}
G_{\rm gh}(p)=\lim_{n \rightarrow 0} \Gamma_{c\bar c}^{-1}(p)
\end{equation}
and
 \begin{equation}
G_T(p)=\lim_{n \rightarrow 0}  \Gamma_T^{-1}(p)\label{eq_gluon_propa},
\end{equation} 
where we have decomposed the gluon propagator $G_{\mu\nu}(p)$ as in \Eqn{eq:decompmunu}. In contrast, the longitudinal component of the gluon propagator involves various vertex functions in the sector $(A_\mu,ih,\Lambda_k)$ (see Appendix~\ref{sec_complete_propagator}):
\begin{widetext}
\beq
\label{eq:long}
G_L(p)=\lim_{n \rightarrow 0}  \frac{\Gamma_{ihih}\left(\Gamma_1+\beta_0\left(n-2\right)\Gamma_3\right)}{(\Gamma_L \Gamma_{ihih}-(\Gamma_{ihA})^2 p^2)\left(\Gamma_1+\beta_0\left(n-2\right)\Gamma_3 \right)-\Gamma_{ihih}p^2 \beta_0 (\Gamma_{4})^2(n-1)}.
\eeq 
 \end{widetext}

Note that in the case $n=1$, one recovers the standard expression of the longitudinal propagator in the CF model
\beq
\label{eq:CFlongprop}
 G_L^{\rm CF}(p)= \left.\frac{\Gamma_{ihih}(p)}{\Gamma_L(p) \Gamma_{ihih}(p)-p^2\left[\Gamma_{ihA}(p)\right]^2}\right|_{n=1}.
\eeq 
Using Slavnov-Taylor identities for the (non-nilpotent) Becchi-Rouet-Stora-Tyutin (BRST) symmetry of the CF model (see Appendix~\ref{sec:appST}), this can be rewritten as
\beq
\label{eq:gfgfgf}
 G_L^{\rm CF}(p)= -\left.\frac{\beta_0\Gamma_{ihih}(p)}{\Gamma_L(p) \Gamma_{c\bar c}(p)}\right|_{n=1}.
\eeq

\begin{figure}[t]
  \centering
  \includegraphics[width=.17\linewidth]{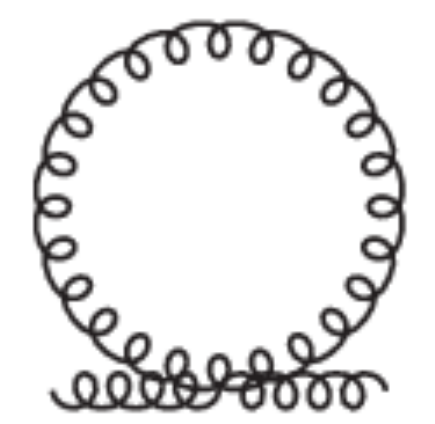}\quad
  \includegraphics[width=.27\linewidth]{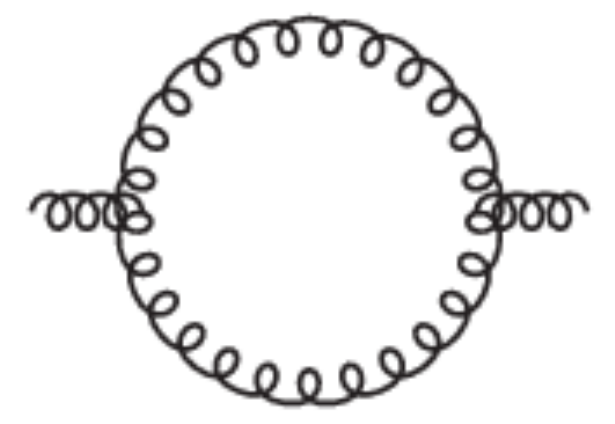}\quad
  \includegraphics[width=.27\linewidth]{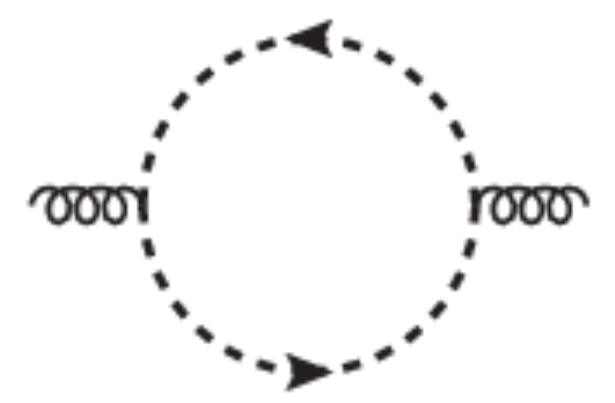}\\\vspace{.3cm}
  \includegraphics[width=.29\linewidth]{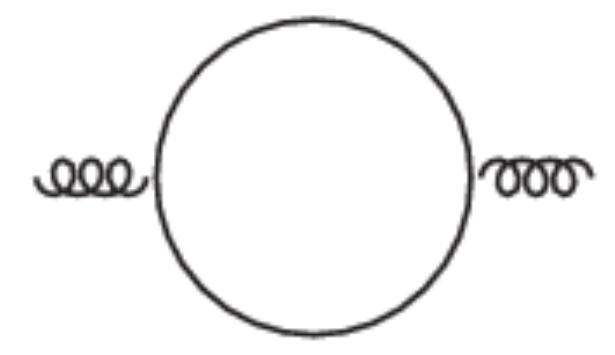}\quad
  \includegraphics[width=.27\linewidth]{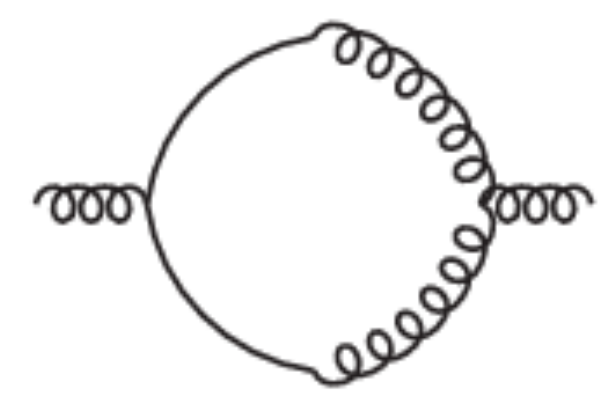}
  \caption{One-loop diagrams for the vertex $\Gamma_{AA}$. The diagrams of the first line are present in the Landau gauge and {\it a fortiori} in the CF model. The diagrams of the second line involve the superfield sector and are thus specific of the present gauge fixing (they are proportional to $n-1$ and thus vanish in the CF model). The first one is proportional to $\beta_0\xi_0$ and the second one to $\beta_0\xi_0^2$.}
  \label{fig_AA}
\end{figure}

\begin{figure}[ht]
  \centering
  \includegraphics[width=.17\linewidth]{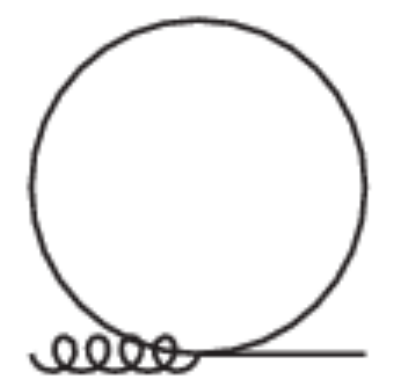}\quad
  \includegraphics[width=.27\linewidth]{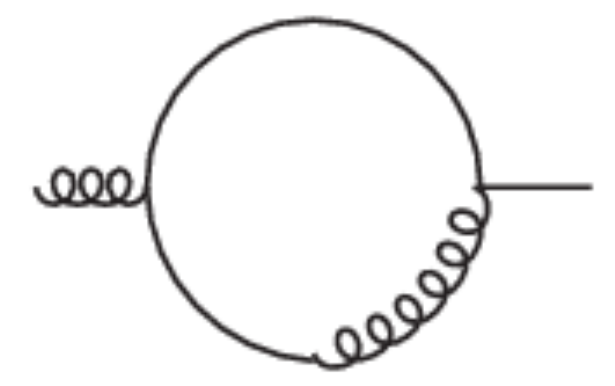}\quad
  \includegraphics[width=.27\linewidth]{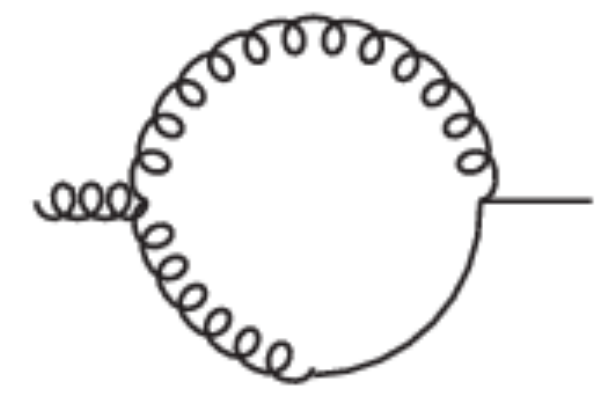}
  \caption{One-loop diagrams for the vertex $\Gamma_{A\Lambda}$.}
  \label{fig_A_lambda}
\end{figure}

We display in Figs.~\ref{fig_AA}\,--\,\ref{fig_A_lambda} the relevant Feynman diagrams for the one-loop calculations. We use the standard graphical conventions for the gluon (wiggly) and ghost (dashed) lines. The plain line represents the superfield correlator \eqn{eq_propagLL}. The second diagram on the second line involves the mixed $A$-$\Lambda$ correlator \eqn{eq_propagLA}. Double lines stand for the field $ih$.

 \begin{figure}[ht]
  \centering
  \includegraphics[width=.27\linewidth]{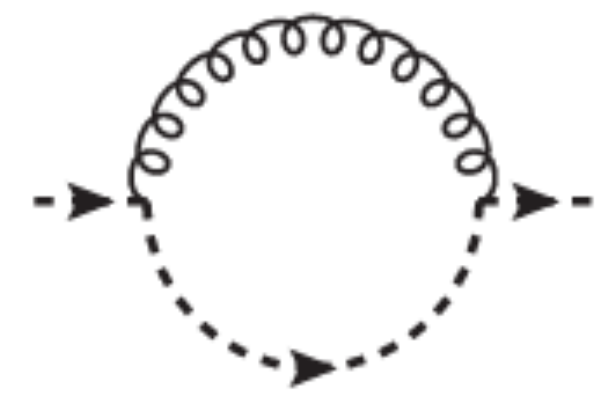}\quad
  \includegraphics[width=.17\linewidth]{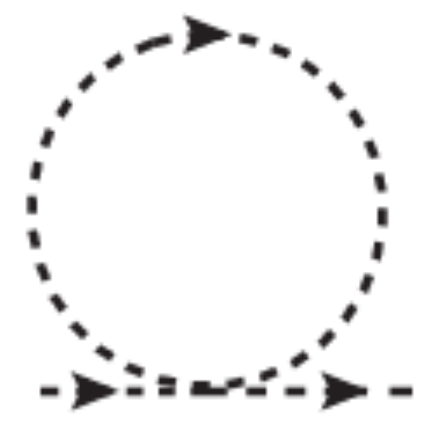}\quad
  \includegraphics[width=.27\linewidth]{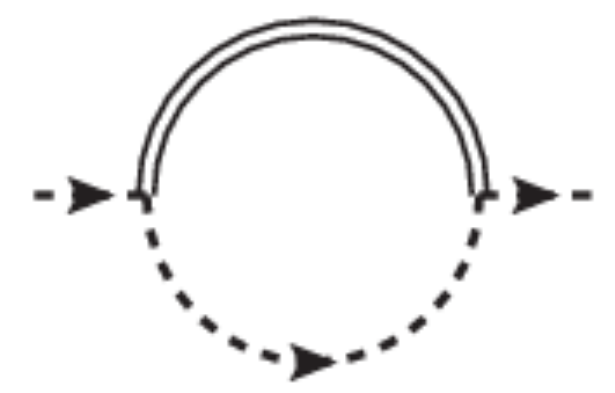}\vspace{.3cm}
  \includegraphics[width=.27\linewidth]{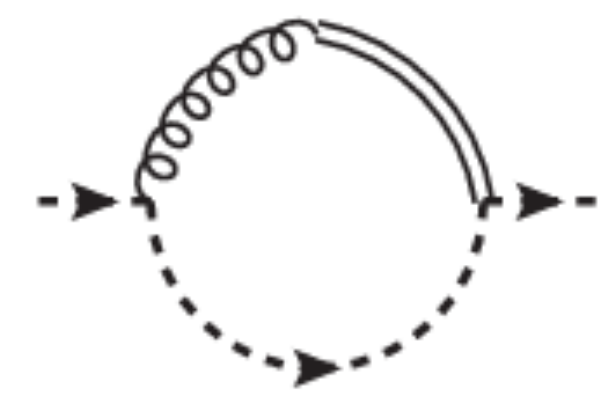}\quad
  \includegraphics[width=.27\linewidth]{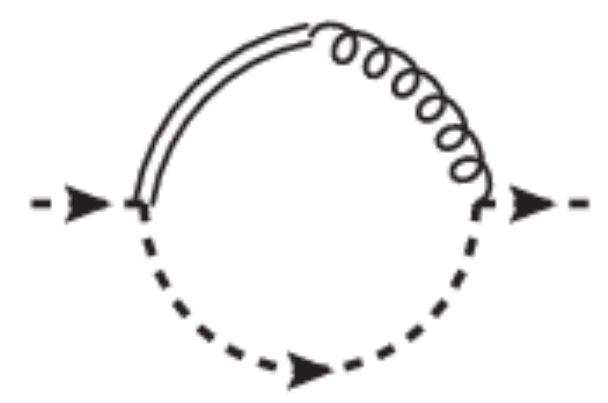} 
  \caption{One-loop diagrams for the vertex $\Gamma_{c\cb}$. Only the first diagram on the first line is present in the Landau gauge. All the others are present in the CF model. Note that the diagrams on the second line involve the mixed $ih$--$A$ correlator \eqn{eq_propagAh}. There is no diagram involving the superfields.}
  \label{fig_ccb}
\end{figure}
 
\begin{figure}[ht]
  \centering
 \includegraphics[width=.27\linewidth]{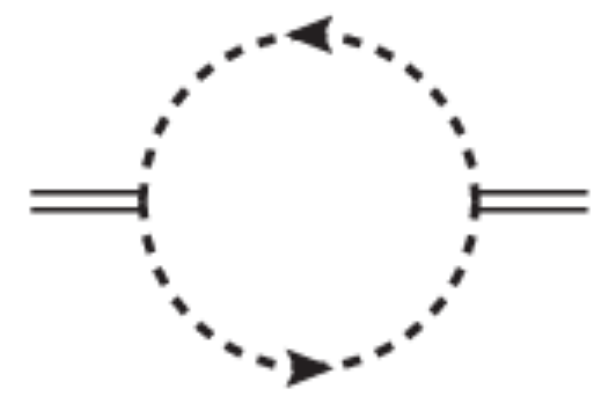}\qquad\qquad
  \includegraphics[width=.27\linewidth]{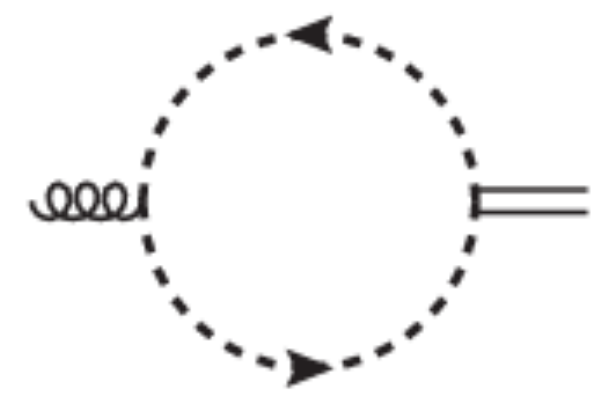}
  \caption{One-loop diagrams for the vertices $\Gamma_{ihih}$ (left) and $\Gamma_{ihA}$ (right). Both diagrams are present in the CF model. There is no diagram involving the superfields.}
  \label{fig_h_h}
\end{figure} 

\begin{figure}[ht]
  \centering
  \includegraphics[width=.17\linewidth]{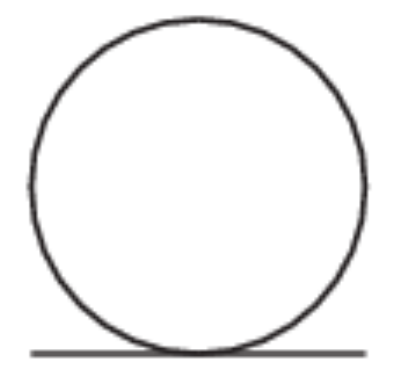}\,\,
  \includegraphics[width=.17\linewidth]{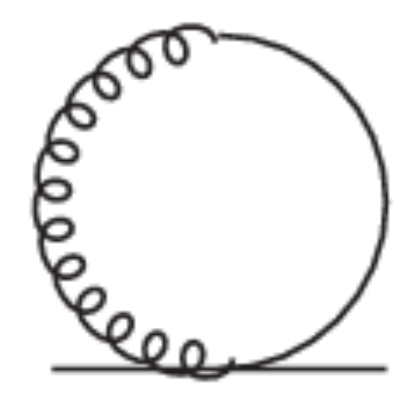}\,
  \includegraphics[width=.27\linewidth]{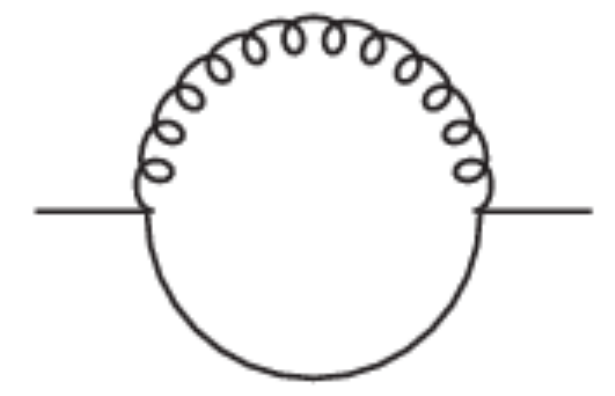}\,
  \includegraphics[width=.27\linewidth]{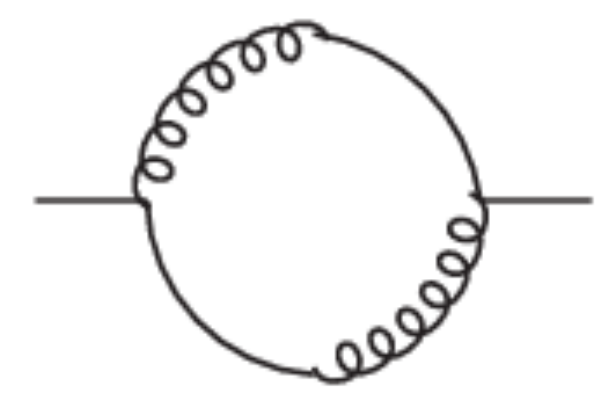}
  \caption{One-loop diagrams for the vertex $\Gamma_{\Lambda\Lambda}$.}
  \label{fig_lambda_lambda}
\end{figure}
 
The calculation is lengthy but straightforward. In order to illustrate a typical example involving the Grassmann subspaces, we treat explicitly the first diagram on the second line of \Fig{fig_AA}, which involves the vertex \Eqn{eq:vertexLLA} and a loop of superfields. According to the Feynman rules, its contribution to the gluon self energy is given by

\begin{equation}
\label{explicit_calc1}
\begin{split}
 \Gamma^{\Lambda-{\rm loop}}_{A^a_\mu A^b_\nu}(p)= \frac{g_0^2}{8} \!&\sum_{i,j=2}^{n} \int_k  \int_{\underline{\theta}_i,\underline{\theta}_j}\!   f^{acd}f^{bdc}( k_\mu +\ell_\mu )  ( k_\nu +\ell_\nu )  \\ 
 &\quad\times
 \left\lbrace \frac{\delta_{ij}\delta (\underline{\theta}_i,\underline{\theta}_j ) }{k^2+\beta_0 \xi_0} +\frac{(1+\delta_{ij})\xi_0}{k^2(k^2+\beta_0 \xi_0)} \right\rbrace  \\
&\quad \times\left\lbrace \frac{\delta_{ij}\delta(\underline{\theta}_i ,\underline{\theta}_j)}{\ell^2+\beta_0 \xi_0} +\frac{(1+\delta_{ij})\xi_0}{\,\ell^2(\ell^2+\beta_0 \xi_0)}\right\rbrace,
\end{split}
\end{equation}
where $a,b$ and $\mu,\nu$ are  respectively the color and Lorentz indices of the external gluon legs carrying momentum $p$ and $\ell_\mu =k_\mu-p_\mu$. We use the notation $\int_k=\mu^\epsilon\int \frac{d^dk}{(2 \pi)^d}$, with $d=4-\epsilon$, where the arbitrary scale $\mu$ is introduced for dimensional reasons. The replica indices $i,j$ are associated with the internal superfield lines.  

The Grassmannian integrals are trivially performed using the identities 
\beq
\int_{\underline{\theta}}\delta(\underline{\theta},\underline{\theta}')f(\underline{\theta})=f(\underline{\theta}')\,,\quad\delta\left(\underline{\theta}, \underline{\theta}\right)=0\quad{\rm and}\quad\int_{\underline{\theta}} 1  = \beta_0.
\eeq
After summing over the replica indices, we obtain the following momentum integral:
\begin{align}
\label{explicit_calc2}
 &\Gamma^{\Lambda-{\rm loop}}_{A^a_\mu A^d_\nu}(p)= -\delta^{ad}(n-1)\frac{g_0^2 N\beta_0 \xi_0}{8}\nn&\times \int_k  \frac{\left( k_\mu +\ell_\mu \right) \left( k_\nu +\ell_\nu \right)}{\left(k^2+\beta_0 \xi_0\right)\left(\ell^2+\beta_0 \xi_0\right)} \left(\frac{2}{k^2}+\frac{2}{\ell^2}+\frac{(n+2)\beta_0 \xi_0 }{k^2\ell^2} \right),
\end{align} 
which is logarithmically divergent in the ultraviolet (UV).
A similar calculation yields, for the second diagram on the second line of \Fig{fig_AA}, also involving the replica sector,
\begin{align}
\label{eq:mixed}
 \Gamma^{\rm mixed}_{A^a_\mu A^d_\nu}(p)&=-\delta^{ad}(n-1)\frac{g_0^2 N\beta_0 \xi_0^2}{d-1}P^T_{\mu\nu}(p)\nn&\times \int_k  \frac{k^2p^2-(k\cdot p)^2}{k^2\ell^2\left(k^2+\beta_0 \xi_0\right)\left(\ell^2+\beta_0 \xi_0\right)}.
\end{align}
It is transverse and UV finite.
As expected, both replica contributions are $\propto n-1$ and vanish identically in the CF model ($n=1$).

The calculation of the other diagrams of Figs.~\ref{fig_AA}\,--\,\ref{fig_A_lambda} goes along the same lines and we are left with one-loop momentum integrals, whose explicit expressions are given in the supplemented material \cite{supp_mat}. These integrals are UV divergent and can be regularized by standard techniques. We use dimensional regularization with $d=4-\epsilon$. It was shown in Ref.~\cite{Serreau:2013ila} that these divergences can be eliminated at all orders of perturbation theory by means of six independent renormalization factors, as we now recall.
 
\section{Renormalization}
\label{sec_renormalization}

We define renormalized fields in the standard way
\beq
A^{a\mu}=\sqrt{Z_A}\,A_R^{a\mu}\,,\quad c^a=\sqrt{Z_c}\,c_R^{a}\,,\quad \bar{c}^a=\sqrt{Z_c}\,\bar{c}_R^{a}\,,
\eeq
and
\beq
h^a=\sqrt{Z_{h}}h_R^{a}\,,\quad\Lambda^a=\sqrt{Z_\Lambda}\Lambda_R^{a}.
\eeq
In the following, we only refer to renormalized fields and correlators and we suppress the index $R$ for simplicity. A nontrivial issue concerns the interplay between renormalization and the limit $n\to0$. As pointed out in Refs.~\cite{Serreau:2012cg,Serreau:2013ila}, part of the explicit $n$ dependence of the bare theory can be absorbed in the definition of renormalized parameters. For instance, one observes from \Eqn{eq_propagAA} that the tree-level square mass of the transverse gluons is $n \beta_0$. Hence, defining renormalized gauge-fixing parameters $\beta$ and $\xi$ as $\beta_0={Z_\beta} \beta$ and $\xi_0={Z_\xi}\xi$ makes the (transverse) gluon massless in the limit $n \rightarrow 0$. In the case of the Landau gauge ($\xi_0 = 0$), the theory thus reduces to the standard Faddeev-Popov Lagrangian \cite{Serreau:2012cg}. Instead, one may define a renormalized squared mass parameter $m^2$ as $n \beta_0 =Z_{m^2} m^2$, such that the transverse gluon propagator remains massive in the limit $n\to0$. For $\xi_0=0$, it was shown in Ref.~\cite{Serreau:2012cg} that the resulting theory is perturbatively equivalent to the Landau limit of the CF model. As already mentioned, the latter admits infrared-safe renormalization-group trajectories \cite{Tissier_10} and provides a perturbative description of Yang-Mills correlation functions in good agreement with lattice results, both in the vacuum \cite{Tissier_10,Pelaez:2013cpa,Pelaez:2014mxa} and at finite temperature \cite{Reinosa:2013twa}. This motivates us to consider similar renormalization schemes in the case $\xi_0\neq0$.

In that case, however, the parameter $\beta_0$ does not only appear in the transverse gluon mass, but also in the combination $\beta_0\xi_0$ in the tree-level square masses of the longitudinal gluon and of the fields $(ih,c,\bar{c},\Lambda_k)$; see Eqs.~\eqn{eq_propagAA}--\eqn{eq_propagLA}. In order to keep this mass finite in the limit $n\to0$, we choose the following definitions of the renormalized mass and gauge-fixing parameters\footnote{This choice is arbitrary. One could also choose to have a diverging mass for these fields, in which case they would decouple and lead to a different phenomenology. Lattice data would be needed to determine the most relevant scenario. We stress that such definitions of the renormalization constants are consistent with the renormalizability of the theory, since the proof of renormalizability given in Ref.~\cite{Serreau:2013ila} holds for generic $n$.}:
\beq
\label{renormalization constant mass and gauge}
 n \beta_0 =Z_{m^2} m^2\quad{\rm and}\quad \frac{\xi_0}{n} = Z_\xi \,\xi,
\eeq
such that $\beta_0\xi_0=Z_\xi Z_{m^2}\xi m^2$. Important consequences of this choice are that the ghost propagator acquires a nonvanishing mass and that various correlators, e.g., which are proportional to $\xi_0$, vanish in the limit $n\to0$ at fixed $\xi$.  For instance, it is easy to show diagrammatically that $\Gamma_{ihih}(p)\propto \xi_0$ up to a function of the parameters $n\beta_0$ and $\beta_0\xi_0$. One thus has $\Gamma_{ihih}(p)\to0$ in the limit $n\to0$ at $n\beta_0$ and $\beta_0\xi_0$ fixed. The function $\Gamma_{ihA}(p)$ being finite in this limit, we conclude that the longitudinal gluon propagator \eqn{eq:long} vanishes identically
\beq
 G_L(p)=0.
\eeq
We stress that this is a consequence of the definitions \eqn{renormalization constant mass and gauge} and that it is independent of the set of renormalization prescriptions discussed below. This is a remarkable property of the present gauge fixing since an exactly transverse gluon propagator is usually expected to be a peculiar feature of the Landau gauge. In particular, the fact that the gluon propagator remains transverse at $\xi\neq0$ is a nontrivial effect of the average over Gribov copies. This is one of the main differences with the CF model, where the longitudinal gluon propagator is nonzero and is given by \Eqn{eq:CFlongprop}.

Finally, we define the renormalized coupling $g$ as 
\beq
\label{eq:couplingrenorm}
 g_0=Z_g\,g.
\eeq
It was shown in Ref.~\cite{Serreau:2013ila} that the combination $Z_{m^2}Z_c/Z_h$ is UV finite so that only six renormalization factors, out of the seven introduced so far, are needed to absorb all the UV divergences. The finite parts are to be determined by means of specific renormalization conditions. In the following subsections, we discuss two sets of prescriptions for the two-point functions and we use a generalization of the Taylor scheme for the coupling. We emphasize that all our prescriptions are set before taking the limit $n\rightarrow 0$, which is done only at the end of the computations.

As an illustration, let us consider the superfield contributions to the transverse gluon two-point vertex in the limit $n\to0$. We have, from \Eqn{explicit_calc2},
\begin{align}
\label{explicit_calc2-ren}
& \lim_{n\to0}\Gamma^{\Lambda-{\rm loop}}_{T}(p)=\frac{g^2 N\xi m^2 }{(d-1)}P^T_{\mu\nu}(p)\nn&\times \int_k  \frac{ k_\mu k_\nu}{\left(k^2+\xi m^2 \right)\left(\ell^2+\xi m^2 \right)}  \left(\frac{1}{k^2}+\frac{1}{\ell^2}+\frac{\xi m^2}{k^2\ell^2} \right),
\end{align} 
whereas the other superfield contribution \eqn{eq:mixed} is proportional to $\beta_0\xi_0^2\propto nm^2\xi^2$ and vanishes in the limit $n\to0$. We stress again that the contribution \eqn{explicit_calc2-ren} is absent in the CF model and is a direct effect of the average over Gribov copies. For instance, the contribution of the superfield sector to the transverse gluon two-point vertex at zero momentum is 
\begin{align}
\label{explicit_calc2-ren-zero}
 \lim_{n\to0}\Gamma^{\Lambda-{\rm loop}}_{T}(p=0)&=\frac{g^2 N\xi m^2}{d} \int_k  \frac{2k^2+\xi m^2}{k^2\!\left(k^2+\xi m^2\right)^2}\nn
 &=\frac{g^2 N\xi m^2}{32\pi^2}\left\{\frac{2}{\epsilon}+1+\ln\left(\frac{\bar \mu^2}{\xi m^2}\right)\right\},
\end{align} 
where we neglected terms ${\cal O}(\epsilon)$ in the last equality and $\bar \mu^2=4\pi e^{-\gamma}\mu^2$, where $\gamma$ is the Euler constant.

Another remarkable consequence of the prescriptions \eqn{renormalization constant mass and gauge} is that despite the presence of a nonzero gauge-fixing parameter $\xi$, the theory possesses some of the properties of the Landau gauge. The fact, already discussed, that the gluon propagator is exactly transverse is one example. But there are other similar features in the loop contributions. In particular, various one-loop diagrams which are $\propto\xi_0$ vanish in the limit $n\to0$, just like the superfield contribution \eqn{eq:mixed} to the gluon two-point vertex discussed above. For instance, it is easy to check that all but the first contributions to the ghost two-point vertex depicted in \Fig{fig_ccb} are proportional to $\beta_0\xi_0^2\propto nm^2\xi^2$ and thus vanish in this limit. The first diagram of \Fig{fig_ccb} is already present in the Landau gauge---with  the difference that, here, the internal ghost and gluon propagators are massive---and is easily shown to be proportional to $p^2$. We conclude that, in the limit $n\to0$, the ghost two-point vertex function at vanishing momentum does not receive correction at one loop, that is,
\beq
\label{eq:nonren1}
 \lim_{n\to0}\Gamma_{c\bar c}(p=0)=\left(\lim_{n\to0}Z_cZ_\xi Z_{m^2}\right)\xi  m^2.
\eeq
A similar argument shows that the one-loop diagrams of \Fig{fig_h_h} also vanish in the limit $n\to0$. More precisely, they are respectively proportional to $\xi_0^2$ and to $\xi_0$. It follows that the $h$ sector of the theory is not renormalized\footnote{We have checked this explicitly at one-loop order but we expect the general argument developed here to be valid at any loop order.}, i.e., 
\beq
\label{eq:nonren2}
 \lim_{n\to0}\Gamma_{ihA}(p)=\lim_{n\to0}\sqrt{Z_AZ_h} 
\eeq
and
\beq
\label{eq:nonren3}
 \lim_{n\to0}\frac{\Gamma_{ihih}(p)}{n\xi}=-\lim_{n\to0}Z_hZ_\xi.
\eeq
The above relations imply that the products $Z_cZ_\xi Z_{m^2}$, $Z_AZ_h$ and $Z_hZ_\xi$ are UV finite in the limit $n\to0$.

 Finally, we employ the same renormalization prescriptions as described above for the CF model. In that case, $n=1$ and there is no issue with the $n$ dependence of the renormalized parameters. Note also that the nonrenormalization relations \eqn{eq:nonren1}--\eqn{eq:nonren3} do not hold in that case. Instead a relation which is known to hold at all orders in perturbation theory in the CF model is that the combination $(Z_\xi^2 Z_c Z_{m^2}/Z_A)_{n=1}$ is finite \cite{Browne:2006uy,Wschebor:2007vh,Tissier_08}.

\subsection{Zero-momentum scheme}\label{sub_sec_zeromomenta}
In this first renormalization scheme, we define the two renormalized masses from the transverse gluon and the ghost two-point vertex functions at zero momentum, that is,
\beq
\label{Rs2 renormalization presciption 1}
\Gamma_{T}(0)=m^2 \quad{\rm and}\quad\Gamma_{c\bar c}(0)=\xi m^2.
\eeq
In order to fix the remaining renormalization factors, we further require that the following vertex functions assume their renormalized tree-level expressions at a scale $\mu^2$:
\begin{eqnarray}
\Gamma_{T}(\mu)&=&m^2+\mu^2 \label{Rs2 renormalization presciption 2},\\
\Gamma_{c\bar c}(\mu)&=&\xi m^2+\mu^2, \\
\Gamma_{1}(\mu)&=&\mu^2, \\
\Gamma_{ih ih}(\mu)&=&-n\xi \label{Rs2 renormalization presciption ih}.
\end{eqnarray}
This leads to the following expressions, defining $\delta Z_\alpha=Z_\alpha-1$:
\begin{eqnarray}
\label{eq:gfgf}
\delta Z_{A}&=&-\frac{\Pi_{T}(\mu)-\Pi_{T}(0)}{\mu^2},\\
\delta Z_{m^2}&=&-\frac{\Pi_{T}(\mu)}{m^2}-\delta Z_{A}\left(1+\frac{\mu^2}{m^2} \right),\\
\delta Z_{c}&=&-\frac{\Pi_{c\bar c}(\mu)-\Pi_{c\bar c}(0)}{\mu^2},\\
\label{eq:ieour}
\delta Z_{\xi}&=&-\frac{\Pi_{c\bar c}(0)}{\xi m^2}-\delta Z_{c}-\delta Z_{m^2},\\
\delta Z_{\Lambda}&=&-\frac{\Pi_{1}(\mu)}{\mu^2},\\
\label{eq:gfgfhh}
\delta Z_{h}&=&\frac{\Pi_{ih ih}(\mu)}{n\xi}-\delta Z_{\xi} ,
\end{eqnarray}
where the various functions $\Pi_T(p)$, $\Pi_{c\bar c}(p)$, etc. denote the loop diagram contributions to the corresponding (renormalized) two-point vertices with the same index.\footnote{For instance we have $\Gamma_T(p)=Z_A(p^2+Z_{m^2}m^2)+\Pi_T(p)$.}  
Extracting the divergent parts $Z_\alpha^{\rm div}\propto1/\epsilon$ and denoting $\kappa = g^2 N/8 \pi^2 \epsilon$, we get \cite{Serreau:2013ila} 
\begin{align}
\label{eq:okzxfm}
 \delta Z^{\rm div}_A&=\left(\frac{13}{6}-\frac{n\xi}{2}\right)\!\kappa\,\,\,{  \to\frac{13}{6}\kappa},\\
  \delta Z^{\rm div}_c&=\left(\frac{3}{4}-\frac{n\xi}{4}\right)\!\kappa\,\,\,{  \to\frac{3}{4}\kappa},\\
  \delta Z^{\rm div}_{m^2}&=\left(-\frac{35}{12}+\frac{n\xi}{4}\right)\!\kappa\,\,\,{  \to-\frac{35}{12}\kappa},\\
  \delta Z^{\rm div}_\xi&=\left(\frac{13}{6}-\frac{n\xi}{4}\right)\!\kappa\,\,\,{  \to\frac{13}{6}\kappa},\\
  \delta Z^{\rm div}_\Lambda&=\left(\frac{3}{4}-\frac{n\xi}{12}\right)\!\kappa\,\,\,{  \to\frac{3}{4}\kappa},\\
\label{eq:oisjdgf}
   \delta Z^{\rm div}_h&=-\frac{13}{6}\kappa,
\end{align} 
where we indicated the various $n\to0$ limits.
We check that the nonrenormalization relations mentioned below Eqs.~\eqn{eq:nonren1}--\eqn{eq:nonren3} are satisfied. We also recover the Landau gauge results of Ref.~\cite{Serreau:2012cg} in this limit. In particular, we have $\delta Z^{\rm div}_\Lambda\to\delta Z^{\rm div}_c$. For $n=1$, we check that the above divergent parts coincide with the known ones of the CF model \cite{deBoer:1995dh,Gracey:2002yt,Serreau:2013ila}, except of course for $Z_\Lambda$.

As for the finite parts of the renormalization factors, note that, in the case $n\to0$, the relation \eqn{eq:nonren1} combined with the second prescription in \Eqn{Rs2 renormalization presciption 1} implies that
\beq
\label{eq:jbsoin1}
 \lim_{n\to0} Z_cZ_\xi Z_{m^2}=1.
\eeq
Similarly, the relation \eqn{eq:nonren3} and the prescription \eqn{Rs2 renormalization presciption ih} yield
\begin{align}
\label{eq:jbsoin2}
 \lim_{n\to0}Z_hZ_\xi=1.
\end{align}
For later use, we also note that the relation \eqn{eq:nonren1} is equivalent to $ \lim_{n\to0}\Pi_{c\bar c}(0)=0$ and thus implies that
\beq
\label{eq:jbsoin3}
 \lim_{n\to0}Z_c=1-\left.\frac{\Pi_{c\bar c}(\mu)}{\mu^2}\right|_{n=0}.
\eeq
We also check from \Eqn{eq:ieour} that \Eqn{eq:jbsoin1} is indeed satisfied at one-loop order. Similarly, the relation \eqn{eq:nonren3} is equivalent to $ \lim_{n\to0}\Pi_{ihih}(p)/(n\xi)=0$ and we check, from the prescription \eqn{eq:gfgfhh}, that \Eqn{eq:jbsoin2} is indeed satisfied.

\subsection{Infrared-safe scheme}\label{sub_sec_IRsafe}

Our second set of renormalization prescriptions is inspired from the infrared-safe scheme put forward in Ref.~\cite{Tissier_10}, that we adapt to the case $\xi \neq 0$. As we shall see below, this also leads to infrared-safe RG trajectories in the present case.
In this scheme, all renormalization prescriptions are defined at the scale $\mu$. We keep the set of prescriptions \Eqn{Rs2 renormalization presciption 2}-\eqn{Rs2 renormalization presciption ih} of the previous scheme, and replace the prescriptions \eqn{Rs2 renormalization presciption 1} at zero momentum by
\begin{eqnarray}
\label{IRs renormalization presciption Zm}
\Gamma_{ih A}(\mu)=1  \label{IRs renormalization presciption ih A} & \mbox{and} &Z_\xi^2 Z_c Z_{m^2}=Z_A .
\end{eqnarray}
This leads to similar expressions as Eqs. \eqn{eq:gfgf}--\eqn{eq:gfgfhh}, which we do not write explicitly. The divergent parts of the counterterms at one-loop order are the same as before [see Eqs.~\eqn{eq:okzxfm}--\eqn{eq:oisjdgf}], as required for renormalizability.
As already mentioned, in the CF model ($n=1$), the second relation in \Eqn{IRs renormalization presciption Zm} is true for the divergent parts at all orders of perturbation theory \cite{Browne:2006uy,Wschebor:2007vh,Tissier_08}. For arbitrary $n$, this relation is satisfied by the divergent parts at one-loop order, as can be checked from Eqs.~\eqn{eq:okzxfm}--\eqn{eq:oisjdgf}. Here, we use this relation to define the finite parts of the renormalization factors. We stress that although we do not know whether this relation is compatible with renormalizability---that is, whether it is satisfied by the divergent parts---at higher orders (except for $n=1$), it is always possible to use it to define the finite parts.\footnote{In general, one can always replace the second relation in \Eqn{IRs renormalization presciption Zm} by $Z_\xi^2Z_c Z_{m^2}=Z_A \hat{Z}$, with $\hat{Z}=1+{\cal O}(g^4)$ chosen such that $Z_{m^2}$, $Z_A$, $Z_c$, and $Z_\xi$ have the correct divergent parts.}
Obviously, in this renormalization scheme, the inverse gluon and ghost propagators at zero momentum are not directly given by the parameters $m^2$ and $\xi m^2$ anymore.

Finally, we note that, in the limit $n\to0$, the relation \eqn{eq:jbsoin2}, which follows from Eqs.~\eqn{eq:nonren3} and \eqn{Rs2 renormalization presciption ih}, remains valid. In the present scheme, the relation \eqn{eq:nonren2} combined with the first prescription in \Eqn{IRs renormalization presciption Zm} implies that
\beq
\label{eq_id2}
 \lim_{n\to0}Z_hZ_A=1.
\eeq
Combining this with \Eqn{eq:jbsoin2} and with the second prescription in \Eqn{IRs renormalization presciption Zm}, we obtain that $\lim_{n\to0}Z_A/Z_\xi=1$ and, hence, that the relation \eqn{eq:jbsoin1} is also valid in the infrared-safe scheme.\footnote{We mention that the combination of the relations derived here yield $\lim_{n\to0}Z_AZ_cZ_{m^2}=1$, which corresponds to the infrared-safe renormalization prescription proposed in the case of the Landau gauge in Refs.~\cite{Tissier_10,Serreau:2012cg}.}

The relations \eqn{eq:nonren2} and \eqn{eq:nonren3} also imply that the renormalization factor $Z_c$ is identical in both schemes. Indeed, in the present scheme we have
\beq
\label{eq:pkpojk}
 \delta Z_c =- \frac{\Pi_{c\bar{c}}(\mu)}{\mu^2}+\frac{\xi m^2}{\mu^2} \left[ \frac{\Pi_{ih ih}(\mu)}{\xi n} +2\Pi_{ih A}(\mu)\right]\\
\eeq
The relations \eqn{eq:nonren2} and \eqn{eq:nonren3} imply $\lim_{n\to0}\Pi_{ihA}(p)=\lim_{n\to0}\Pi_{ihih}(p)=0$ and \Eqn{eq:jbsoin3} holds here too. The (combination of) renormalization factors $Z_c$ and $Z_cZ_\xi Z_{m^2}$---the only ones which enter the calculation of the ghost propagator at one-loop order---are thus identical (in the limit $n\to0$) in the two renormalization schemes considered here. We conclude that the latter, as obtained from strict perturbation theory at one-loop order, is identically the same in the two schemes provided one uses the same value for the parameters $m^2$, $\xi$, and $g$. Strictly speaking, this remark does not hold when RG improvement is taken into account since the running of the various parameters are different in the two schemes.

\subsection{Renormalization of the coupling constant in the Taylor scheme}

\begin{figure}[ht]
  \centering
  \includegraphics[width=.3\linewidth]{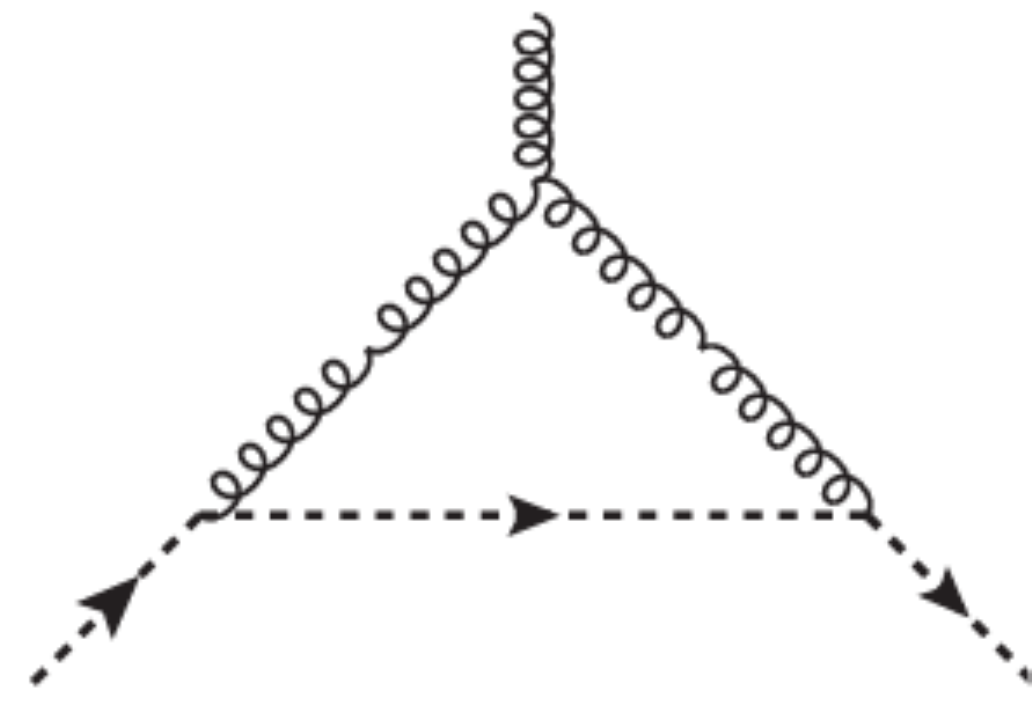}\quad
  \includegraphics[width=.3\linewidth]{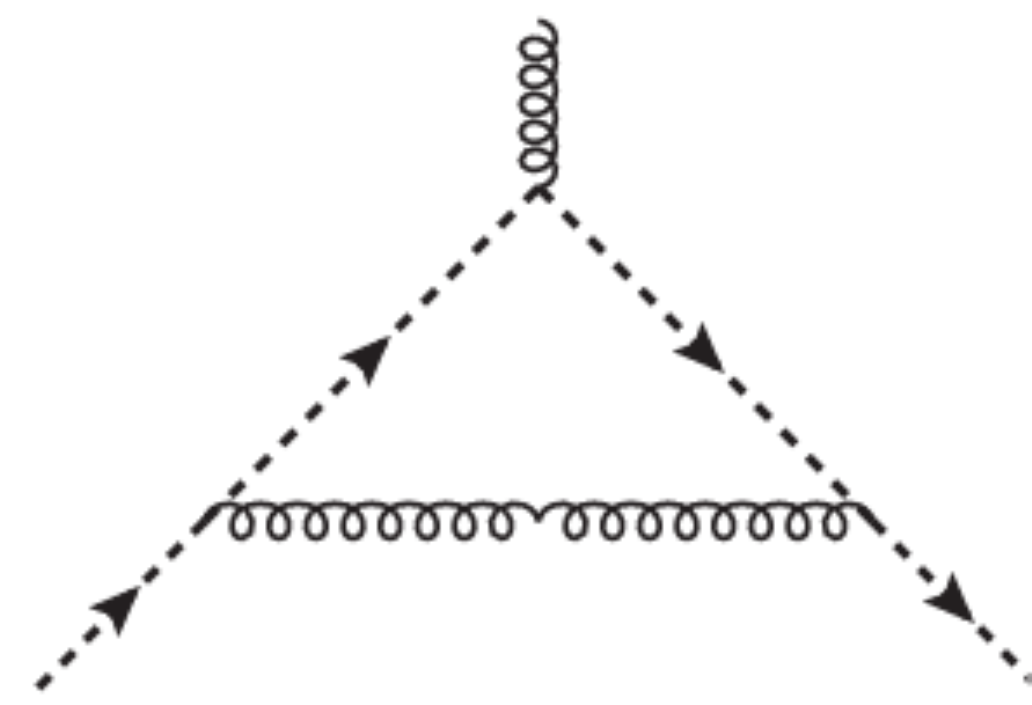}\\
  \vspace{.2cm}
  \includegraphics[width=.12\linewidth]{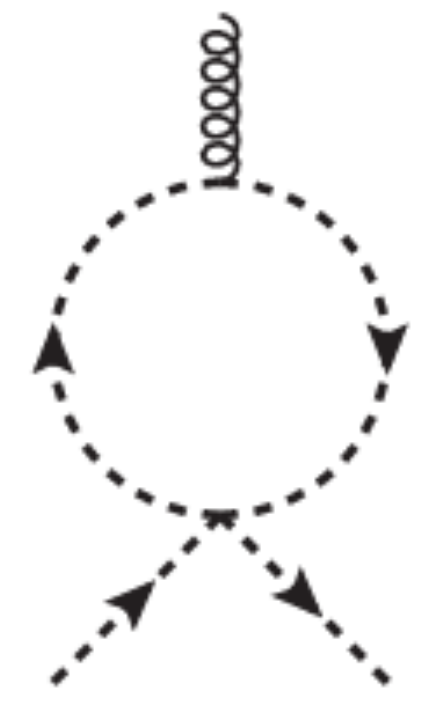}
  \includegraphics[width=.28\linewidth]{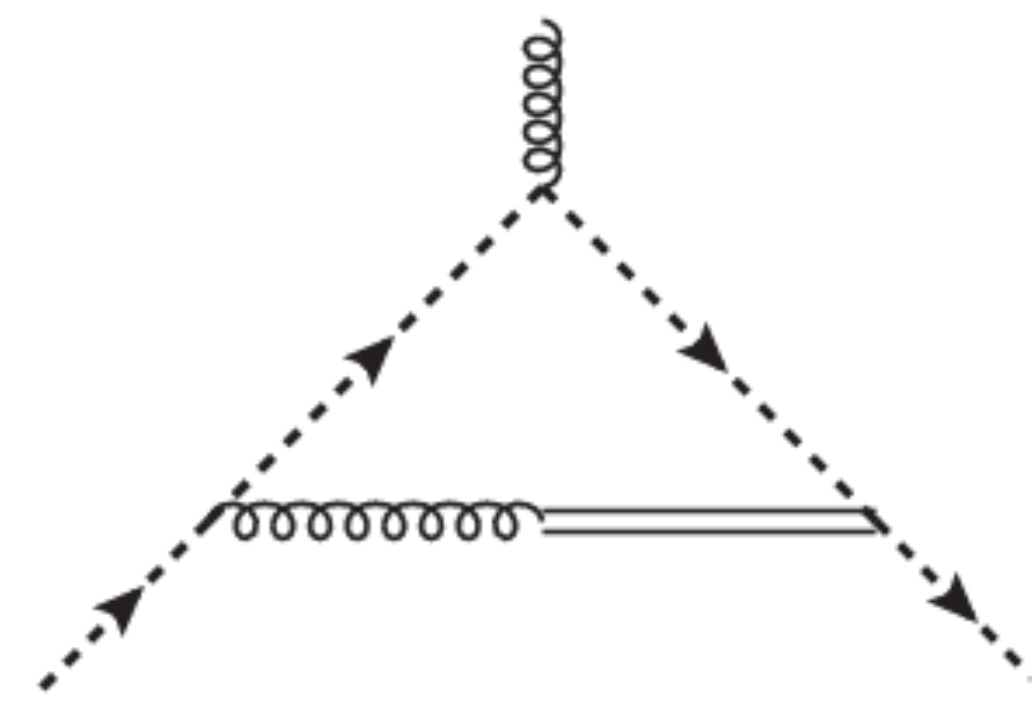}
    \includegraphics[width=.28\linewidth]{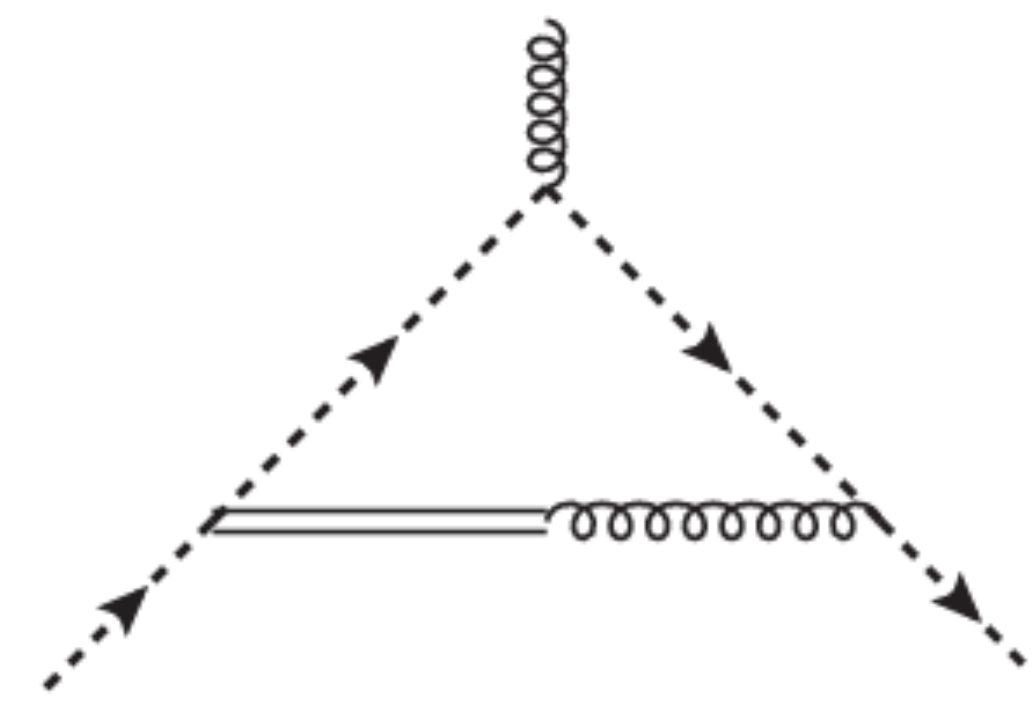}
    \includegraphics[width=.28\linewidth]{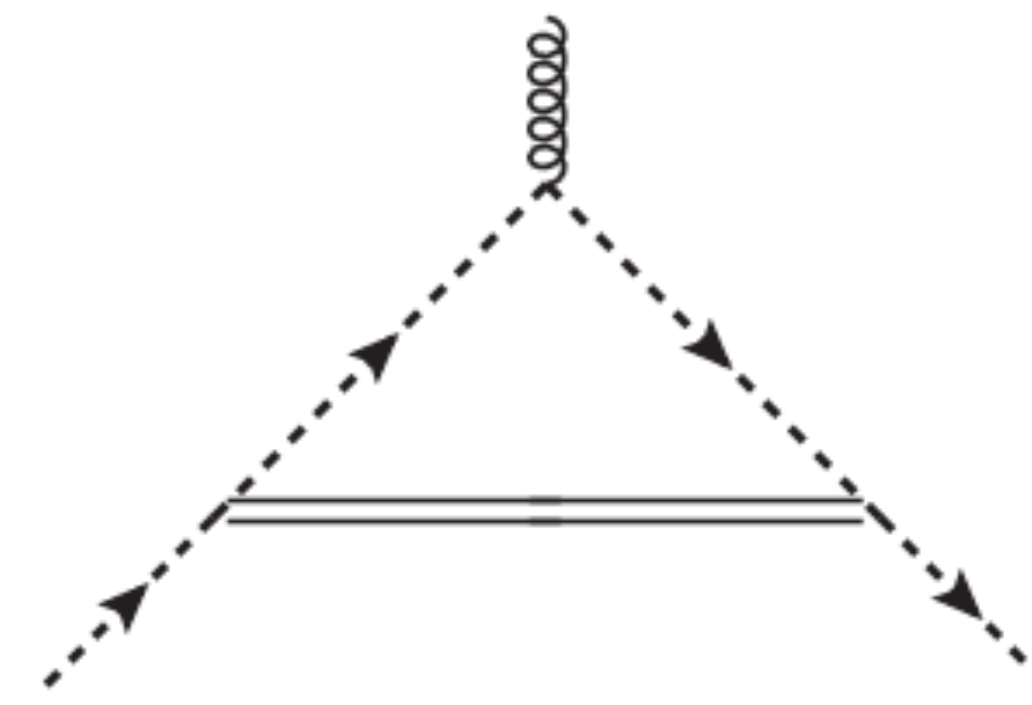}\\
  \vspace{.2cm}
  \includegraphics[width=.3\linewidth]{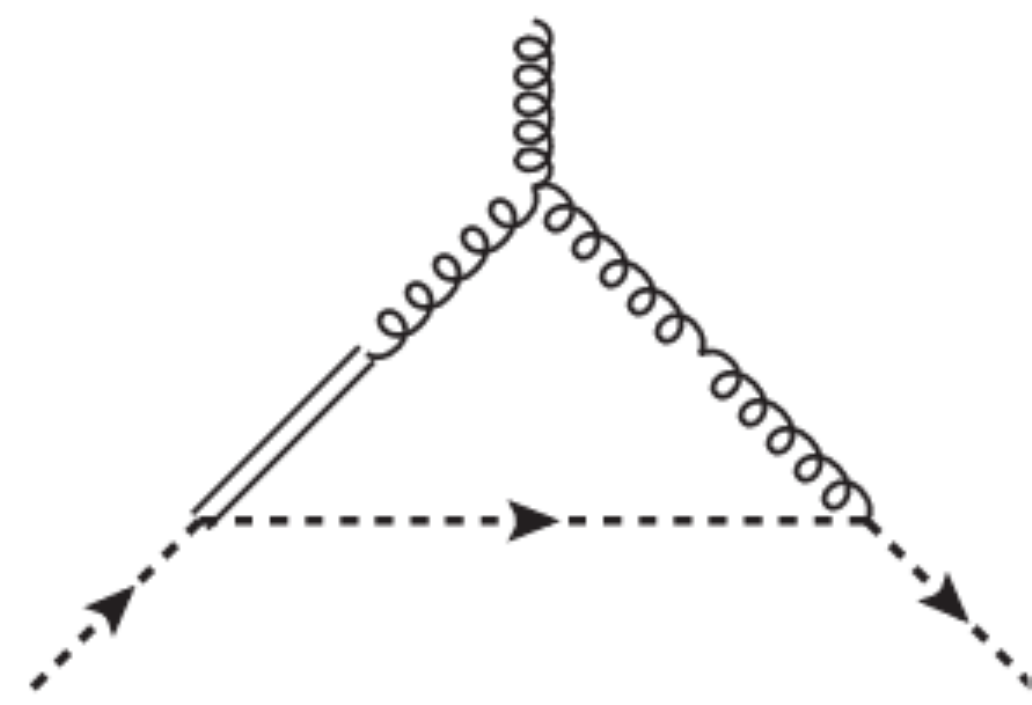}\,\,
  \includegraphics[width=.3\linewidth]{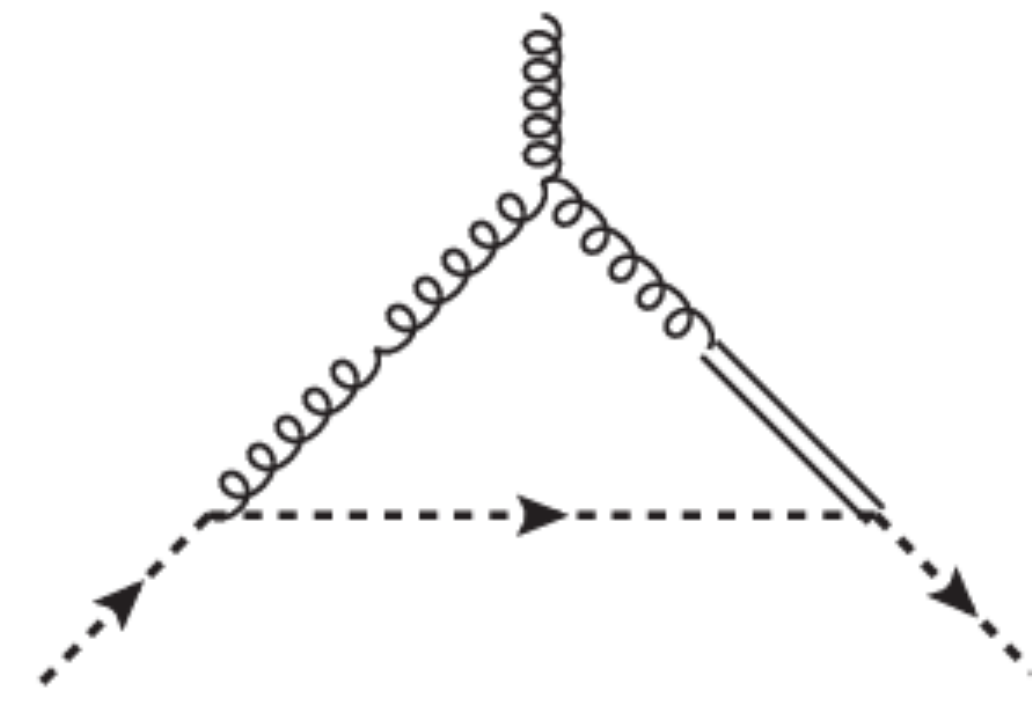}\,\,
  \includegraphics[width=.3\linewidth]{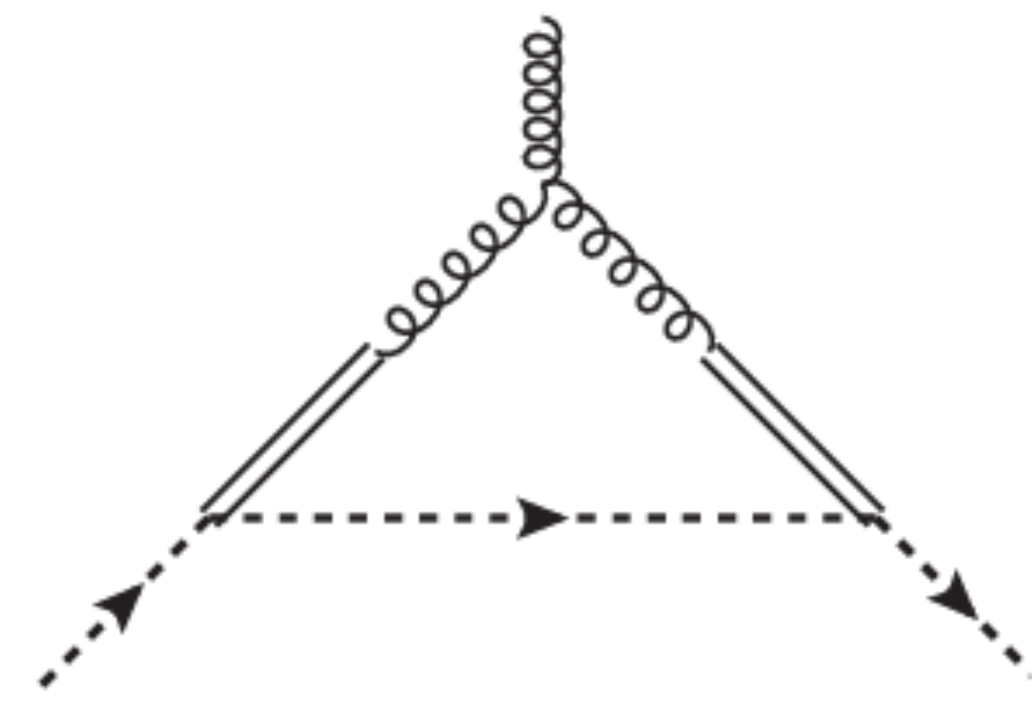}\\
  \vspace{.2cm}
  \includegraphics[width=.3\linewidth]{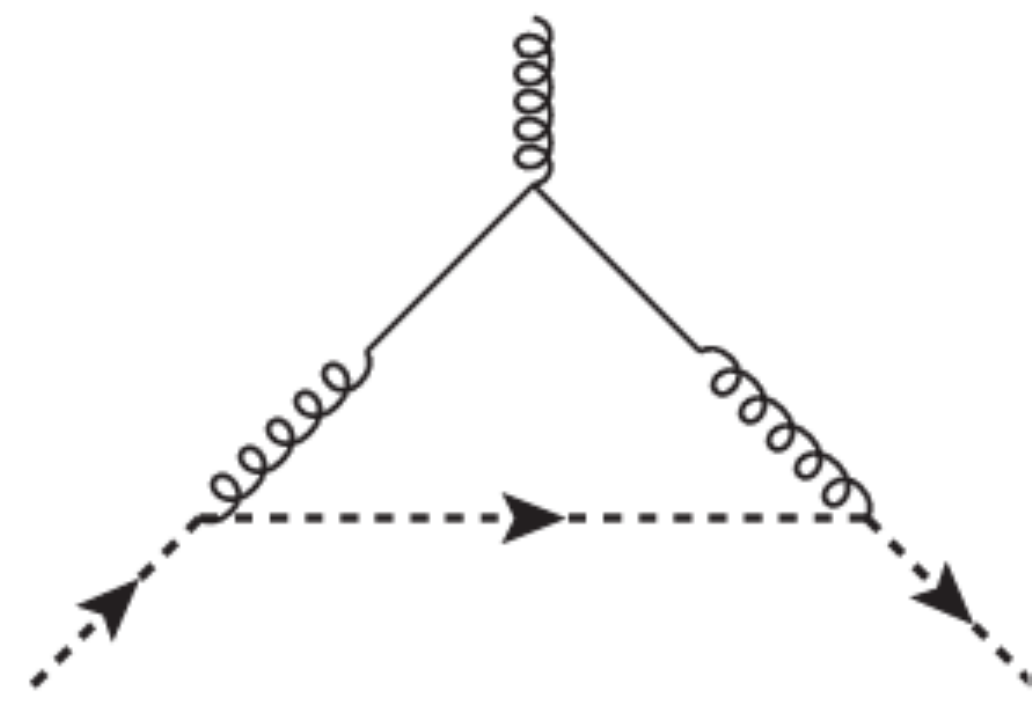}
  \caption{One-loop contributions to the ghost-antighost-gluon vertex function $\Gamma_{Ac\bar{c}}$. The diagrams on the first line are present in the standard Landau gauge---here with massive gluon and ghost propagators---and are proportional to the antighost external momentum. The topologies on the second and third lines are present in the standard CFDJ gauge and give contributions proportional to $\xi_0$. The replicated superfield sector only contributes in the last diagram, which is proportional to $\xi_0$ and to the antighost external momentum.}
  \label{fig_vertex}
\end{figure}

 We follow the general strategy of Ref. \cite{Taylor:1971ff} to renormalize the coupling constant. We fix the value of $Z_g$ [\Eqn{eq:couplingrenorm}] from the ghost-antighost-gluon vertex at vanishing ghost momentum, $\Gamma_{Ac\bar{c}}(p,0,-p)$. The Feynman diagrams contributing to  $\Gamma_{Ac\bar{c}}$ at one-loop order are shown in \Fig{fig_vertex}.
The two usual Yang-Mills diagrams (first line of \Fig{fig_vertex}) involve, at the vertex where the external ghost leg is attached, a Lorentz contraction between a gluon propagator and the (internal) antighost momentum. For vanishing external ghost momentum, both the antighost and the gluon propagators in the loop carry the same momentum (up to a sign). It follows that only the longitudinal gluon propagator  contributes to these diagrams which are, thus, proportional to $\xi_0\propto n\xi$---up to a function of $\beta_0\xi_0=Z_\xi Z_{m^2}\xi m^2$ [see \Eqn{eq_propagAA}]---and vanish in the limit $n\to0$. It is easy to check that all the other diagrams are also proportional to $\xi_0\propto n\xi$ up to a function of $\beta_0\xi_0$ and also vanish in this limit. We conclude that the vertex $\Gamma_{Ac\bar{c}}(p,0,-p)$ does not receive any radiative correction at one-loop order in the limit $n\to0$. The divergent parts of the renormalization factors thus satisfy the relation
\begin{equation}\label{eq_taylor}
\lim_{n \to 0} Z_g Z_c\sqrt{Z_A} =1.
\end{equation}
Defining the renormalized coupling constant from the vertex $\Gamma_{Ac\bar{c}}(p,0-p)$, this relation holds to the finite parts as well, which completely fix the factor $Z_g$. Unlike in the Landau gauge, where the relation \eqn{eq_taylor} holds to all orders \cite{Taylor:1971ff}, in the present case we have only checked that this is so at one-loop order. In particular, we do not know whether \Eqn{eq_taylor} is compatible with the renormalizability of the theory in general. But, as in the case of the second relation \eqn{IRs renormalization presciption Zm}, we can always impose the prescription \Eqn{eq_taylor} for the finite parts only. As discussed in Ref. \cite{Serreau:2013ila}, the explicit calculation of the divergent part of $Z_g$ at one-loop order shows that the latter is both $n$ and $\xi$ independent and thus coincides with the standard result 
\beq
\label{eq:cccccc}
 \delta Z_g^{\rm div}=-\frac{11}{6}\kappa,
\eeq
yielding the usual universal one-loop beta function, as it should. More generally, we show in Appendix~\ref{appsec:UV} that the present model reproduces the known one-loop UV behavior of Yang-Mills theories in the Landau gauge.

The identity \eqn{eq_taylor} is valid only in the limit $n \to 0$ with the definitions \eqn{renormalization constant mass and gauge}. It does not hold in the CF case ($n=1$), where the contributions to the ghost-antighost-gluon vertex proportional to $\xi_0$ do not vanish. For the purpose of comparing the $n\to 0$ and $n=1$ results, we employ a renormalization scheme for $n=1$ as close as possible to that used for $n\to0$. It is based on a nonrenormalization relation \cite{Tissier_08}\footnote{The definition of the renormalization factor $Z_\xi$ used here differs from that of Ref. \cite{Tissier_08}.}
\begin{equation}
\label{eq_coupling_CF}
Z_g Z_c Z_\xi^2=Z_A^{3/2},
\end{equation}
which is valid for the divergent parts. As discussed above, we extend this equality to the finite parts, which gives us a definition of the renormalization coefficient $Z_g$. It can be checked that this definition coincides with \Eqn{eq_taylor} in the limit $\xi\to0$.

\section{One-loop results}

\label{sec_perturbativ_results}

We present our results for the gluon and the ghost propagators at one-loop order in the SU($3$) theory\footnote{Results for SU($2$) are qualitatively similar.} within the two renormalization schemes discussed above. 
In the following, we set the scale $\mu=1$~GeV and we use the values of the parameters $m$ and $g$ which provide the best fits to the lattice results in the Landau gauge in Ref.~\cite{Tissier_10}, that is, $m=0.54$~GeV and $g=4.9$ in the zero-momentum scheme and $m=0.39$~GeV and $g=3.7$ in the infrared-safe scheme. Notice that the relevant expansion parameter is $3g^2/(16\pi^2)\lesssim1$. We have no reason, {\it a priori}, to exclude a dependence of the parameters $m$ and $g$ at a certain scale with the value of $\xi$ at the same scale. Such dependences would have to be inferred from fits to lattice data. In the absence of such data, we assume fixed values of $m$ and $g$ adjusted from lattice data at $\xi=0$.

\subsection{Zero-momentum scheme}
\begin{figure}[ht]
  \centering
  \includegraphics[width=1\linewidth]{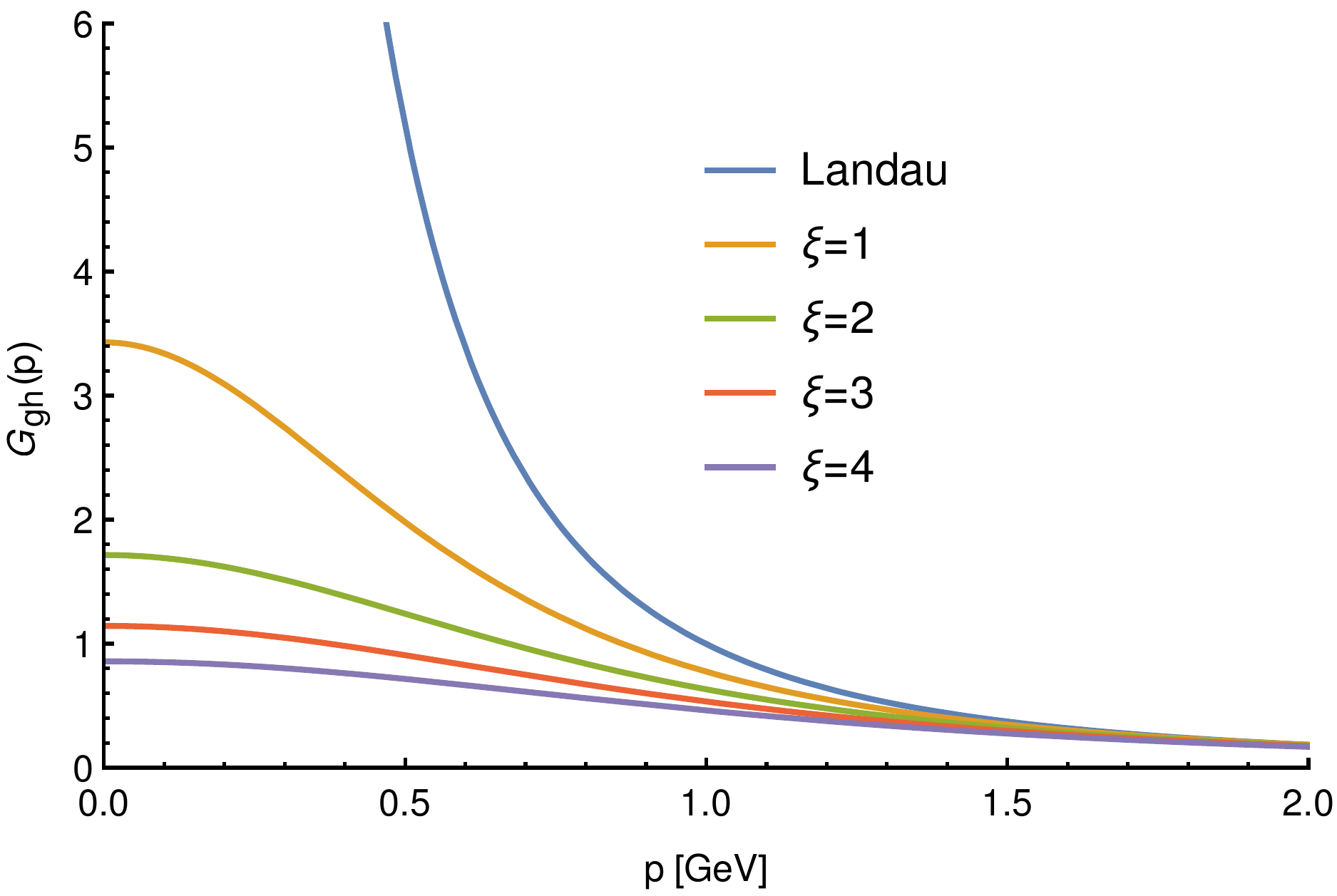}\\
  \includegraphics[width=1\linewidth]{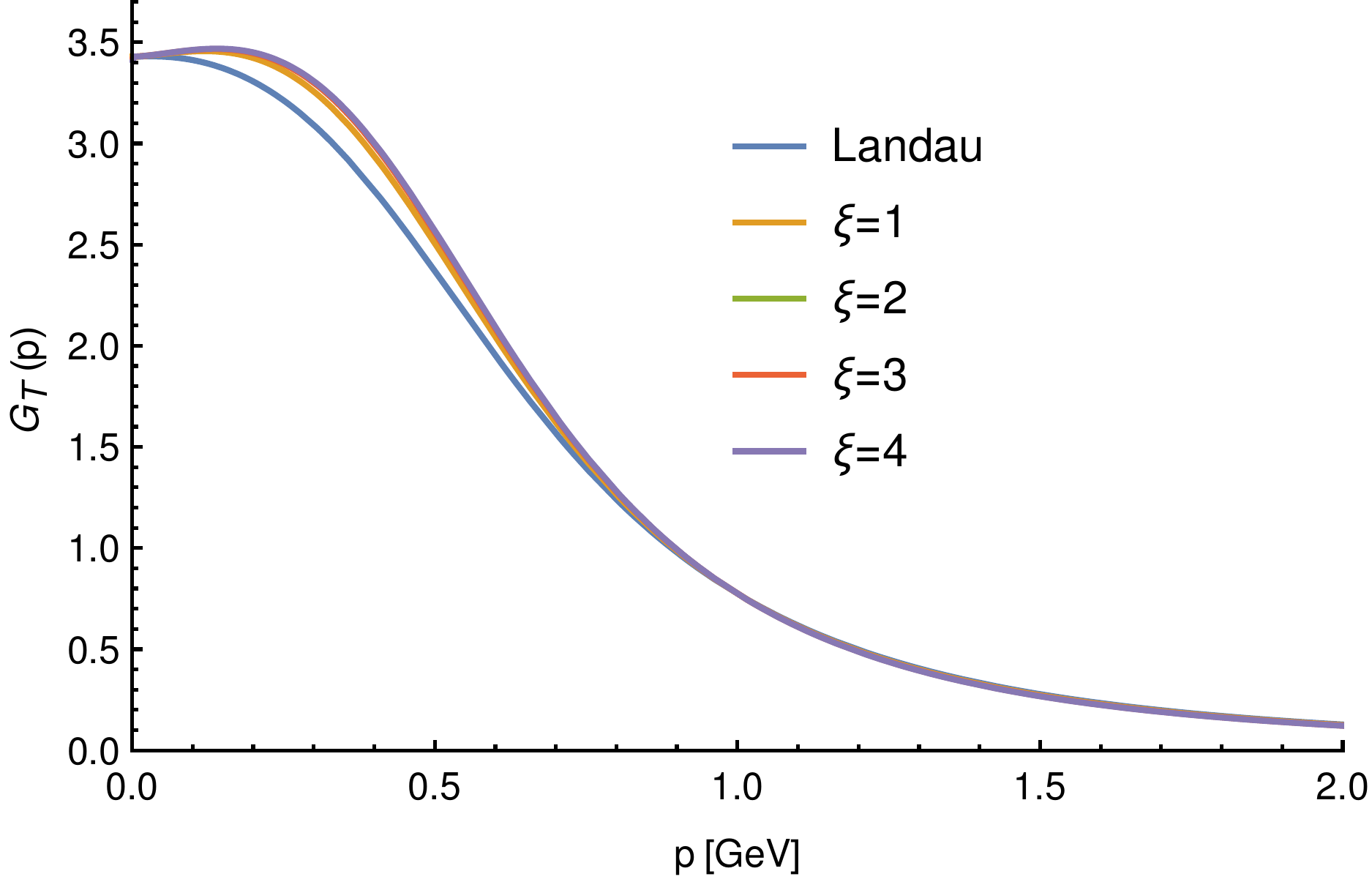}
  \caption{The ghost (top) and the transverse gluon (bottom) propagators as functions of momentum for various values of $\xi$ in the zero-momentum renormalization scheme with $m=0.54$~GeV and $g=4.9$.}
  \label{fig_gluon_propagator_scheme2}
\end{figure}

Figure \ref{fig_gluon_propagator_scheme2} shows the ghost and transverse gluon propagators renormalized within the zero-momentum prescriptions \eqn{Rs2 renormalization presciption 1}--\eqn{Rs2 renormalization presciption ih} for different values of the gauge-fixing parameter $\xi$. At small $\xi$, we recover the Landau gauge results of Ref.~\cite{Tissier_10}, where, in particular, the ghost propagator diverges at low momenta but the ghost dressing function $p^2G_{\rm gh}(p)$ remains finite. 
The renormalization condition \eqn{Rs2 renormalization presciption 1} imposes that for $\xi\neq0$ the ghost propagator itself remains finite in the infrared, where we see that it essentially behaves as a massive propagator. The strong $\xi$ dependence at low momentum is governed by the second condition \eqn{Rs2 renormalization presciption 1}. As for the transverse gluon propagator, its values at $p=0$ and $p=\mu$ are constrained by our choice of renormalization conditions \eqn{Rs2 renormalization presciption 1} and \eqn{Rs2 renormalization presciption 2}, which do not involve $\xi$, thus leaving little room for a possible $\xi$ dependence. Indeed, in this scheme, the latter only appears through loop effects and is thus small. Still, we observe that when $\xi$ grows, the gluon propagator flattens near its $p=0$ value. In fact, the lattice results in the Landau gauge show that the transverse gluon propagator is nonmonotonous at small $p$. For $\xi=0$, we reproduce the one-loop behavior of Ref.~\cite{Tissier_10}: $G_T^{-1}(p)=m^2+g^2N/(192\pi^2)p^2\ln(p^2/\mu^2)+{\cal O}(p^2)$. We observe in \Fig{fig_gluon_propagator_scheme2} that this nonmonotonous behavior is more pronounced for increasing $\xi$.

\begin{figure}[ht]
  \centering
  \includegraphics[width=1\linewidth]{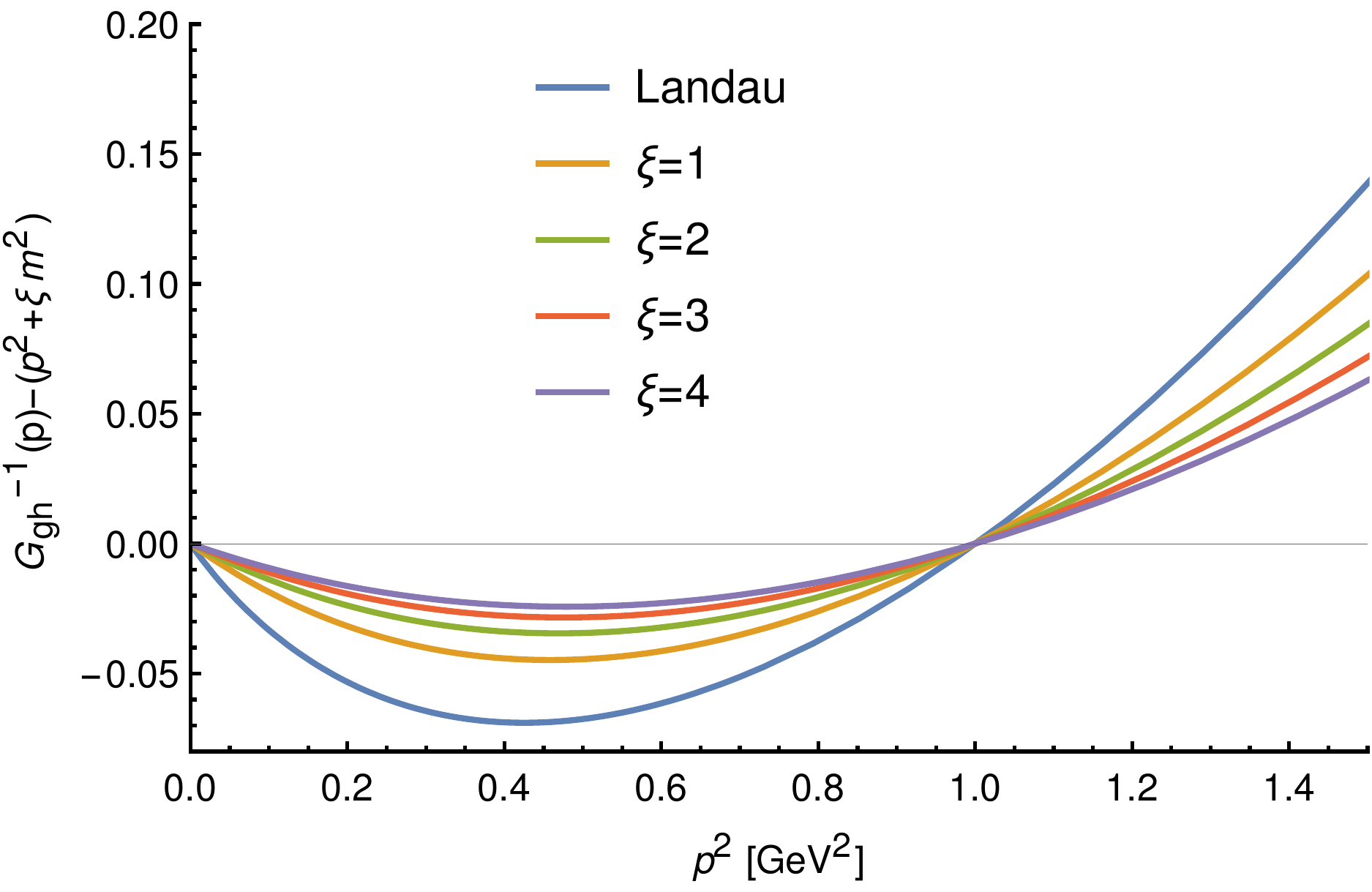}\\
  \includegraphics[width=1\linewidth]{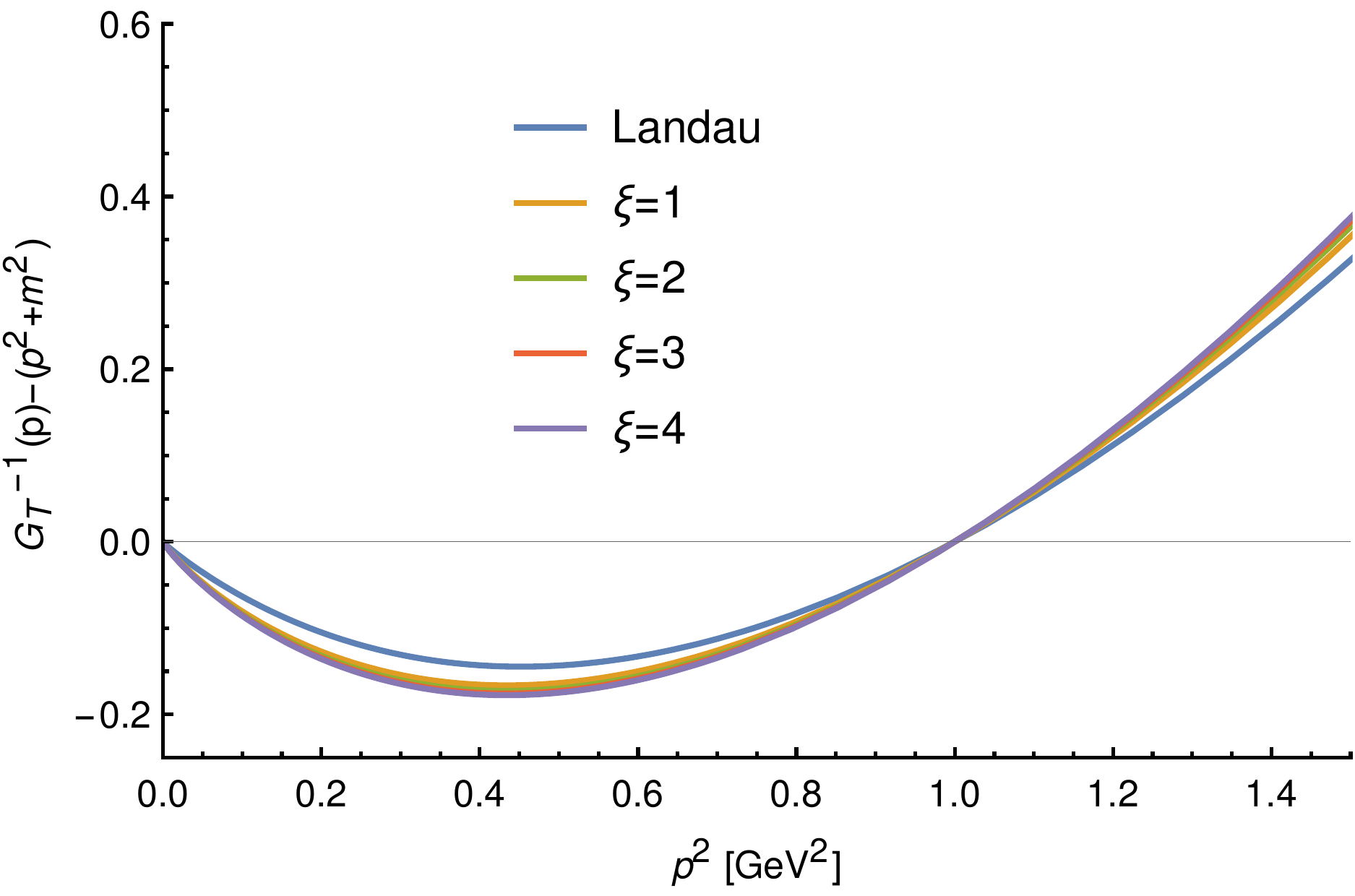}
  \caption{The loop contributions $G^{-1}_{\rm gh}(p)-(p^2+\xi m^2)$ (top) and $G_T^{-1}(p)-(p^2+m^2)$ (bottom) to the ghost and the transverse gluon inverse propagators as functions of $p^2$ in the zero-momentum scheme with $m=0.54$~GeV and $g=4.9$.}
  \label{fig:loopzero}
\end{figure} 

The above results show that the (renormalized) propagators are essentially governed by their tree-level expressions, respectively
\beq
 G^{\rm tree}_{\rm gh}(p)=\frac{1}{p^2+\xi m^2}\quad{\rm and}\quad G^{\rm tree}_T(p)=\frac{1}{p^2+m^2}.
\eeq
In order to emphasize the size and the shape of loop corrections, we plot in \Fig{fig:loopzero} the loop contribution to the inverse propagators as a function of $p^2$. We see that loop corrections are at the percent level for the ghost and at the ten percent level for the transverse gluon.

\subsection{Infrared-safe scheme}

\begin{figure}[ht]
  \centering
  \includegraphics[width=1\linewidth]{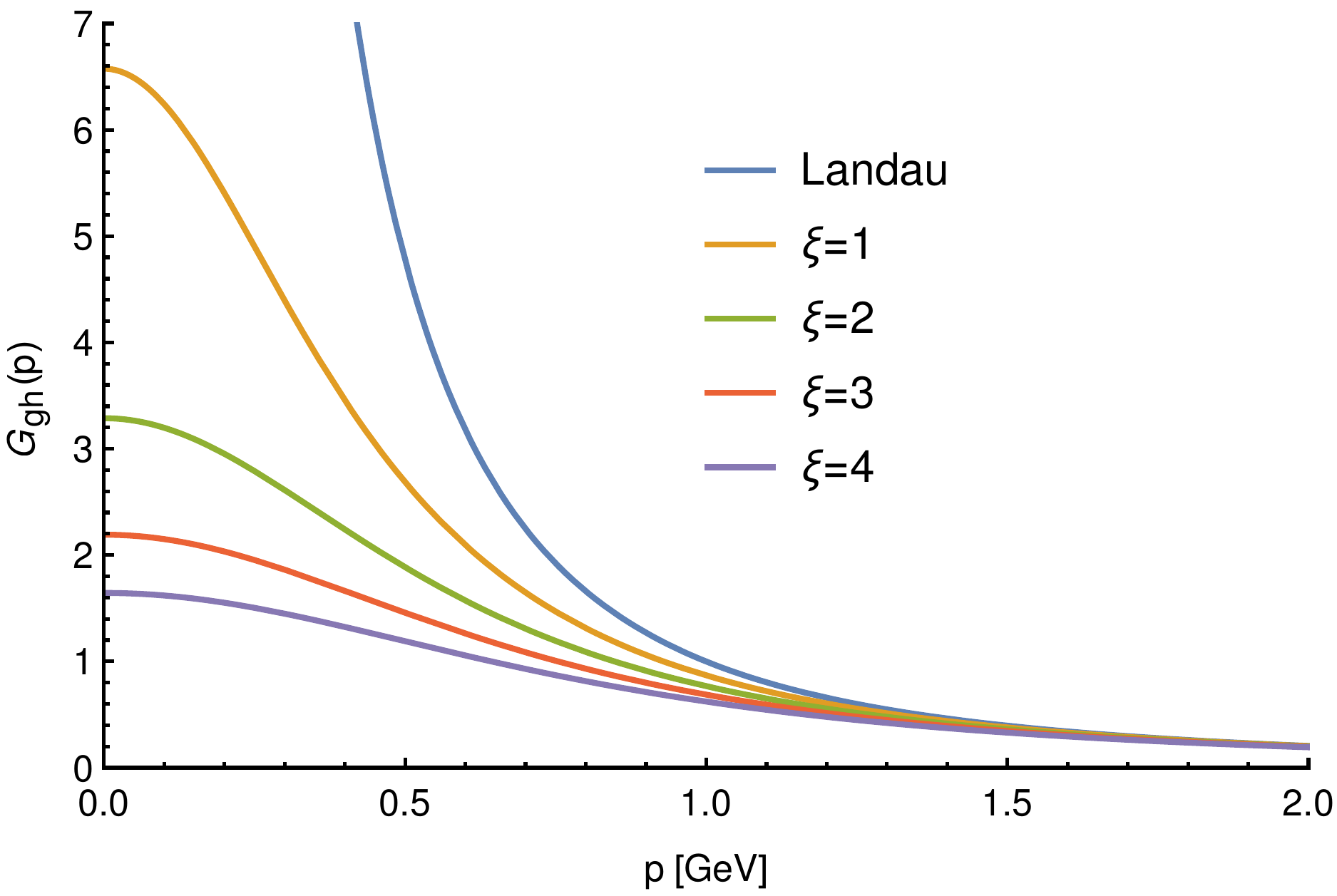}\\
  \includegraphics[width=1\linewidth]{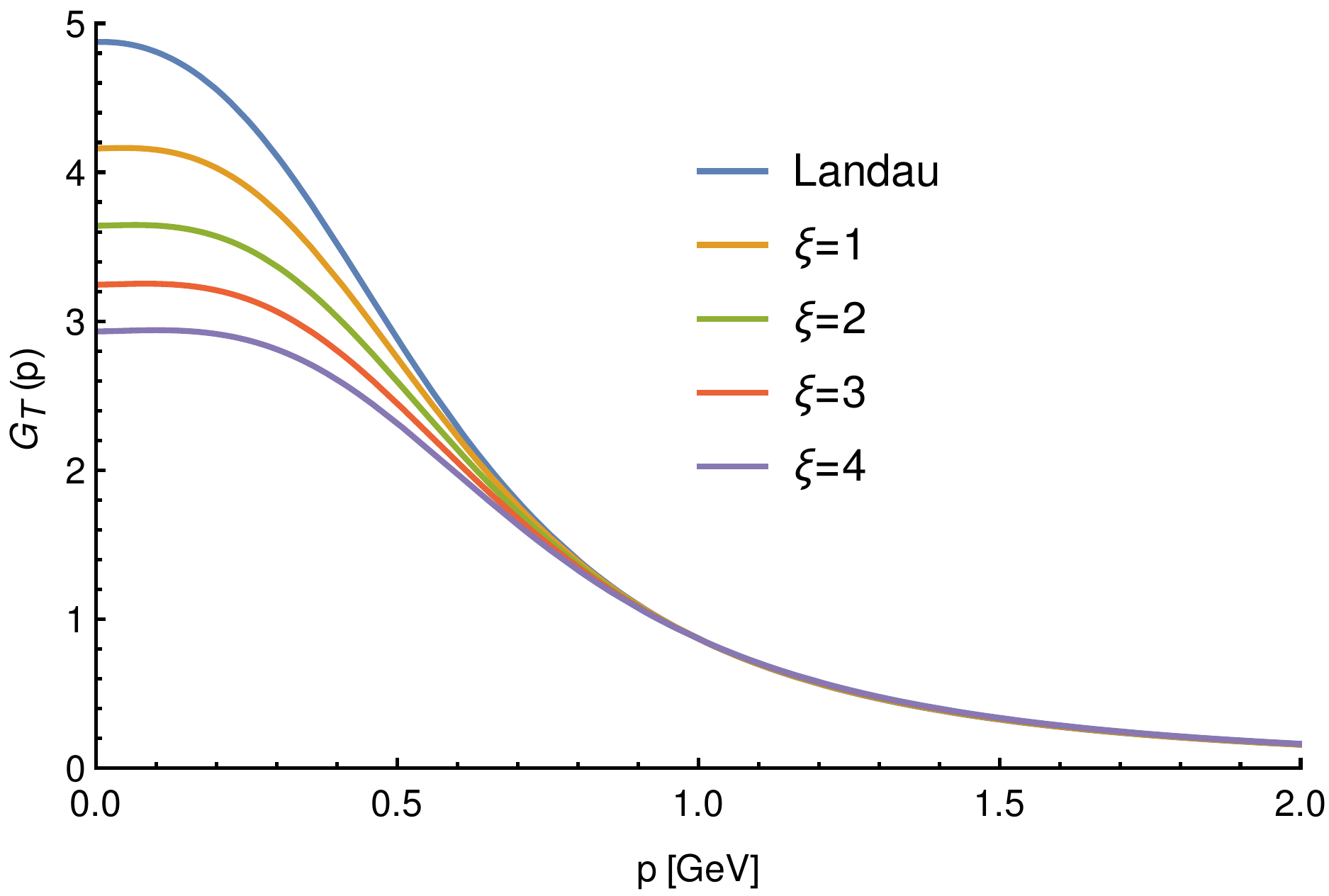}
  \caption{The ghost (top) and the transverse gluon (bottom) propagator as a function of momentum for various values of $\xi$ in the infrared-safe renormalization scheme with $m=0.39$~GeV and $g=3.7$.}
  \label{fig_gluon_propagator_schemeIRsafe}
\end{figure}

The ghost and the transverse gluon propagators in the infrared-safe scheme are shown in \Fig{fig_gluon_propagator_schemeIRsafe}. 
\begin{figure}[ht]
  \centering
  \includegraphics[width=1\linewidth]{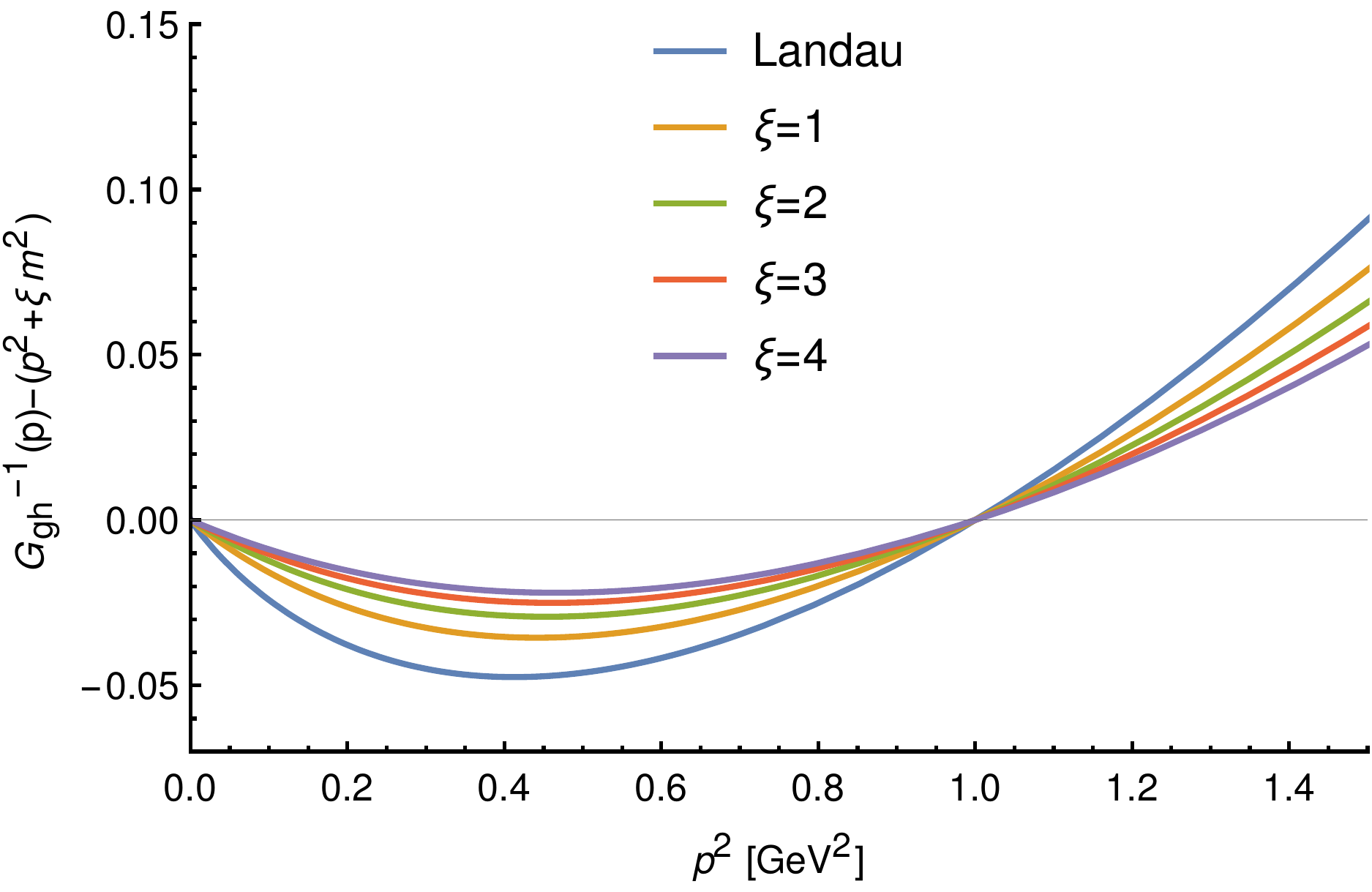}\\
  \includegraphics[width=1\linewidth]{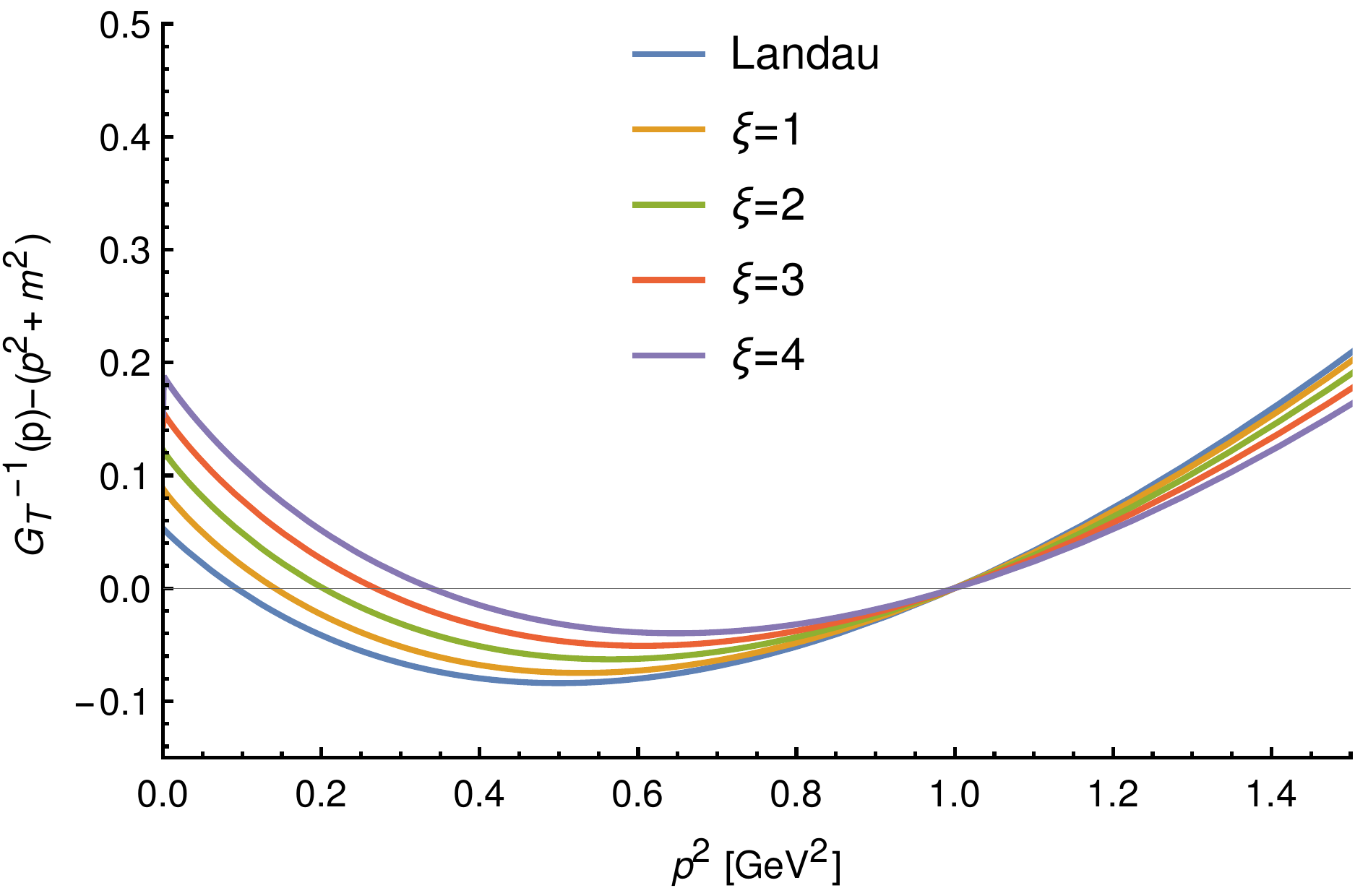}
  \caption{The loop contributions $G^{-1}_{\rm gh}(p)-(p^2+\xi m^2)$ (top) and $G_T^{-1}(p)-(p^2+m^2)$ (bottom) to the ghost and the transverse gluon inverse propagators as functions of $p^2$ in the infrared-safe scheme with $m=0.39$~GeV and $g=3.7$. We observe that the loop contributions to the ghost inverse propagator vanish at $p=0$, as a consequence of the nonrenormalization theorem discussed in the text below \Eqn{eq:pkpojk}; see also \Eqn{eq:nonren1}.}
  \label{fig:loopgluonIR}
\end{figure} 
Again, we check that we recover the results of Ref. \cite{Tissier_10} (in the appropriate renormalization scheme) for small values of $\xi$. In the present case, the value of the gluon propagator at vanishing momentum is not fixed and varies strongly with $\xi$. This can be traced to the fact that $\xi $ influences the propagator at tree level through the second equation in \Eqn{IRs renormalization presciption Zm}. As in the previous case, we observe that the approach to the $p=0$ value flattens as $\xi$ is increased. We plot the loop contribution to the inverse propagators in \Fig{fig:loopgluonIR}. These are again at the percent level for the ghost and at the ten percent level for the gluon.

\subsection{Comparison with the CF model}

We compare the results of the previous section to those of the CF model, obtained by setting the number of replica $n=1$. As already mentioned, this allows us to pinpoint the peculiar effects of our treatment of the Gribov ambiguities. In particular, taking the limit $n\to0$ is a crucial step for our approach to correspond to a bona fide  gauge-fixing procedure. We have already discussed the fact that an important difference with the case $n=1$---already present at tree level---is the fact that the gluon propagator is exactly transverse in the case $n\to0$, in contrast to the case $n=1$. In the present section, we discuss the differences between the two cases arising from the one-loop contributions to the ghost and the gluon propagators.

\begin{figure}[ht]
  \centering
  \includegraphics[width=.45\linewidth]{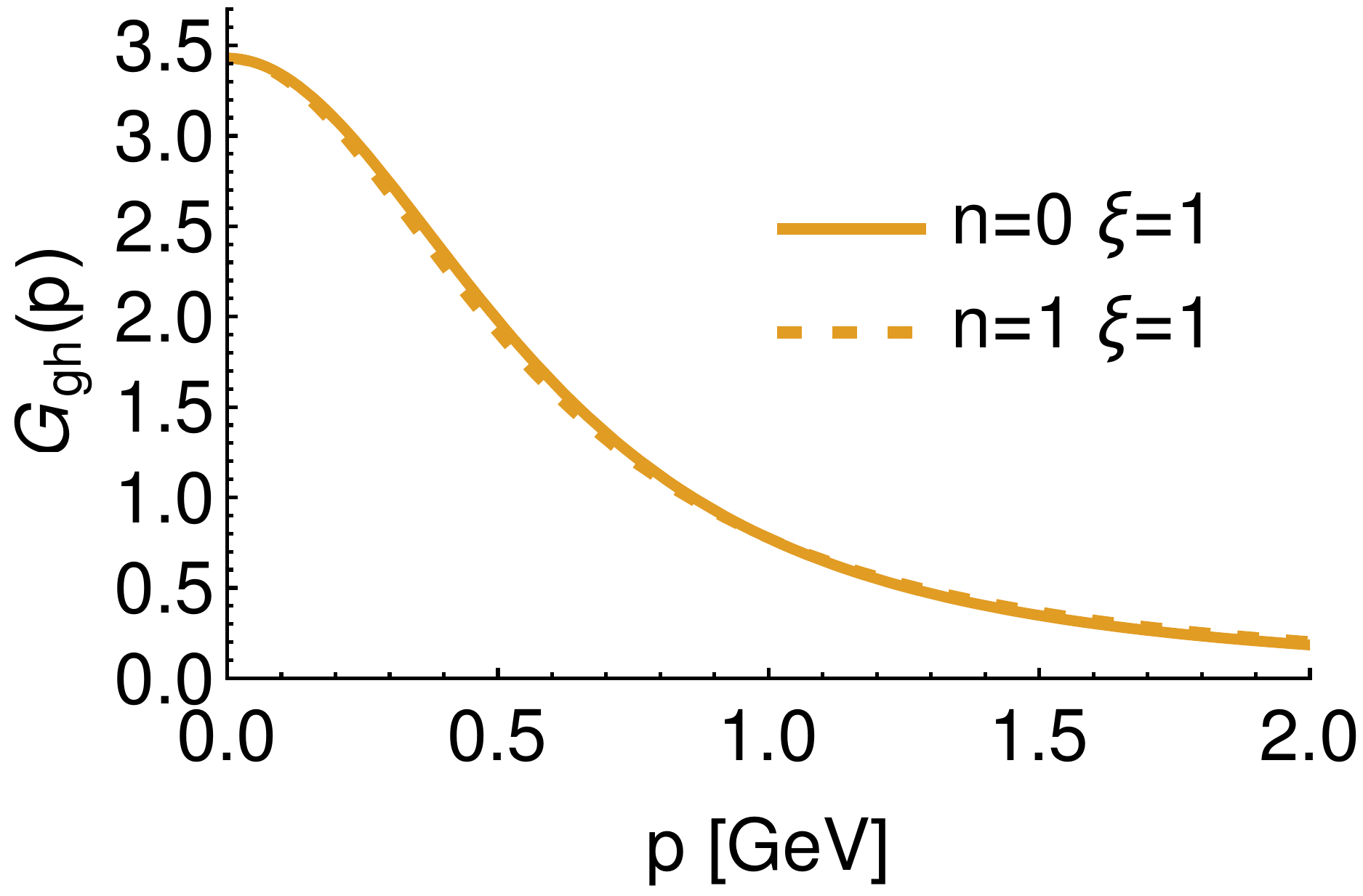}\quad
  \includegraphics[width=.45\linewidth]{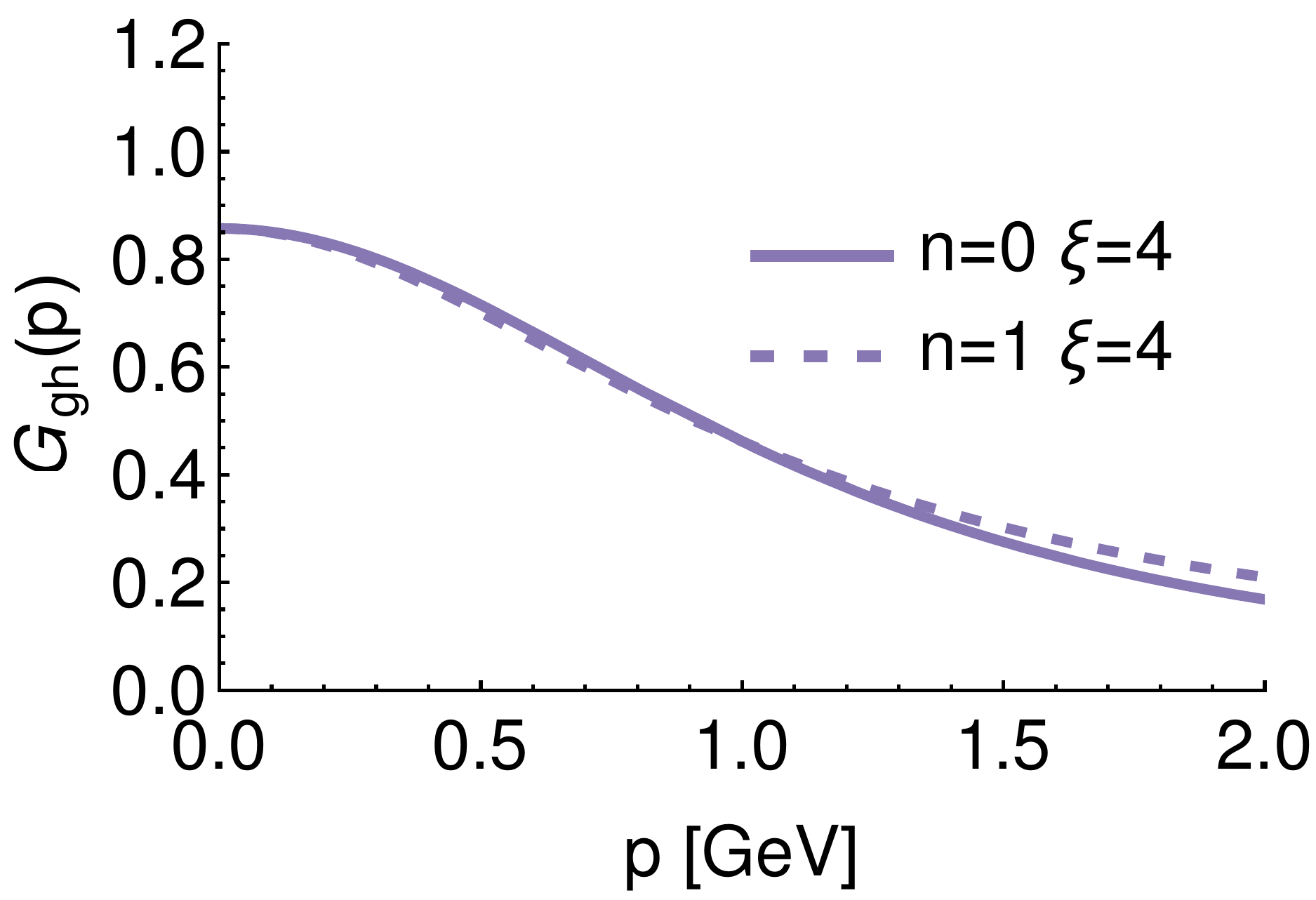}\\
  \includegraphics[width=.45\linewidth]{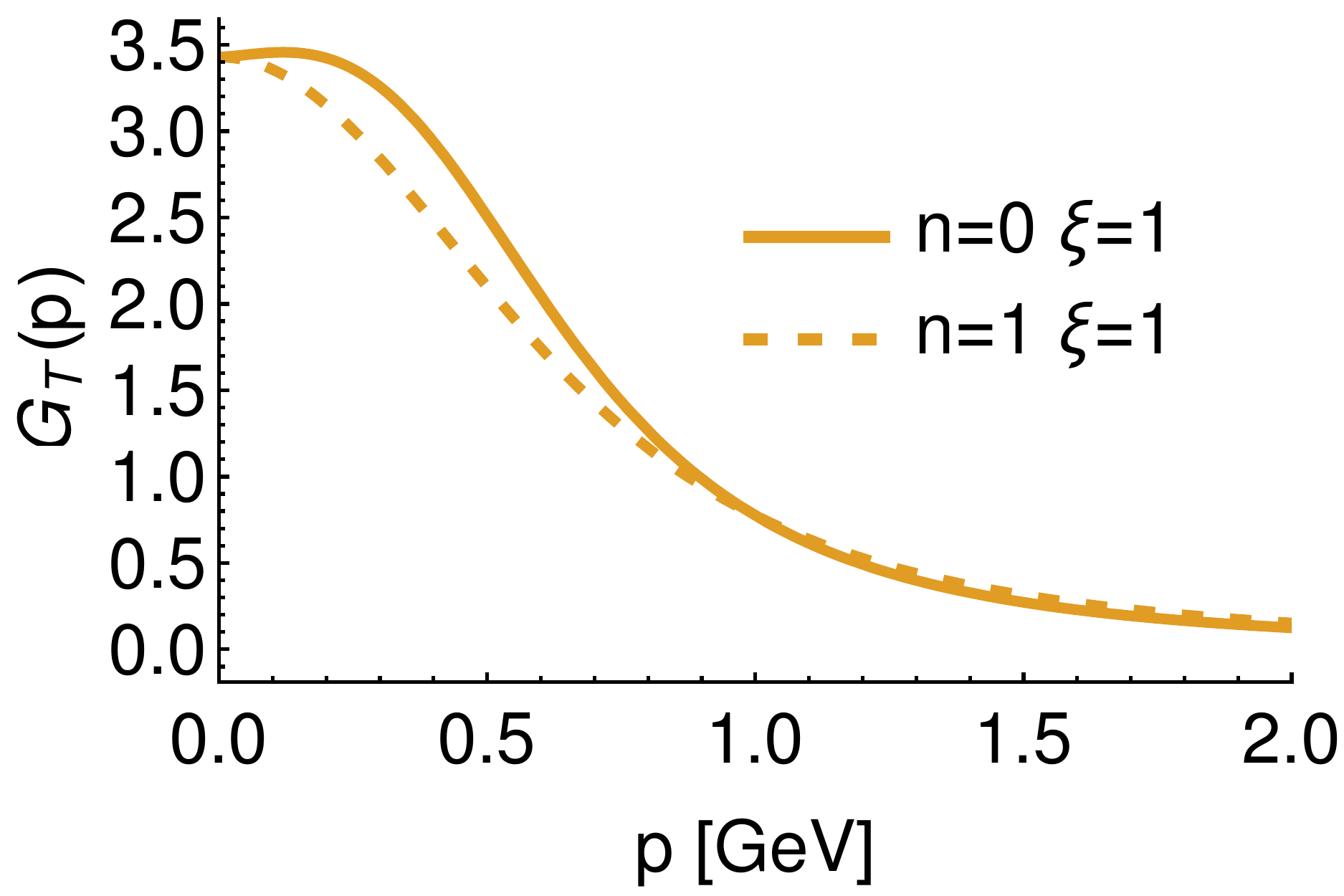}\quad
  \includegraphics[width=.45\linewidth]{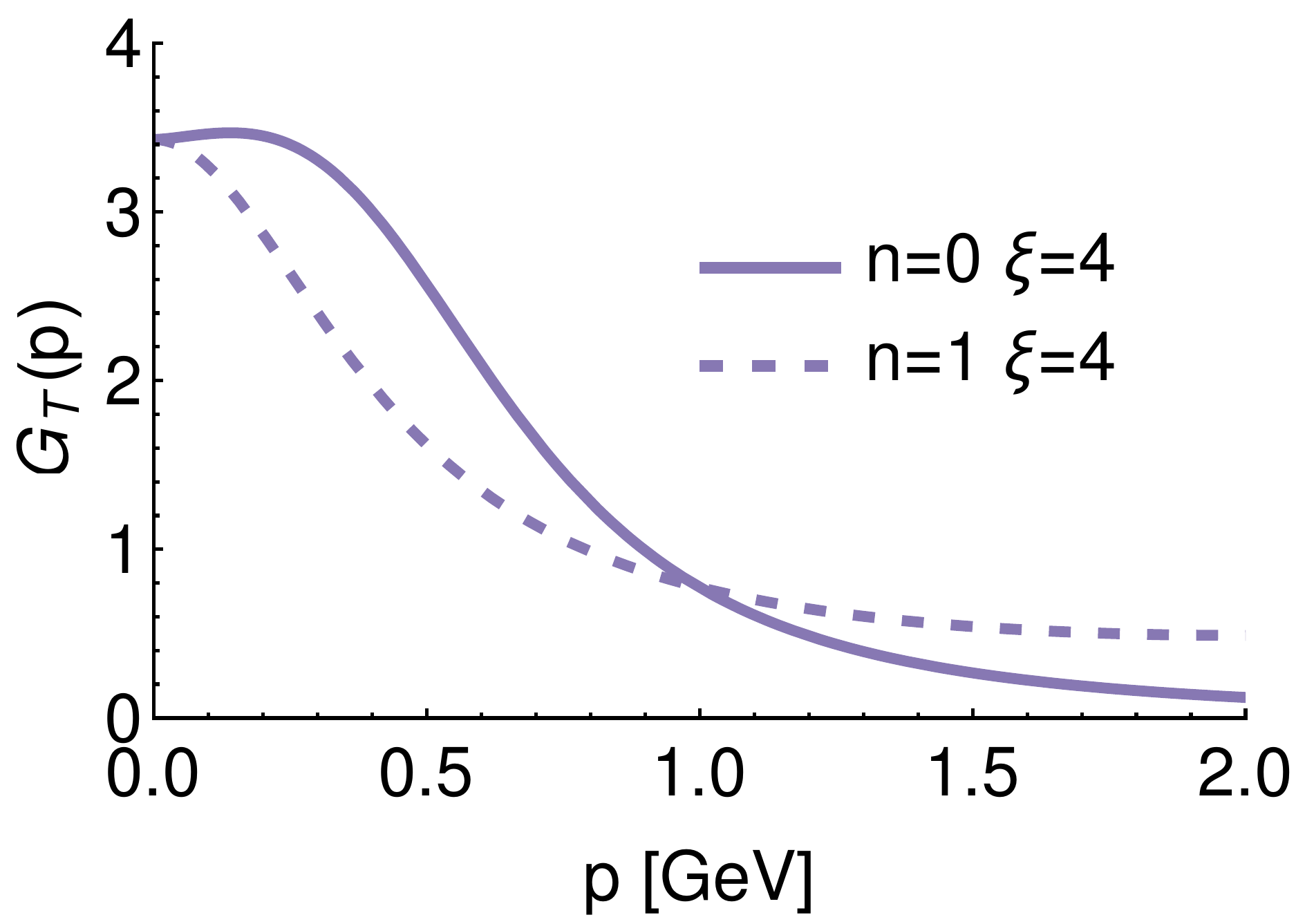}
  \caption{The ghost (top) and transverse gluon (bottom) propagators computed in either the present gauge fixing ($n\to0$) or the CF model ($n=1$) in the zero-momentum scheme with $m=0.54$~GeV and $g=4.9$,  for $\xi=1$ (left) and $\xi=4$ (right).}
  \label{fig:CFzero}
\end{figure}

\begin{figure}[ht]
  \centering
  \includegraphics[width=1\linewidth]{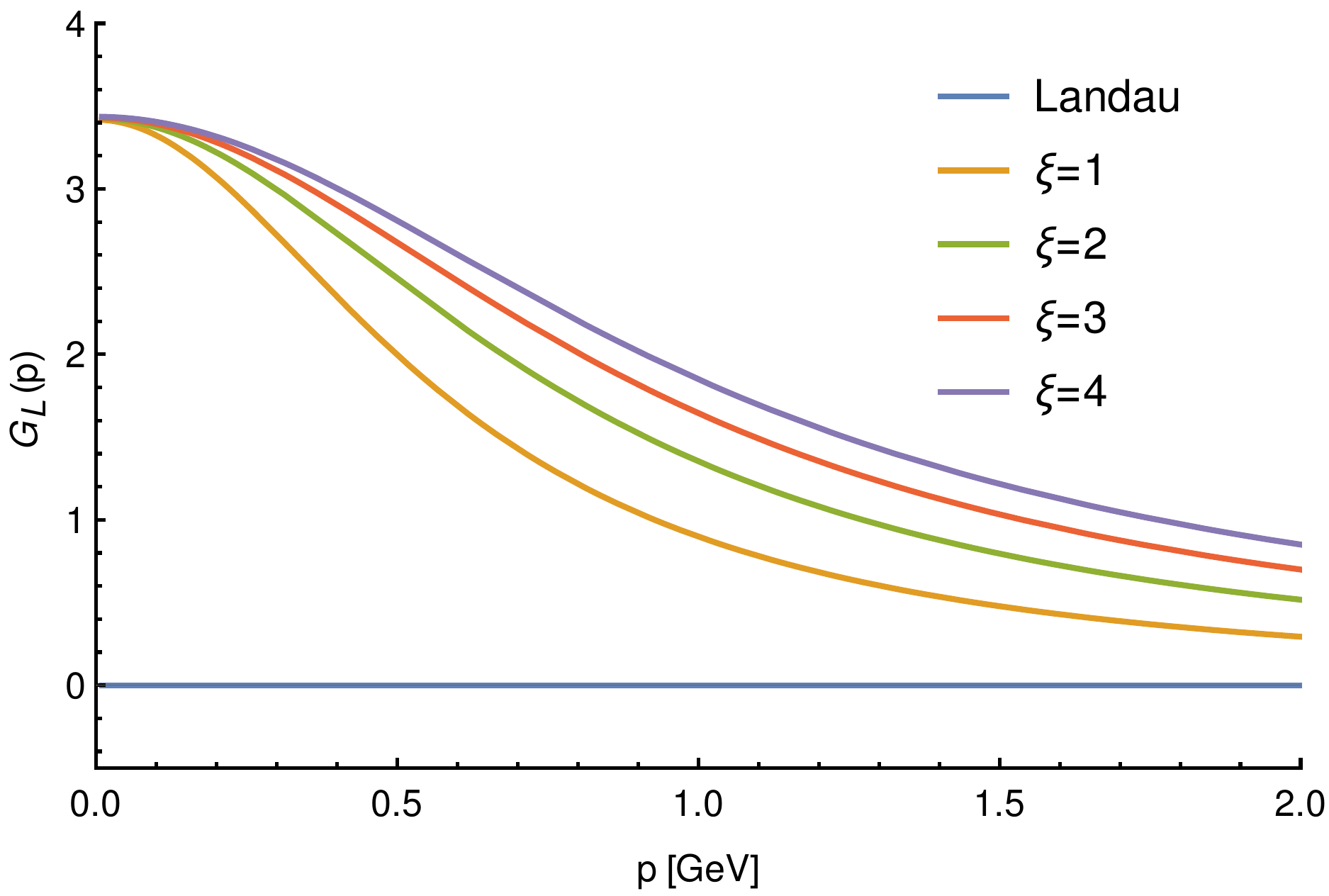}
  \caption{The longitudinal gluon propagator in the CF model ($n=1$) as a function of momentum for various values of $\xi$ in the zero-momentum scheme with $m=0.54$~GeV and $g=4.9$.}
  \label{fig:longCF1}
\end{figure}

The case of the zero-momentum scheme is illustrated in \Fig{fig:CFzero}. We observe that the ghost propagators for $n\to0$ and $n=1$ are essentially the same while the transverse gluon propagators are qualitatively different. Contrarily to the former, the latter receives a direct contribution from a loop of superfield, proportional to $n-1$, as discussed previously; see \Eqn{explicit_calc2}. We see that the flattening near $p=0$ described above in the case $n\to0$ is absent in the CF model. 
For completeness, we show the momentum dependence of the longitudinal gluon propagator in the CF model in \Fig{fig:longCF1}.

\begin{figure}[ht]
  \centering
  \includegraphics[width=.45\linewidth]{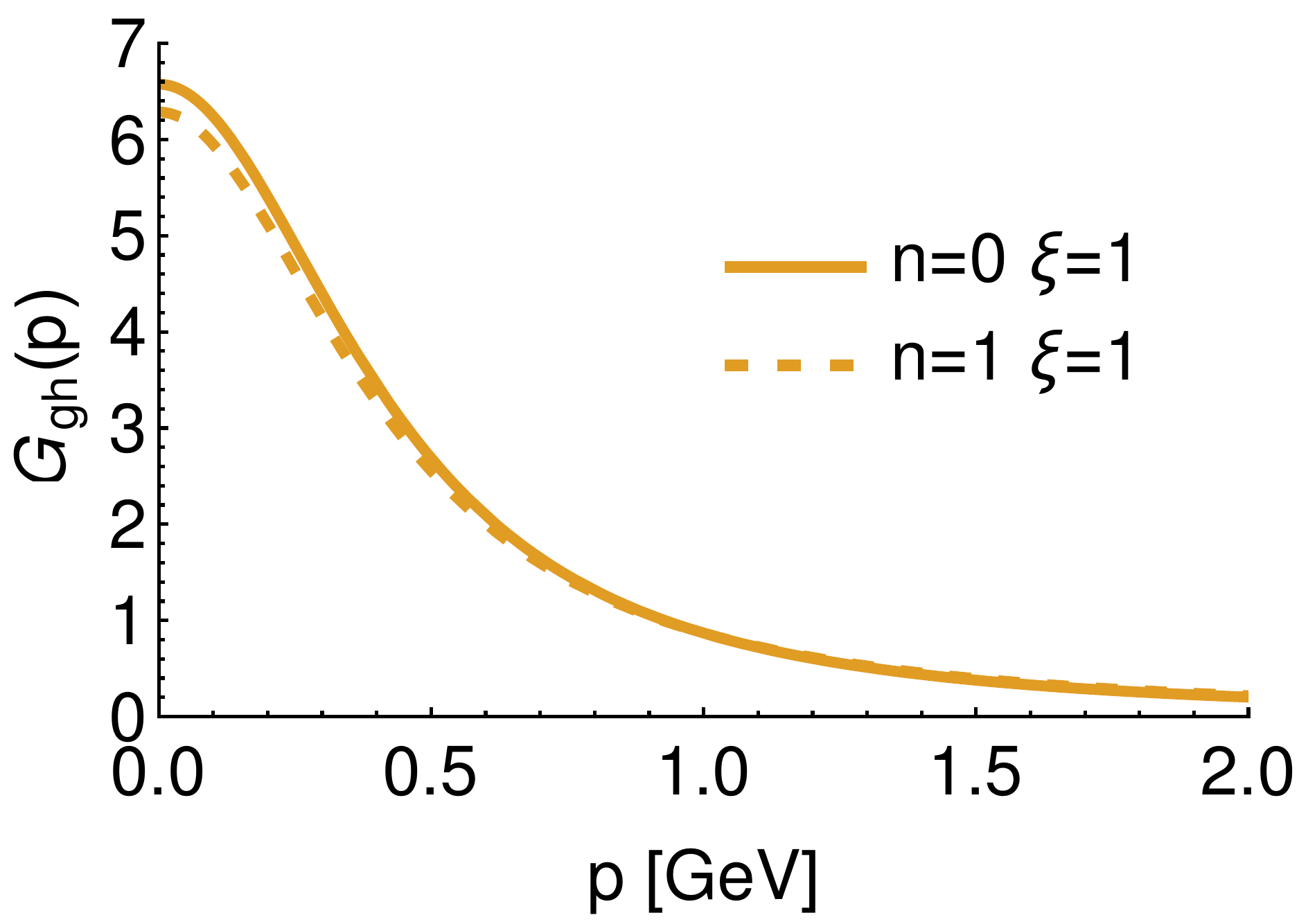}\quad
  \includegraphics[width=.45\linewidth]{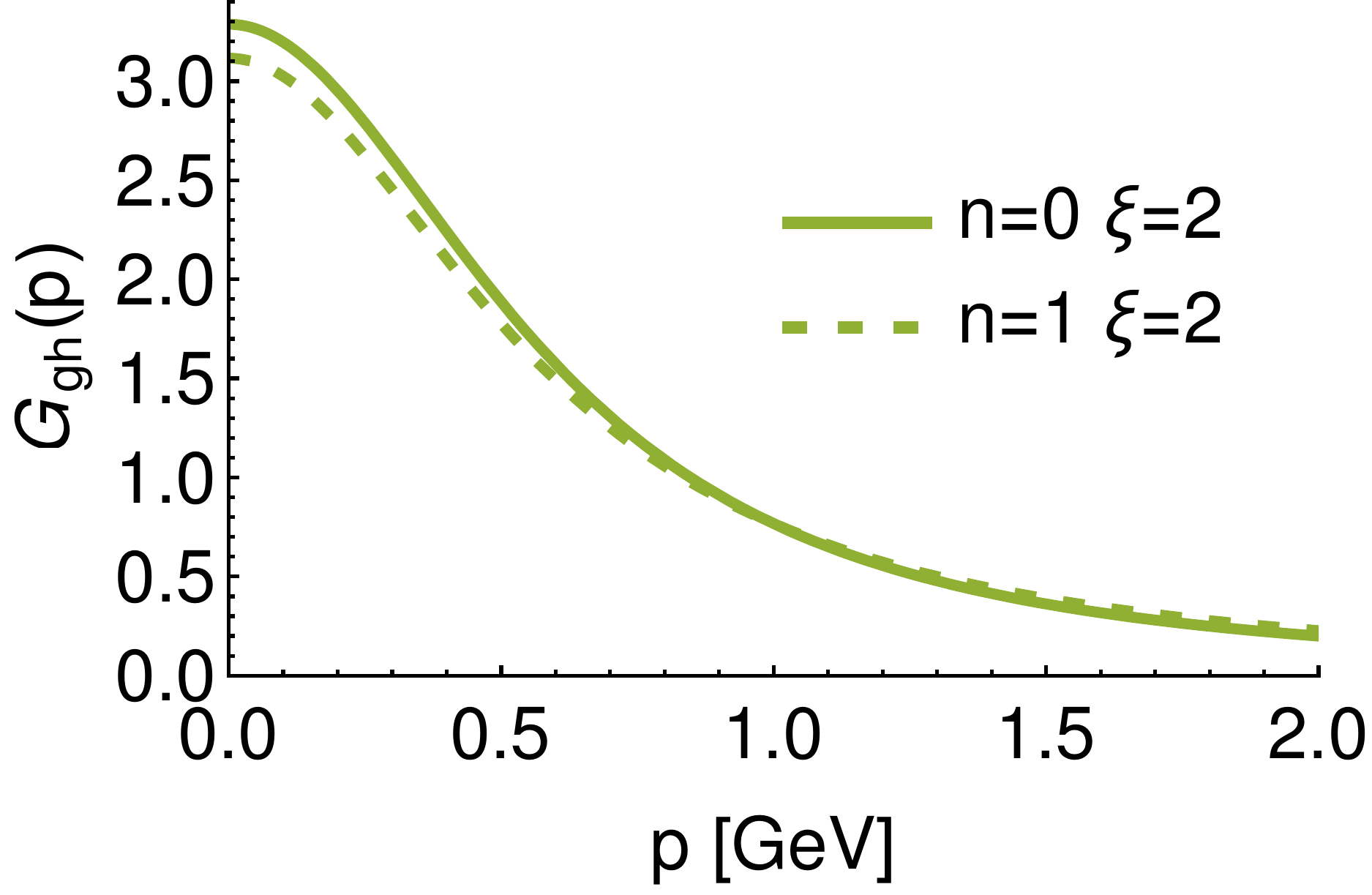}\\
  \includegraphics[width=.45\linewidth]{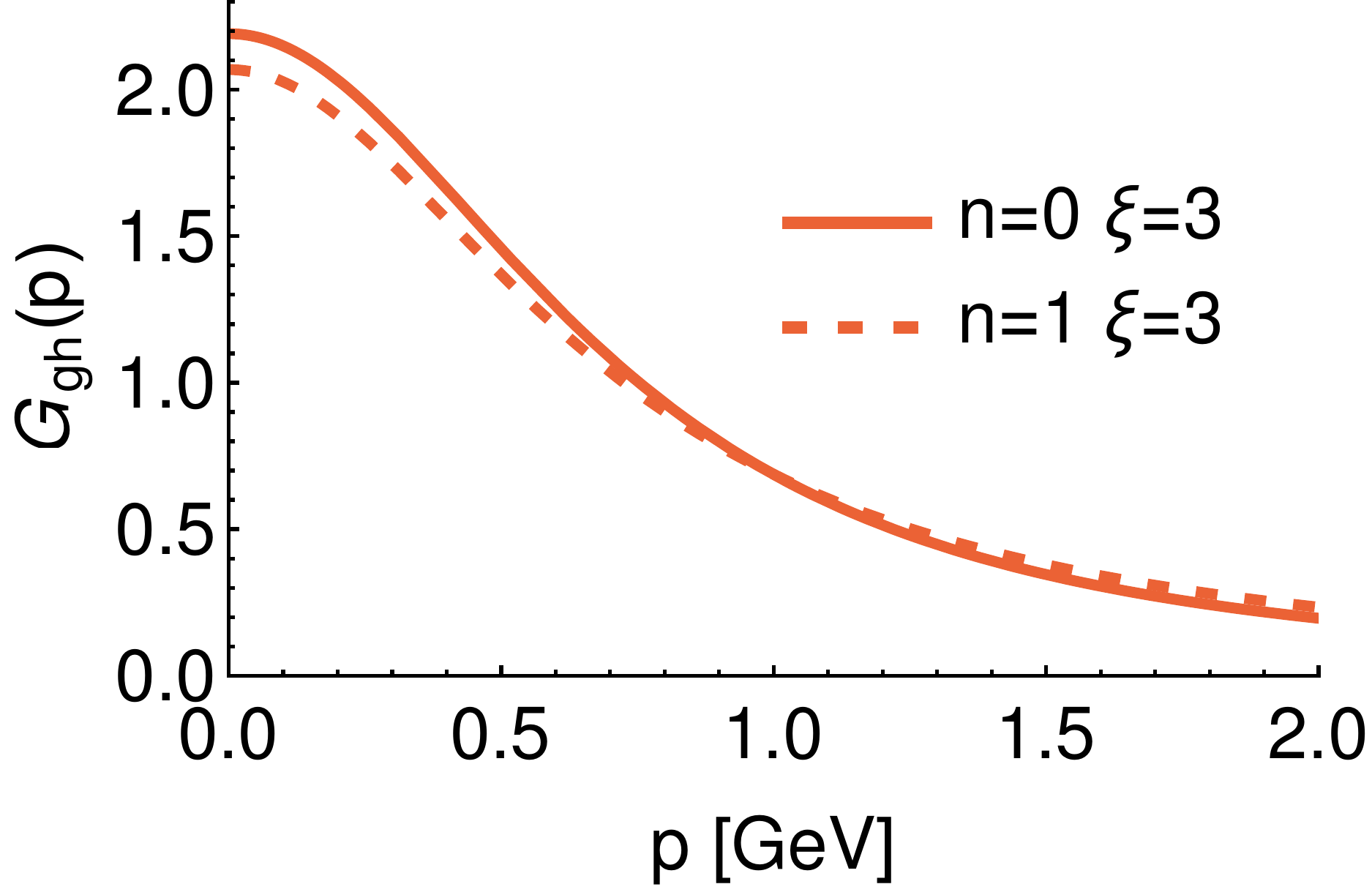}\quad
  \includegraphics[width=.45\linewidth]{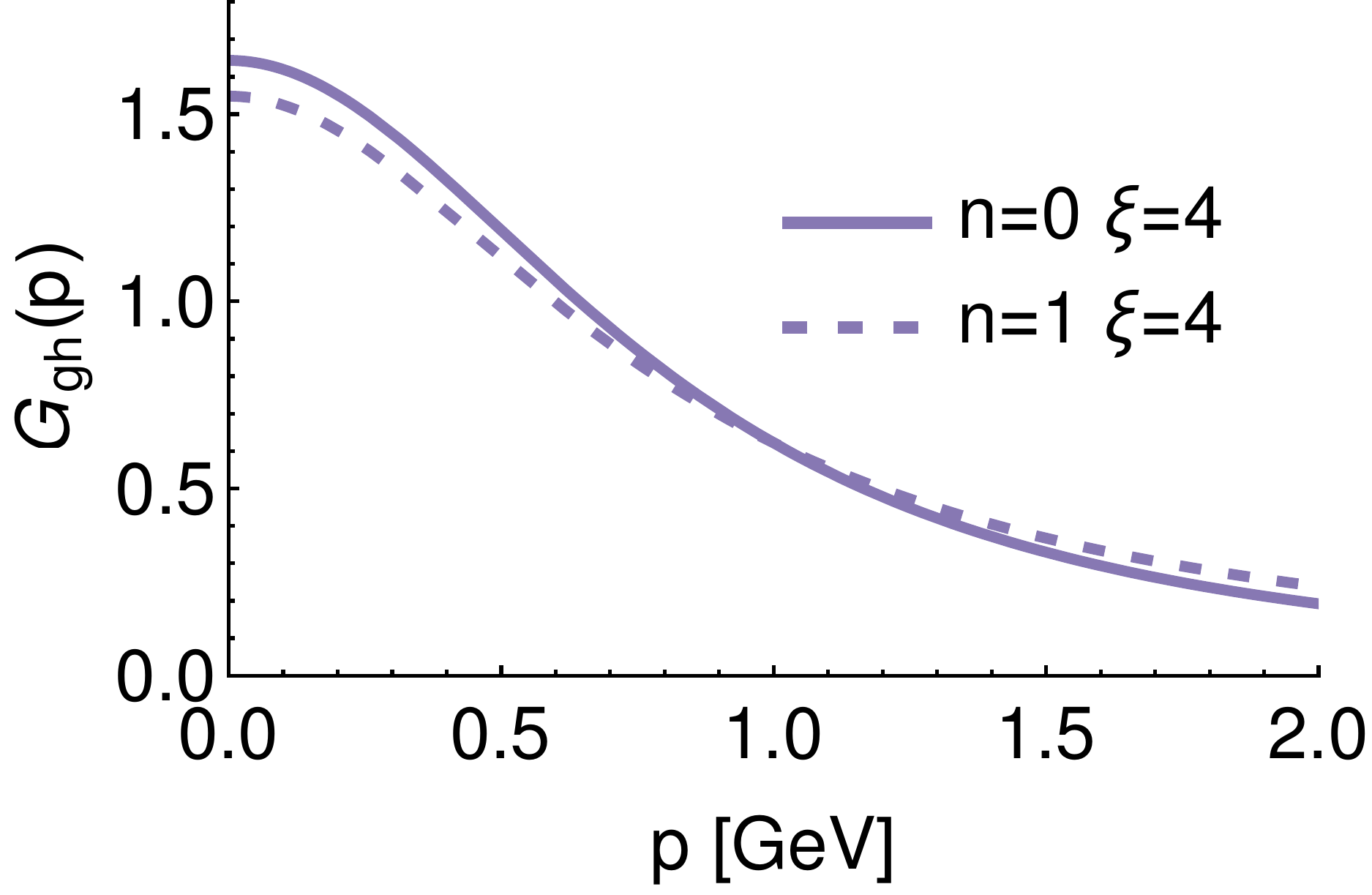}
  \caption{The ghost propagator computed in either the present gauge fixing ($n\to0$) or the CF model ($n=1$) in the infrared-safe scheme with $m=0.39$~GeV and $g=3.7$ for various values of $\xi$.}
  \label{fig:CFIRsafe-gh}
\end{figure}

\begin{figure}[ht]
  \centering
  \includegraphics[width=.45\linewidth]{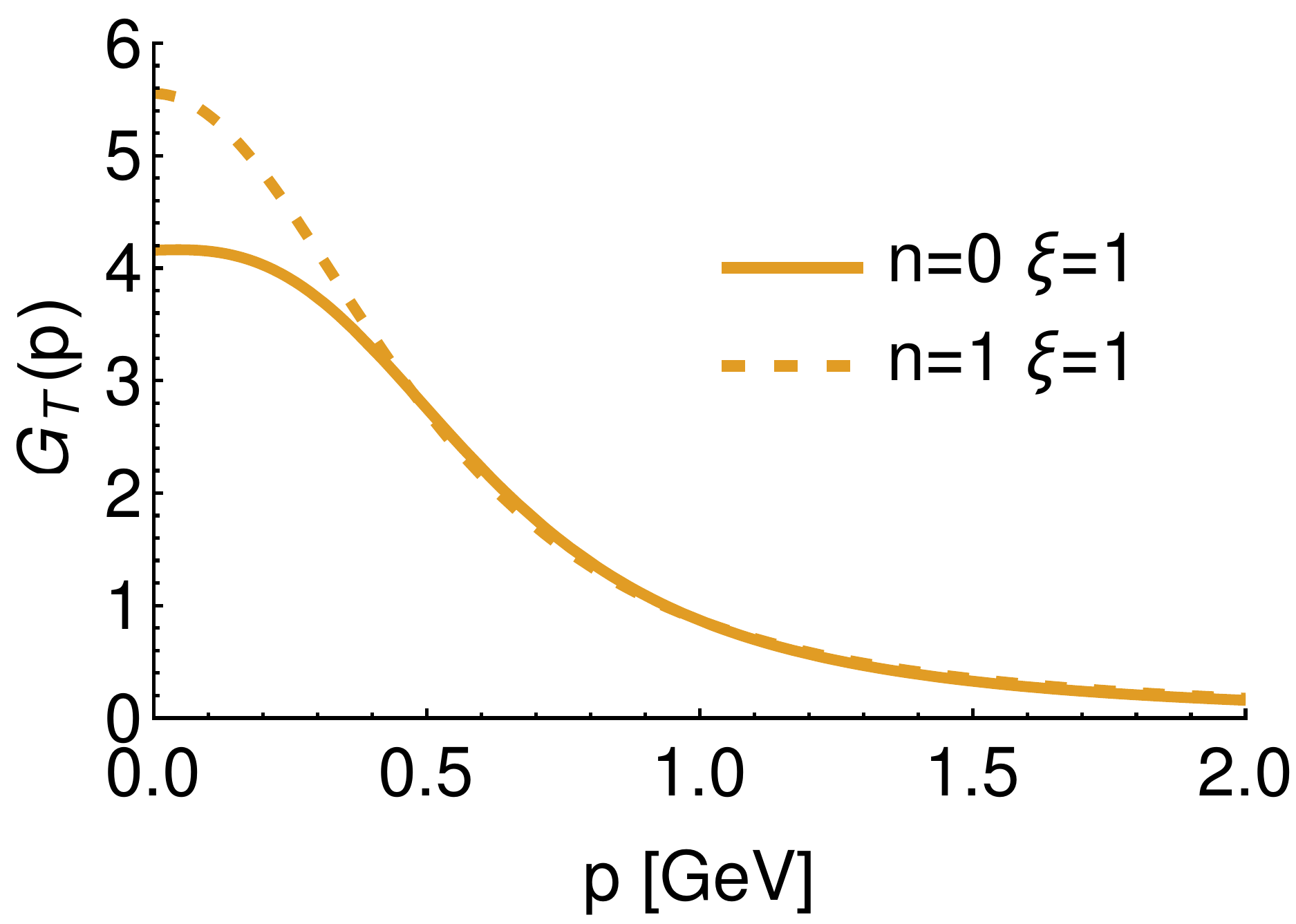}\quad
  \includegraphics[width=.45\linewidth]{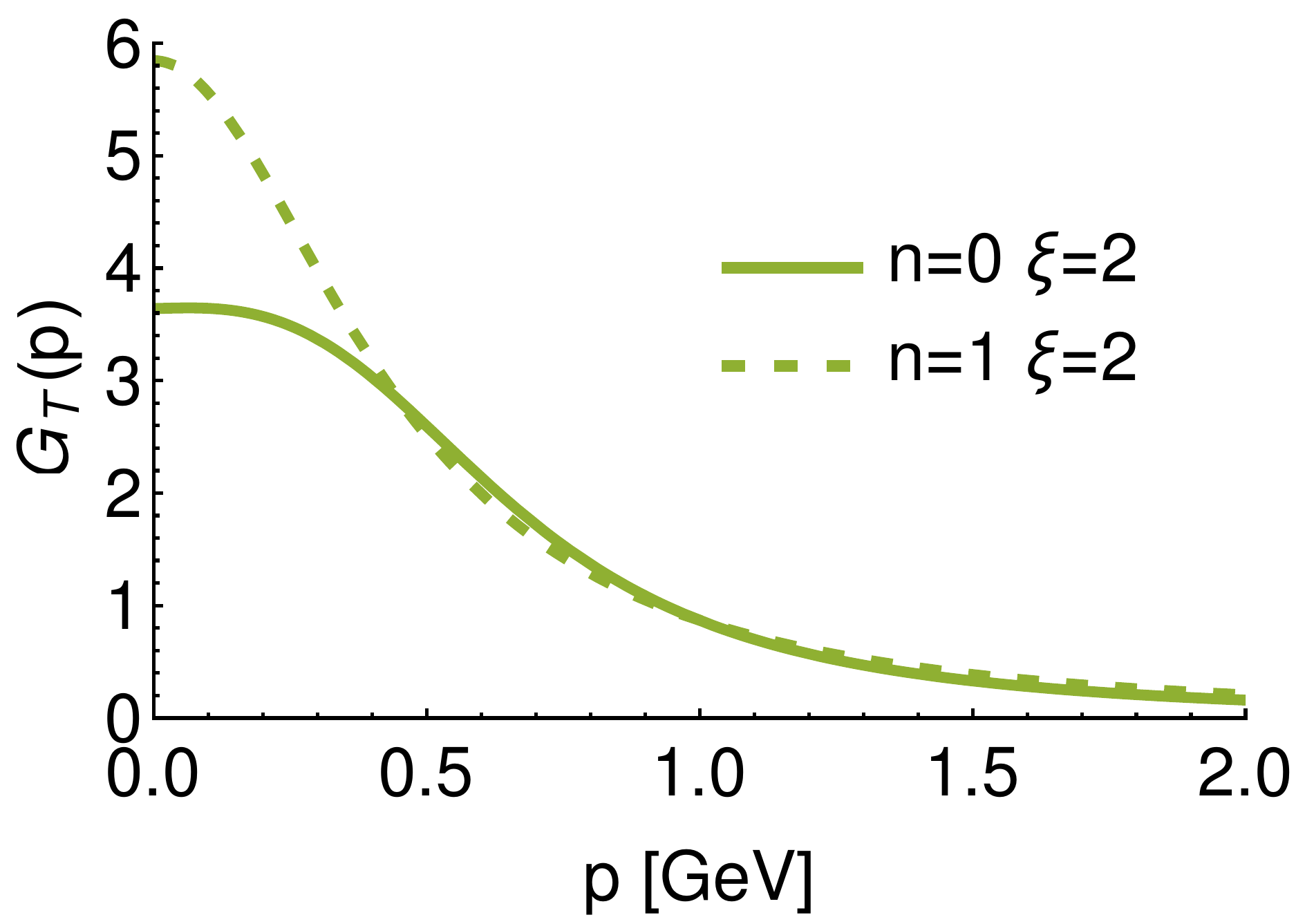}\\
  \includegraphics[width=.45\linewidth]{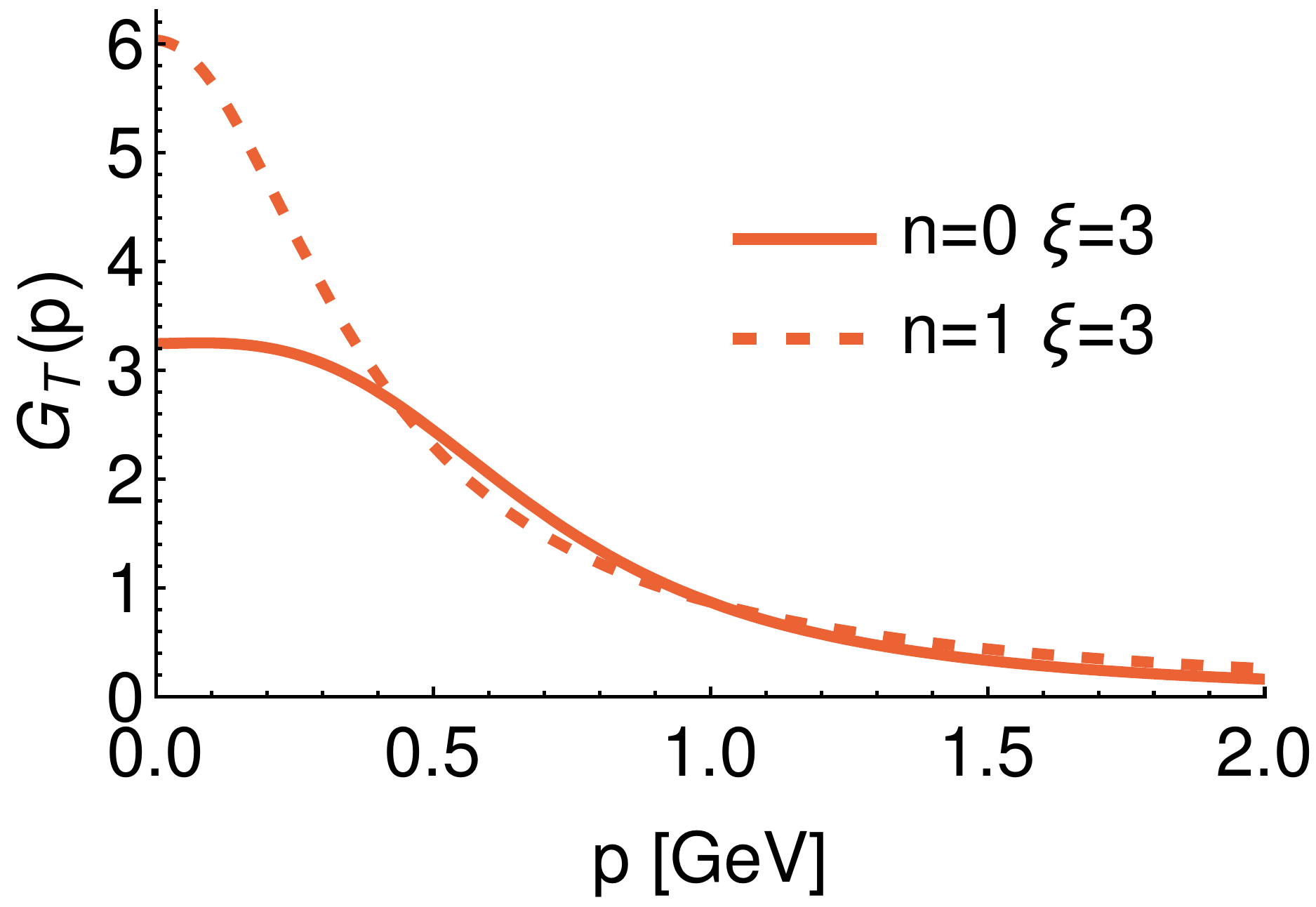}\quad
  \includegraphics[width=.45\linewidth]{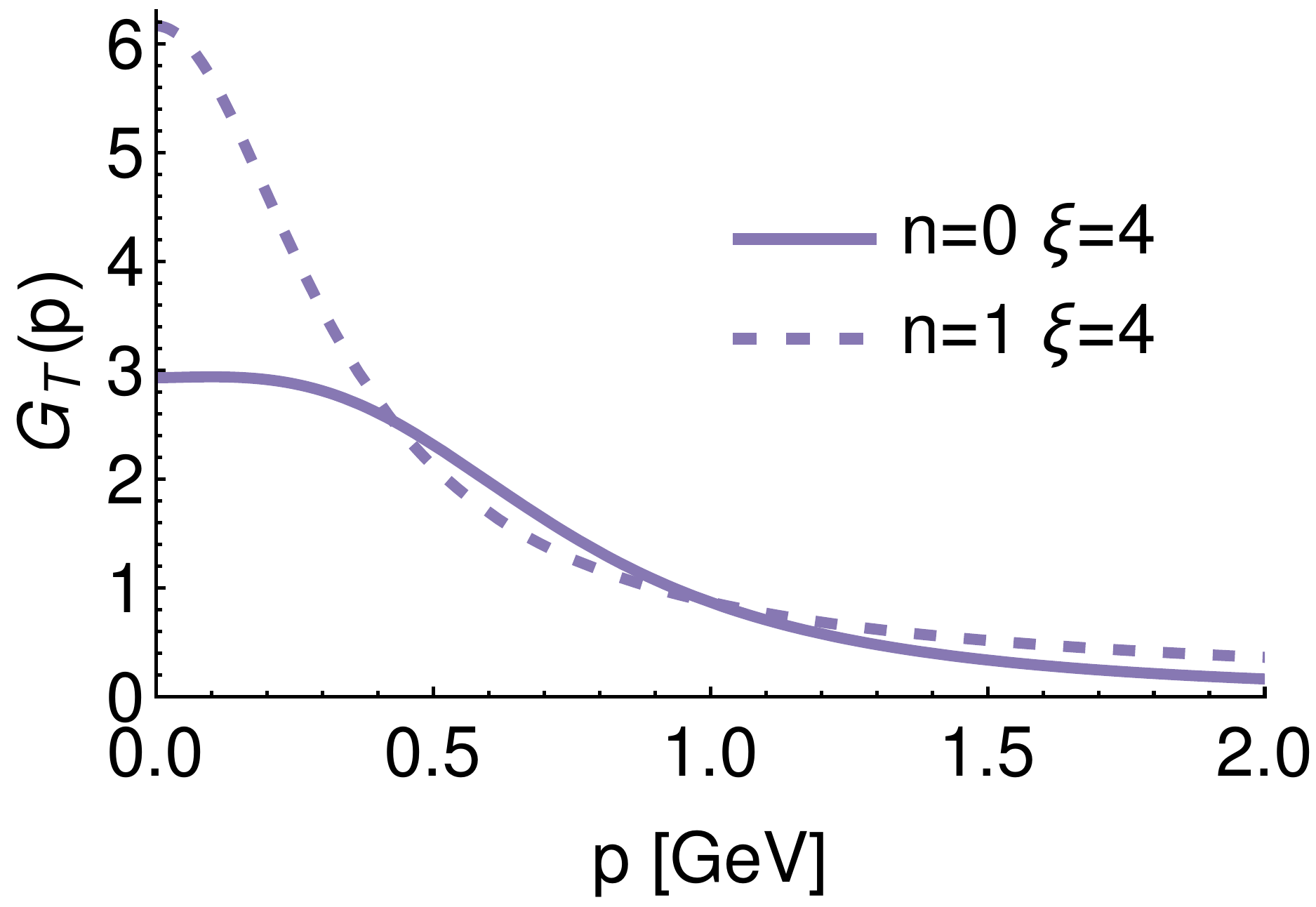}
  \caption{The transverse gluon propagator computed in either the present gauge fixing ($n\to0$) or the CF model ($n=1$) in the infrared-safe scheme with $m=0.39$~GeV and $g=3.7$ for various values of $\xi$.}
  \label{fig:CFIRsafe-gl}
\end{figure}

\begin{figure}[ht]
  \centering
  \includegraphics[width=1\linewidth]{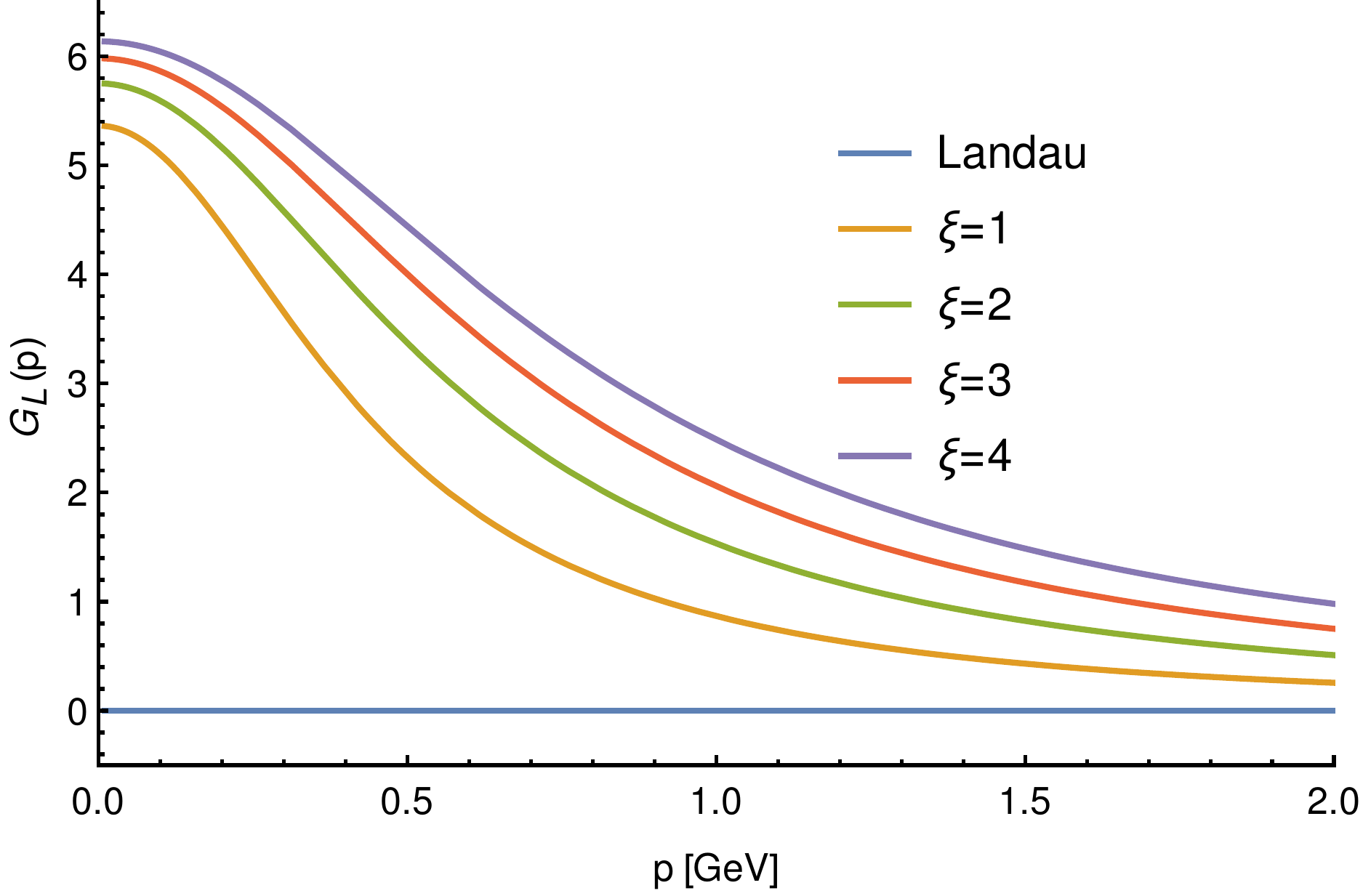}
  \caption{The longitudinal gluon propagator in the CF model ($n=1$) as a function of momentum for various values of $\xi$ in the infrared-safe scheme.}
  \label{fig:longCF2}
\end{figure}

A similar comparison is made in Figs.~\ref{fig:CFIRsafe-gh} and \ref{fig:CFIRsafe-gl} for the infrared-safe scheme. The previous observations hold here too where the differences between $n\to0$ and $n=1$ are even more pronounced. For this reason, we show more values of $\xi$ than in the previous case. Again we observe that the flattening of the gluon propagator near $p=0$ observed previously is absent in the CF model. 
Also we have checked that the nonrenormalization property \eqn{eq:nonren1} of the ghost vertex at zero momentum indeed holds in the case $n\to0$ but not for $n=1$.
Finally, the one-loop longitudinal gluon propagator of the CF model in the infrared-safe scheme is shown in \Fig{fig:longCF2}. 

\section{Renormalization-group improvement}
\label{sec_RG}

We now take into account the RG running of the parameters $m^2$, $g$, and $\xi$ and we investigate the RG trajectories in the two renormalization schemes discussed above. 
As usual, we define the beta functions  $\beta_\alpha$ for the parameters $\alpha=m^2,g,\xi$ as
\beq
\beta_\alpha=\left. \frac{d\,\alpha}{d\ln\mu} \right|_0=-\alpha\left. \frac{d \ln Z_\alpha}{d \ln\mu} \right|_0,\label{Rg_eq_1}
\eeq
and the gluon anomalous dimension as
\beq
\gamma_A =\left. \frac{d \ln Z_A}{d \ln\mu} \right|_0,
\eeq
where the subscript $0$ means that the right-hand side is evaluated at fixed bare quantities. Note that the prescription \eqn{eq_taylor} at one-loop order implies that the ghost anomalous dimension is given by
\begin{equation} \label{anomalous_ghost}
\gamma_c=\left. \frac{d \ln Z_c}{d \ln \mu} \right|_0=-\frac{1}{2} \gamma_A +\frac{\beta_g}{g} .
\end{equation}
Similarly, the relation \eqn{eq:jbsoin1} which, as discussed above, is valid in the two renormalization schemes studied here, implies
\beq
\label{eq:relationgammac}
 \gamma_c=\frac{\beta_\xi}{\xi}+\frac{\beta_{m^2}}{m^2}.
\eeq
These RG functions are obtained from the expressions of the renormalization factors $Z_\alpha$ in each of the renormalization schemes described previously, that is from Eqs.~\eqn{eq:gfgf}--\eqn{eq:gfgfhh} in the case of the zero-momentum scheme or from the corresponding ones in the infrared-safe scheme.
We integrate numerically the flow equations \eqn{Rg_eq_1}--\eqn{anomalous_ghost} with initial conditions at the scale $\mu_0=1$~GeV. We use as initial conditions the values of the mass and coupling parameters of the previous section, namely $m(\mu_0)=0.54$~GeV and $g(\mu_0)=4.9$ for the zero-momentum scheme and $m(\mu_0)=0.39$~GeV and $g(\mu_0)=3.7$ for the infrared-safe scheme, and we vary the gauge-fixing parameter $\xi(\mu_0)$.

\subsection{Zero-momentum scheme}

\begin{figure}[t]
  \centering
  \includegraphics[width=1\linewidth]{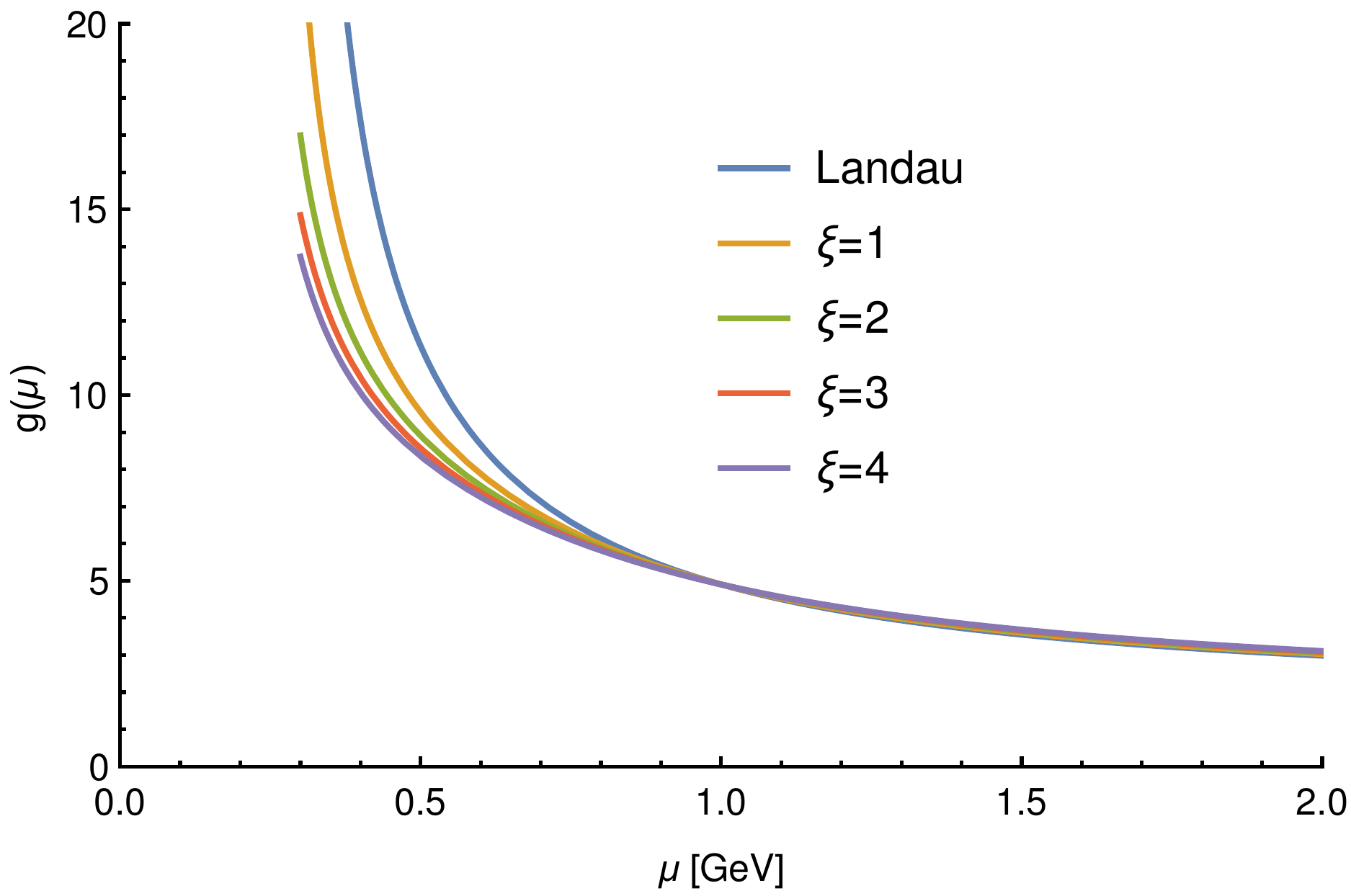}\\
  \includegraphics[width=1\linewidth]{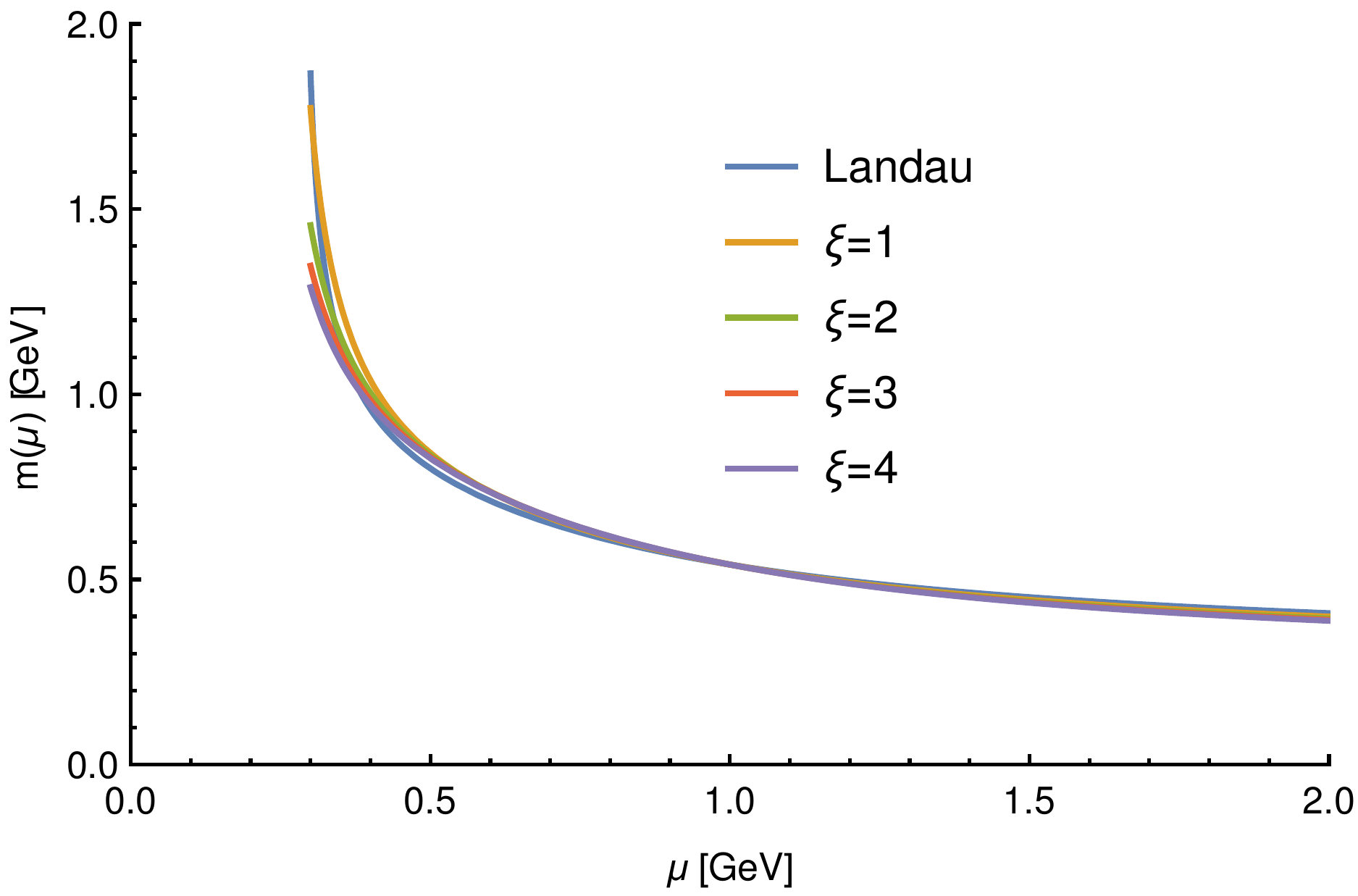}\\
  \includegraphics[width=1\linewidth]{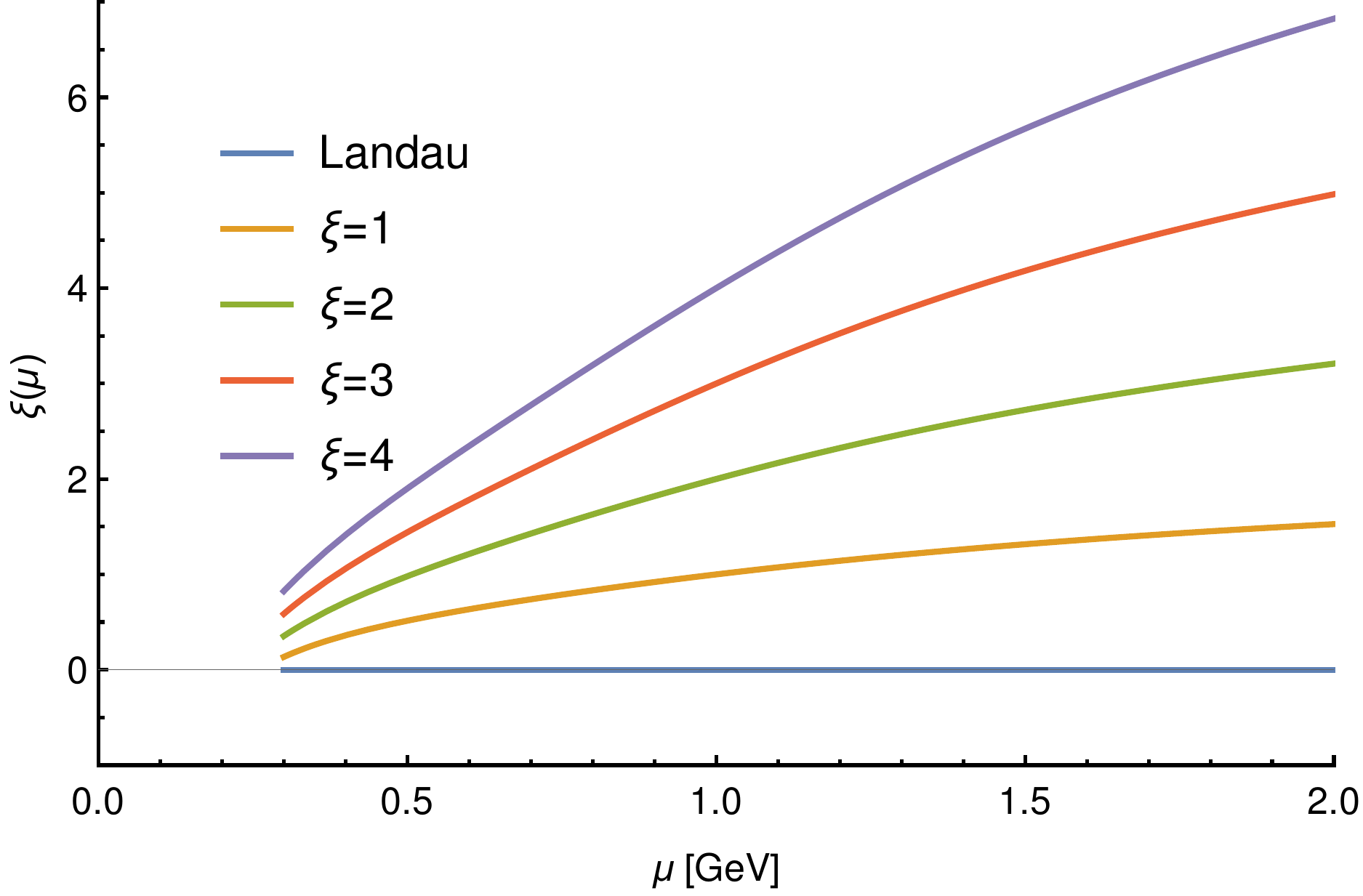}
  \caption{The running of the parameters $g(\mu)$, $m(\mu)$, and $\xi(\mu)$ in the zero-momentum scheme, for various values of $\xi (\mu_0) \equiv \xi$. We observe a Landau pole at a finite scale, where $g$ and $m$ diverge, and the RG flow stops.}
  \label{fig_RG_flow_coupling_constant_scheme2}
\end{figure}

\begin{figure}[ht]
  \centering
  \includegraphics[width=1\linewidth]{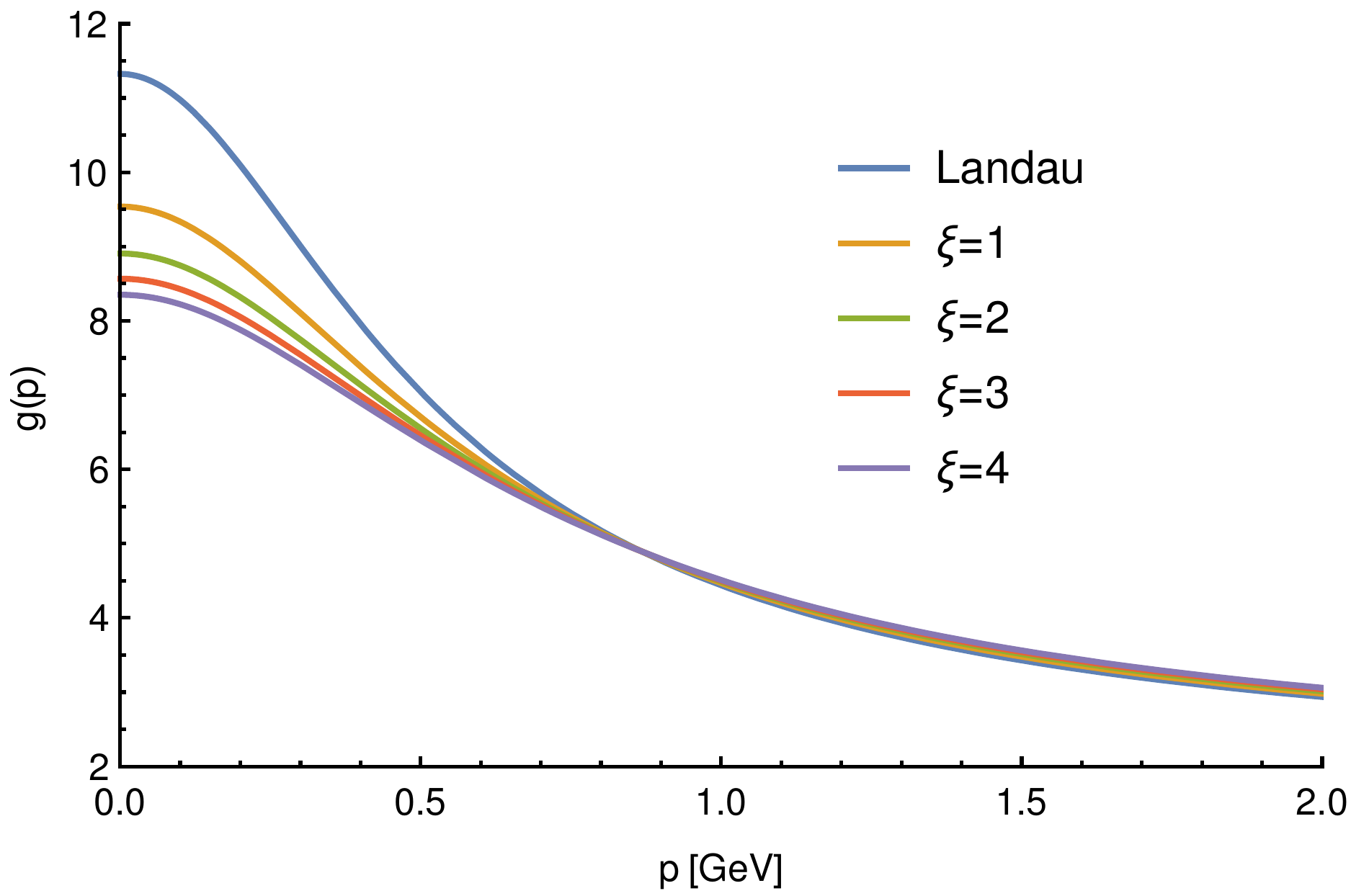}\\
  \includegraphics[width=1\linewidth]{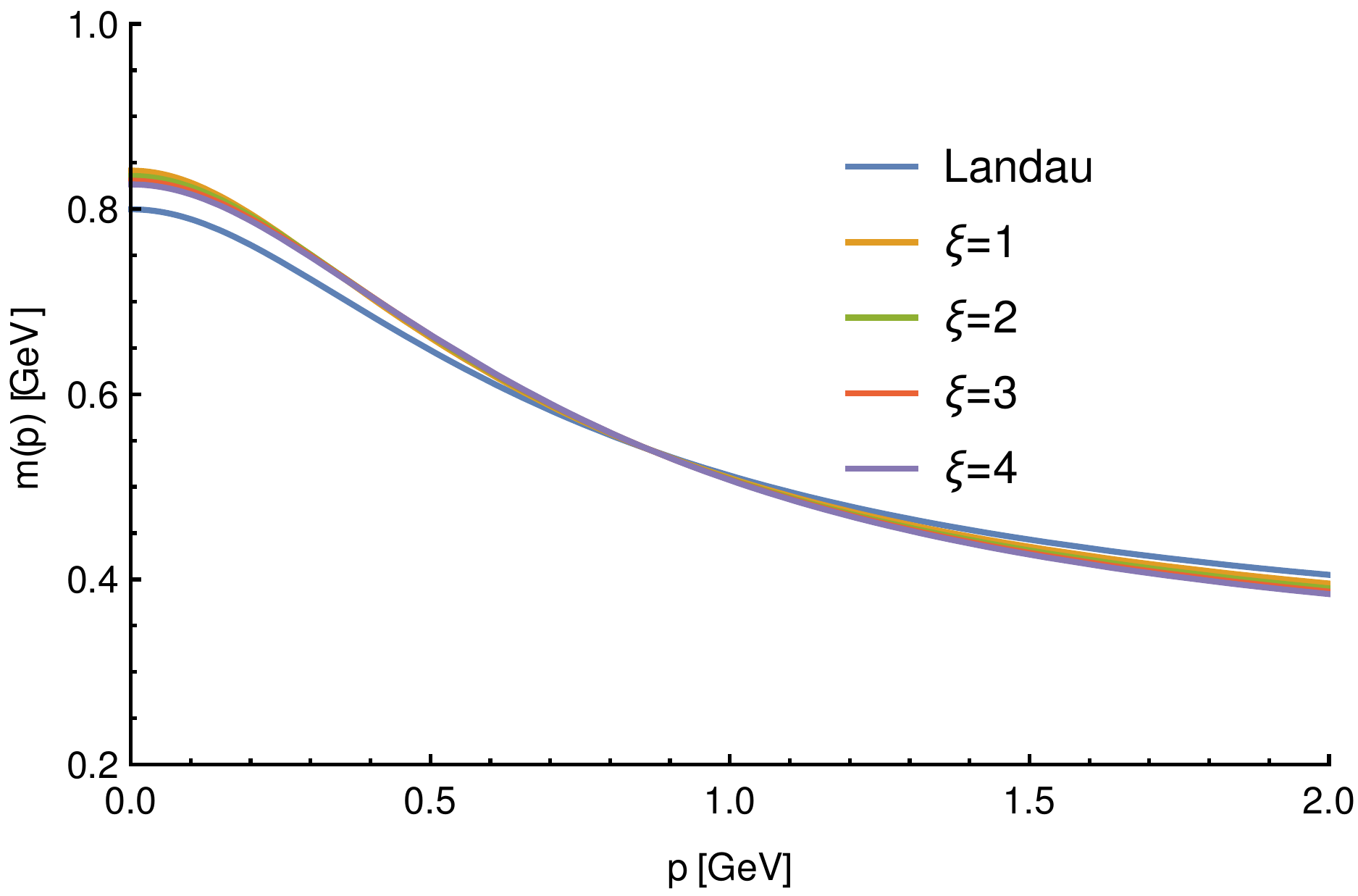}\\
  \includegraphics[width=1\linewidth]{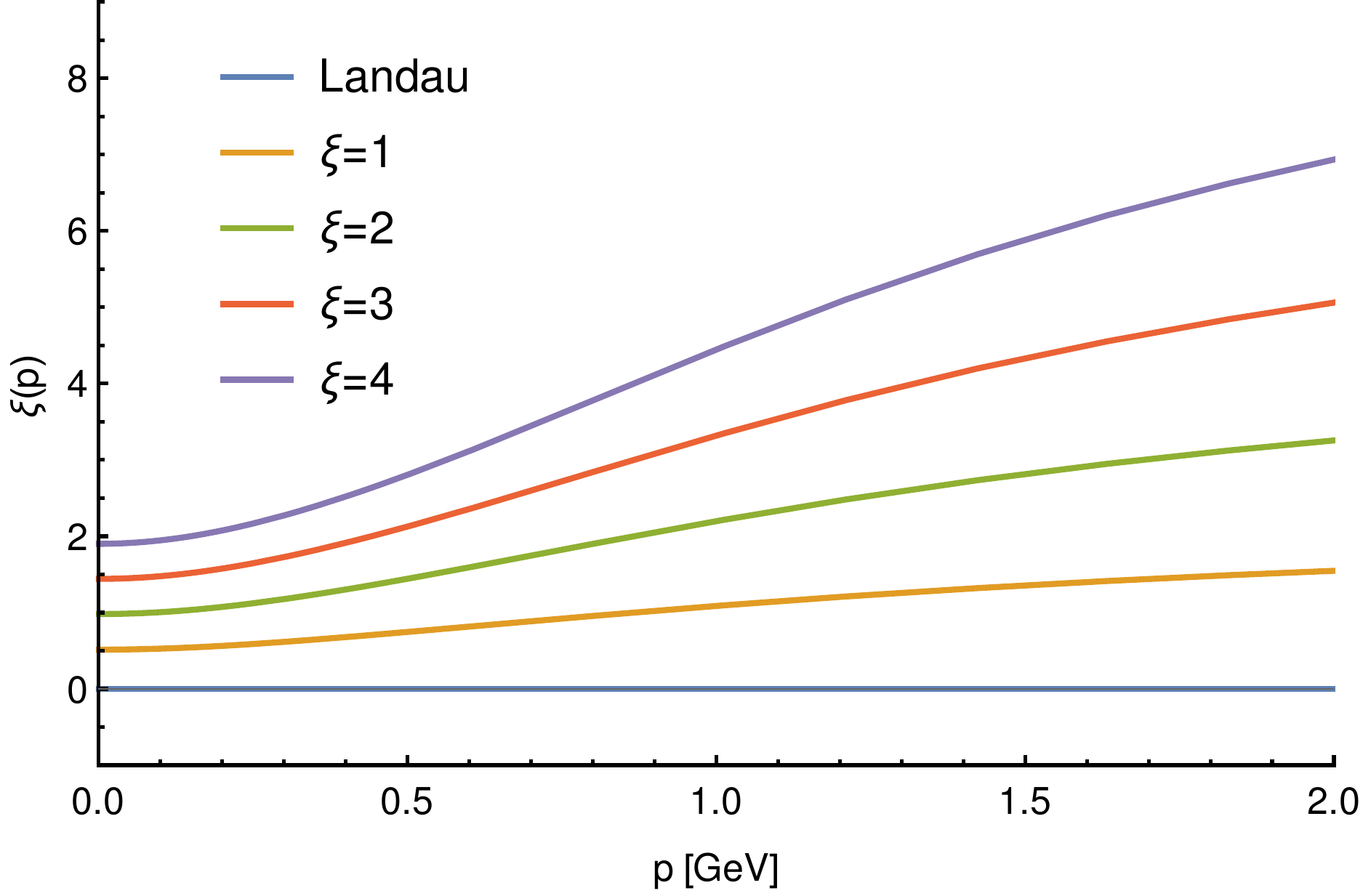}
  \caption{The parameters $g(\mu)$, $m(\mu)$, and $\xi(\mu)$ as functions of the momentum $p$, with $\mu =\sqrt{p^2+M^2}$ and $M=0.5$~GeV in the zero-momentum scheme, for various values of $\xi (\mu_0) \equiv \xi$. The RG flow is effectively frozen for $p\lesssim M$, which avoids the Landau pole.}
  \label{fig_RG_flow_coupling_constant_scheme22}
\end{figure}

In this scheme, the first condition \eqn{Rs2 renormalization presciption 1} implies the further relation
\beq
 \gamma_A=\frac{\beta_{m^2}}{m^2},
\eeq
which, when combined with Eqs.~\eqn{anomalous_ghost} and \eqn{eq:relationgammac}, yields
\beq
 \frac{\xi(\mu)}{\xi(\mu_0)}=\frac{g(\mu)}{g(\mu_0)}\left(\frac{m(\mu_0)}{m(\mu)}\right)^{3}.
\eeq
The running of the parameters $g$, $m$, and $\xi$, obtained from the direct integration of the flow equations in the zero-momentum scheme, is shown in \Fig{fig_RG_flow_coupling_constant_scheme2}. We find a Landau pole at a finite scale, where the parameters $g$ and $m$ diverge, for all values of $\xi(\mu_0)$. In the case $\xi=0$, it was argued in Ref.~\cite{Tissier_10} that this originates from the fact that, at small $\mu$, the renormalization conditions \eqn{Rs2 renormalization presciption 1} and \eqn{Rs2 renormalization presciption 2}, which imply that $G_T(0)>G_T(\mu)$ conflict with the gluon propagator computed at one-loop order, which is an increasing function of $p$ at small $p$. The Landau pole prevents one from following the RG flow down to deep infrared scales. The proposal of Ref.~\cite{Tissier_10} to bypass this problem is to freeze by hand the RG flow under a certain arbitrary scale $M$, chosen above the Landau pole such that the latter is avoided. In the case $\xi=0$, treated in Ref.~\cite{Tissier_10}, this could be justified by a detailed analysis of higher loop contributions in the infrared. We shall not attempt to generalize this discussion to the case $\xi\neq0$ here. Still, we present for the sake of illustration, a modified zero-momentum scheme where we freeze the RG flow by hand under a certain scale, in the spirit of Ref.~\cite{Tissier_10}. 
In practice, we use an RG scale $\mu=\sqrt{p^2+M^2}$, where $p$ is the actual momentum scale at which we evaluate correlation functions. In what follows, we use $M=0.5$~GeV, which is of the order of the typical value of the mass parameter $m$. The corresponding flow of the coupling constant as a function of $p$ is shown in \Fig{fig_RG_flow_coupling_constant_scheme22}.

\begin{figure}[ht]
  \centering
  \includegraphics[width=1\linewidth]{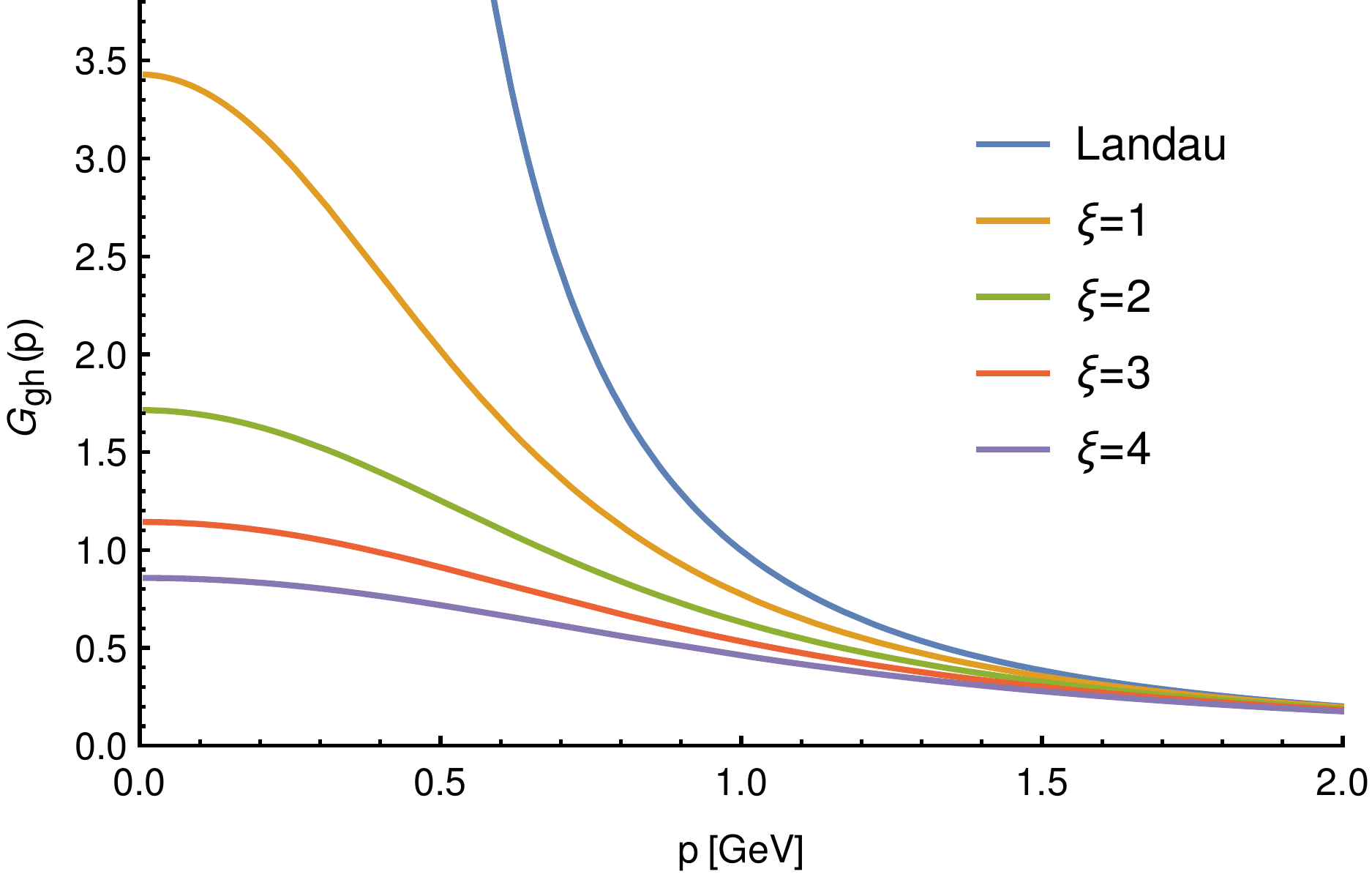}\\
  \includegraphics[width=1\linewidth]{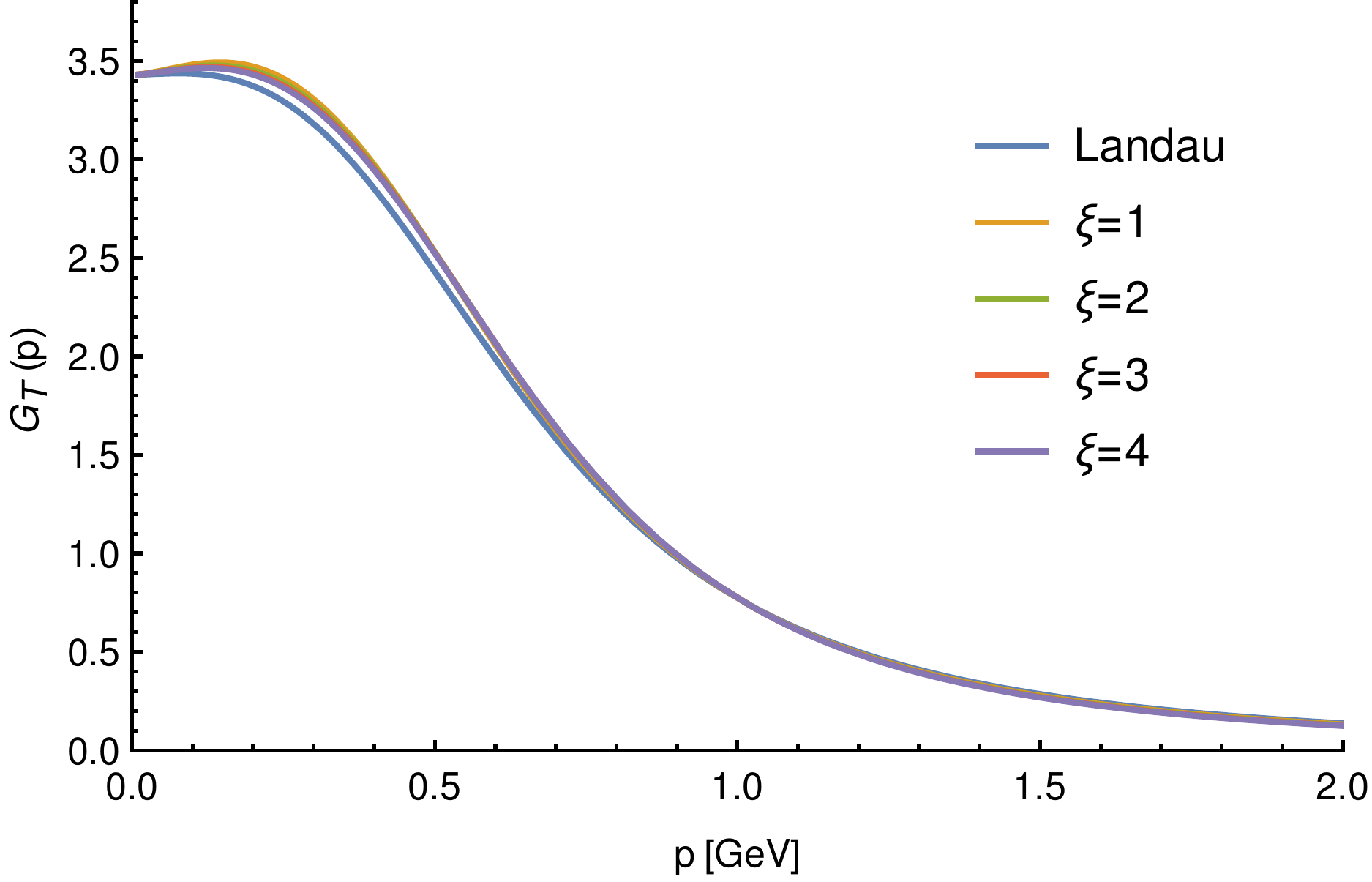}
  \caption{RG-improved ghost (top) and transverse gluon (bottom) propagators as functions of momentum in the zero-momentum prescription scheme with $\mu =\sqrt{p^2+M^2}$, for various values of $\xi (\mu_0) \equiv \xi$. }
  \label{fig_gluon_propagator_scheme2RG}
\end{figure}

We implement this modified RG scheme to compute the flow of all parameters from which we evaluate the RG-improved ghost and gluon propagators, shown in \Fig{fig_gluon_propagator_scheme2RG}. We see that, despite the relatively important flow of the parameters, the propagators are almost unaffected as compared to those of strict perturbation theory in this scheme; see \Fig{fig_gluon_propagator_scheme2}. 

Although the above procedure allows one to avoid the Landau pole, the need to introduce the arbitrary scale $M$ is somewhat uncomfortable. Moreover, despite the fact that most excitations in the present theory are massive, there remain massless degrees of freedom in the replicated superfield sector (which is only absent for the CF model, $n=1$) and it is not clear that the freezing of the RG flow, implemented by hand in the above procedure, is justified. It is the purpose of the infrared-safe scheme to provide a more systematic procedure which avoids the Landau pole \cite{Tissier_10}, as we now discuss. 

\subsection{Infrared-safe scheme}

The relaxation of the first condition \eqn{Rs2 renormalization presciption 1}---replaced by the second condition \eqn{IRs renormalization presciption Zm} in the infrared-safe scheme---avoids the conflicting prescriptions at small $\mu$ discussed above. As a consequence, we find that the infrared-safe RG flow can be integrated down to arbitrarily small scales $\mu$, depending on the choice of initial conditions\footnote{For instance, it is clear that in the case $m(\mu_0)=\xi(\mu_0)=0$ [which implies $m(\mu)=\xi(\mu)=0$ for all $\mu$ and thus corresponds to the standard (Faddeev-Popov) Landau gauge] one gets a Landau pole.}, as was first pointed out in Refs.~\cite{Tissier_10,Serreau:2012cg} in the case $\xi=0$.

\begin{figure}[ht]
\begin{center}
\includegraphics[scale=0.4]{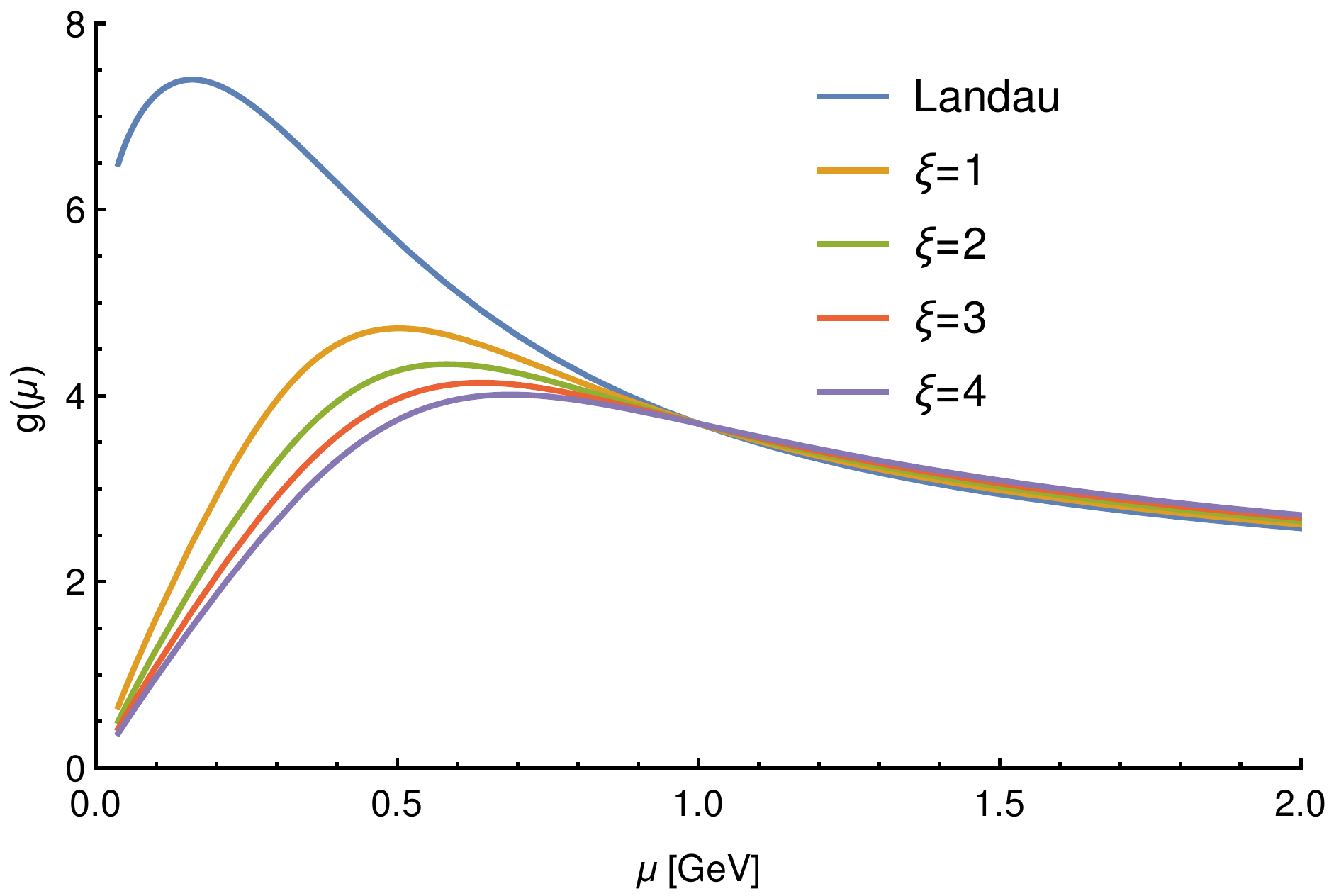}\\
\includegraphics[scale=0.4]{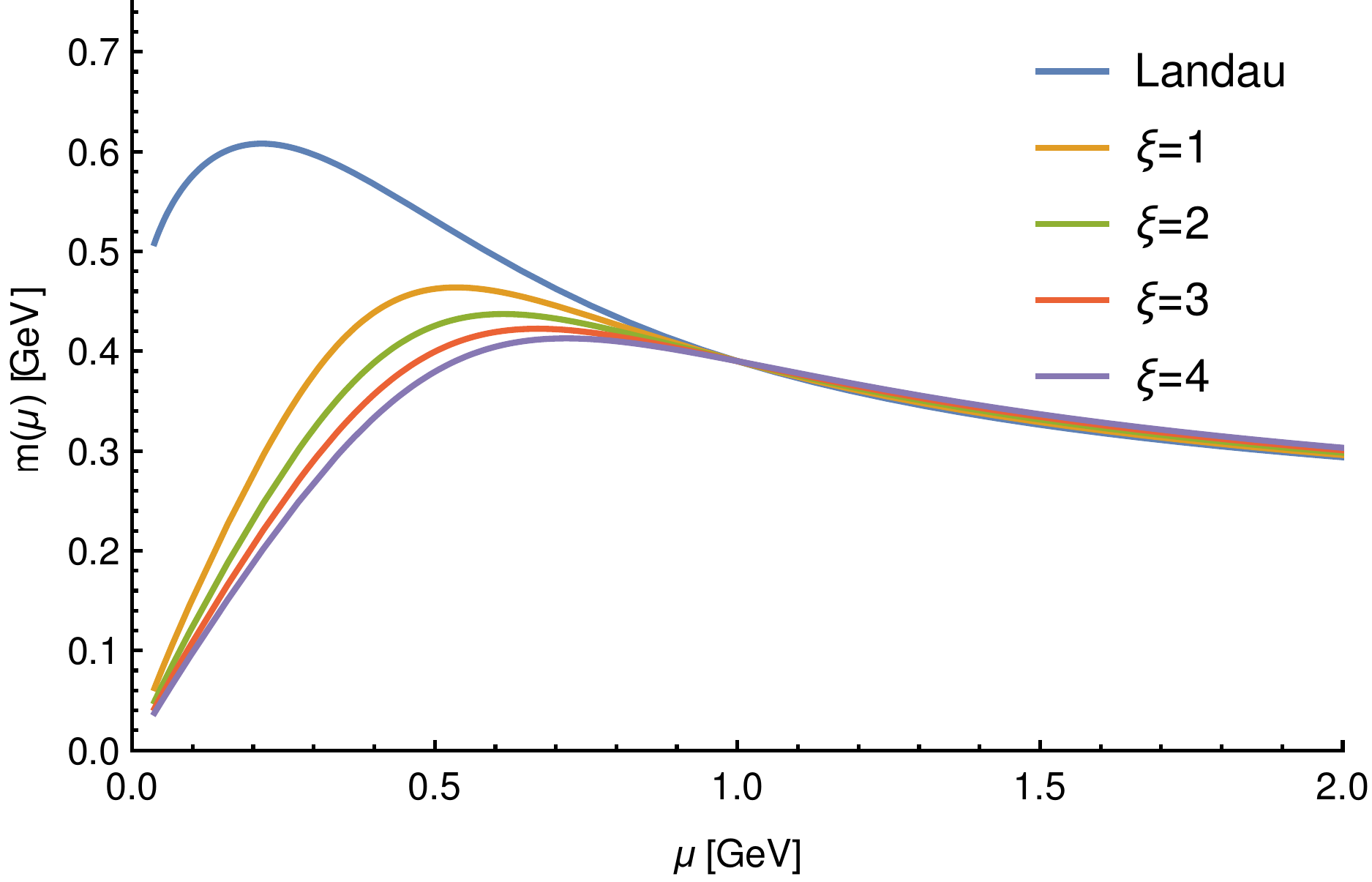}\\
\includegraphics[scale=0.4]{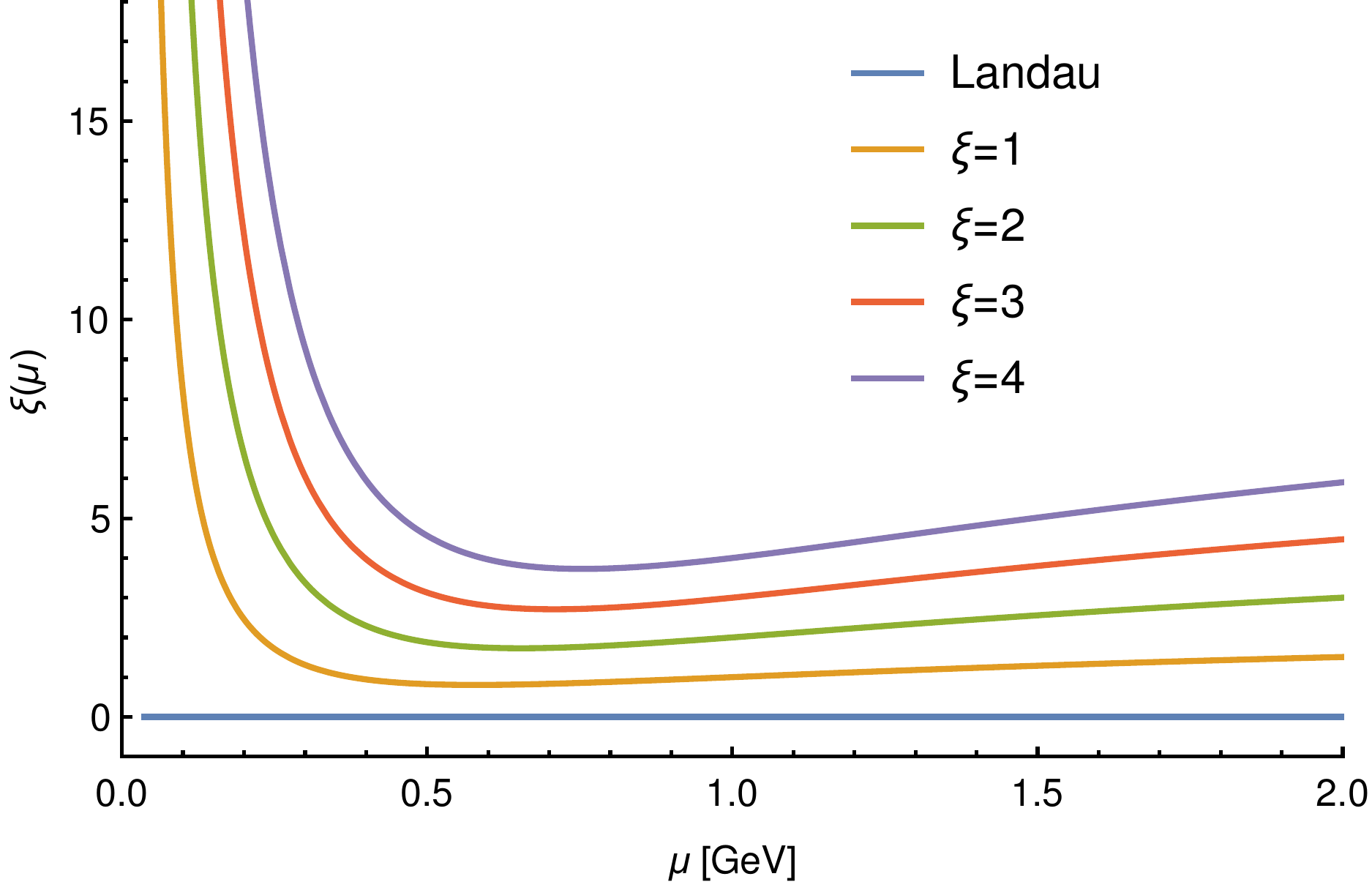}
\end{center}
\caption{Running of the parameters $g(\mu)$, $m(\mu)$, and $\xi(\mu)$ in the infrared-safe scheme, for various values of $\xi (\mu_0) \equiv \xi$.}
\label{gauge_parameter_RG_flow_IRsafe}
\end{figure}

In this scheme, the relation $\lim_{n\to0}Z_A/Z_\xi=1$ discussed above implies that 
\beq
\label{eq:gammaAirsafe}
 \gamma_A=-\frac{\beta_\xi}{\xi},
\eeq
from which we get, using \Eqn{anomalous_ghost} and \eqn{eq:relationgammac},
\beq
\label{eq:xirunir}
  \frac{\xi(\mu)}{\xi(\mu_0)}=\frac{g^2(\mu)}{g^2(\mu_0)}\left(\frac{m(\mu_0)}{m(\mu)}\right)^{4}.
\eeq 
Figure \ref{gauge_parameter_RG_flow_IRsafe} shows the RG flows of the parameters $g$, $m$ and $\xi$ for various values of $\xi(\mu_0)$. We observe that both the coupling and the mass first increase for decreasing $\mu$ and are then attracted towards zero in the infrared. We also see that the maximal values of both parameters decrease with increasing $\xi(\mu_0)$ and are therefore, maximal in the Landau gauge. In all cases we have considered, the coupling remains small enough for perturbation theory to be (qualitatively) meaningful: we recall that the relevant expansion parameter is $3g^2/(16\pi^2)$. Finally, we observe that the gauge-fixing parameter is first attracted towards zero as $\mu$ decreases but eventually diverges in the limit $\mu\to0$. In particular, we find that the Landau gauge fixed point ($\xi=0$) is unstable in the infrared. 
\begin{figure}[ht]
\begin{center}
\includegraphics[scale=0.4]{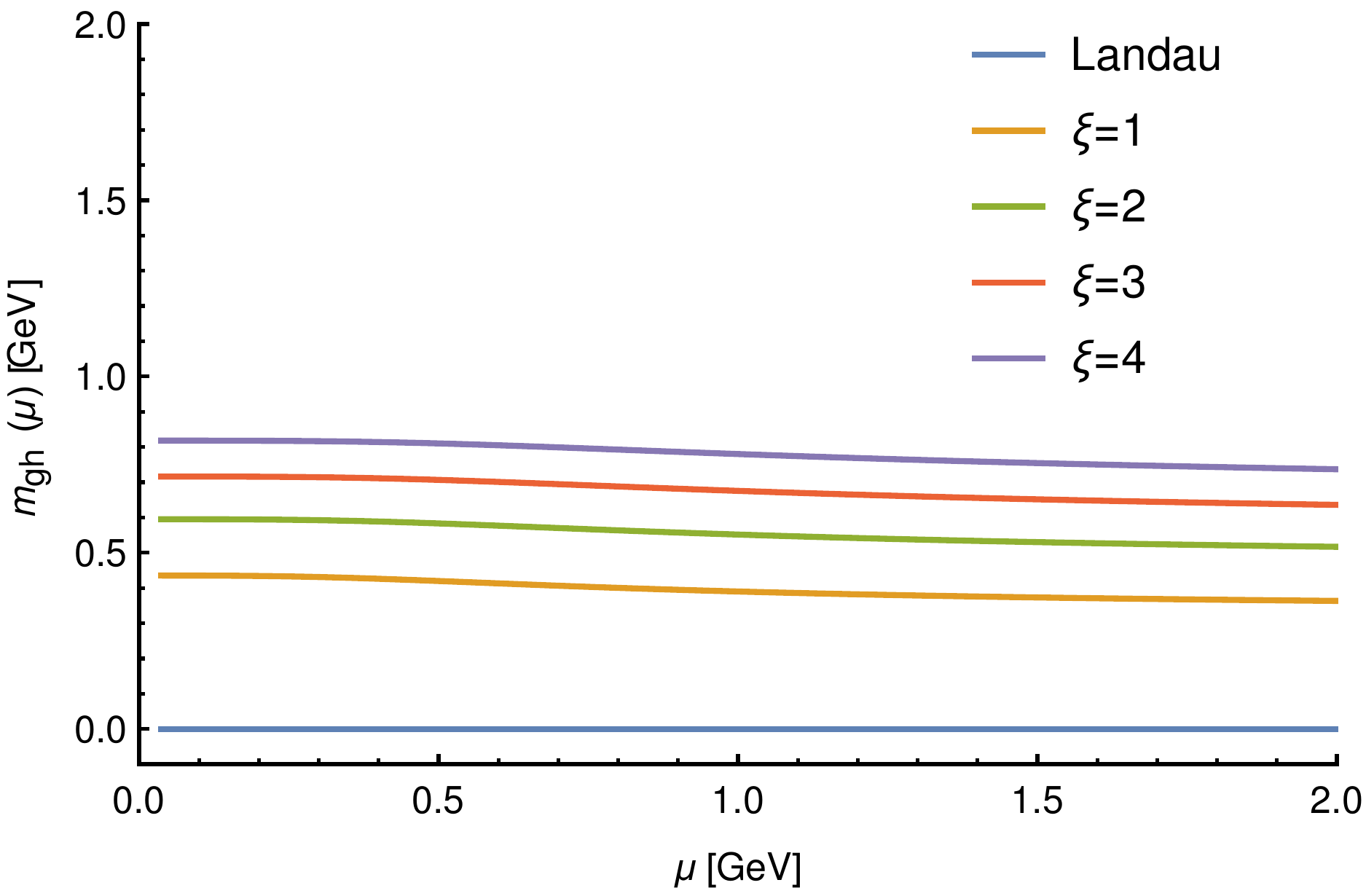}
\end{center}
\caption{Running of the ghost mass parameter $m_{\rm gh}(\mu)=m(\mu)\sqrt{\xi(\mu)}$ in the infrared-safe scheme, for various values of $\xi (\mu_0) \equiv \xi$.}
\label{ghost_mass_RG_flow_IRsafe}
\end{figure}
It is interesting to check the running of the ghost mass parameter,
\beq
\label{eq:ghmass}
 m_{\rm gh}(\mu)=m(\mu)\sqrt{\xi(\mu)},
\eeq
shown in  \Fig{ghost_mass_RG_flow_IRsafe}. Using \Eqn{eq:xirunir}, we have
\beq
 \frac{m_{\rm gh}(\mu)}{m_{\rm gh}(\mu_0)}=\frac{g(\mu)}{g(\mu_0)}\frac{m(\mu_0)}{m(\mu)}.
\eeq
We observe that $m_{\rm gh}(\mu)$ is attracted towards a nontrivial fixed point in the infrared, which shows that $g(\mu)\sim m(\mu)$ and $\xi(\mu)\sim 1/m^2(\mu)\sim1/g^2(\mu)$ when $\mu\to0$.

An important remark to be made here is that the RG flows of the independent parameters $g$ and $m$ never freeze out despite the fact that most degrees of freedom of the theory are massive. This is due to the fact that there remains, in fact, massless excitations in the superfield sector, as can be seen from the tree-level propagators \eqn{eq_propagLL} and \eqn{eq_propagLA}. This is a particular feature of the present theory and, more precisely, of the way Gribov copies are handled. As we shall see below the RG flow of the CF model, where all degrees of freedom are indeed massive, is qualitatively different. This shows that the freezing of the flow introduced by hand in the previous scheme is unrealistic. In fact, the nonfreezing flow observed in the present scheme leads to dramatic RG corrections on the gluon propagator, as we now discuss.

\begin{figure}[ht]
  \centering
  \includegraphics[width=1\linewidth]{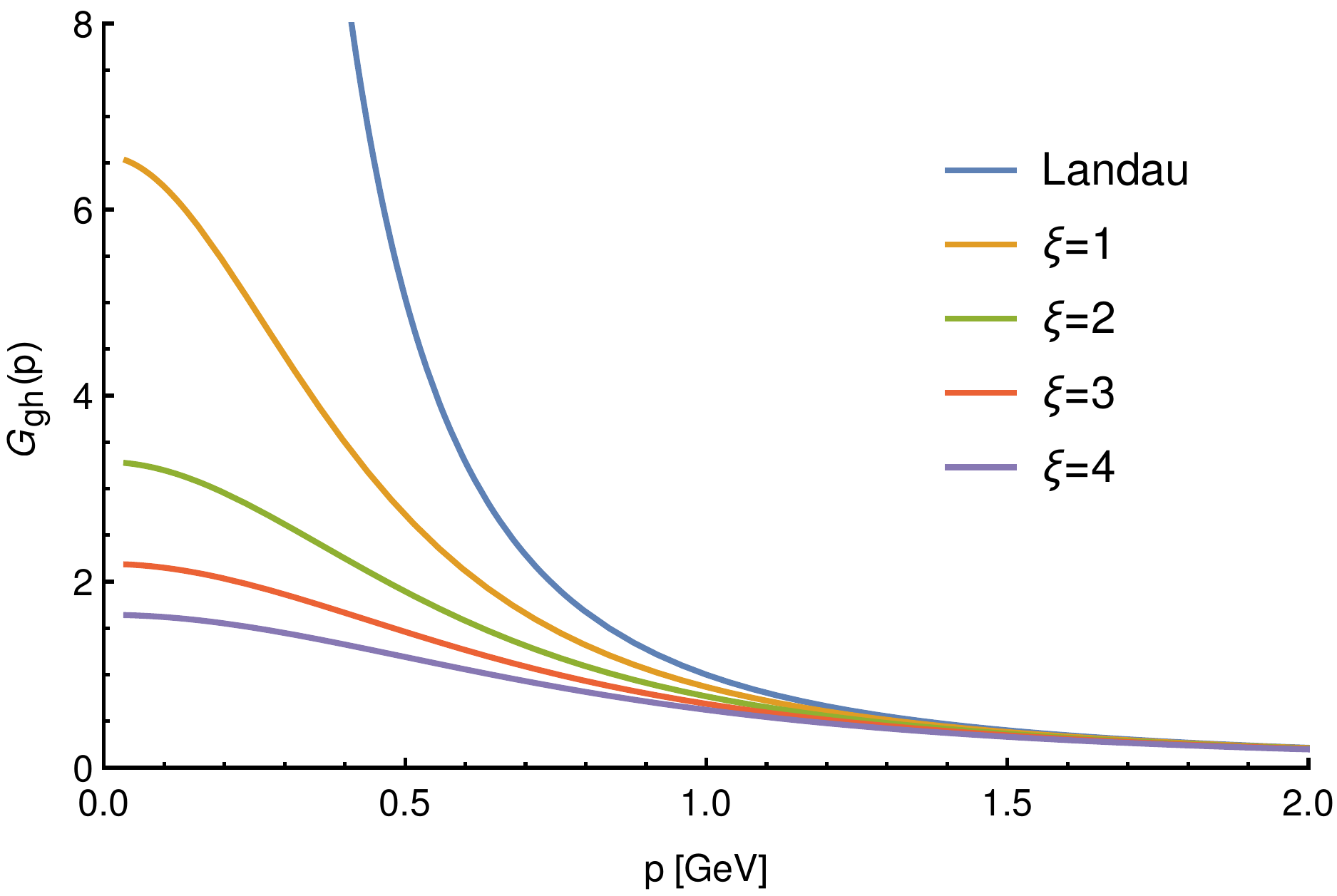}\\
 \includegraphics[width=1\linewidth]{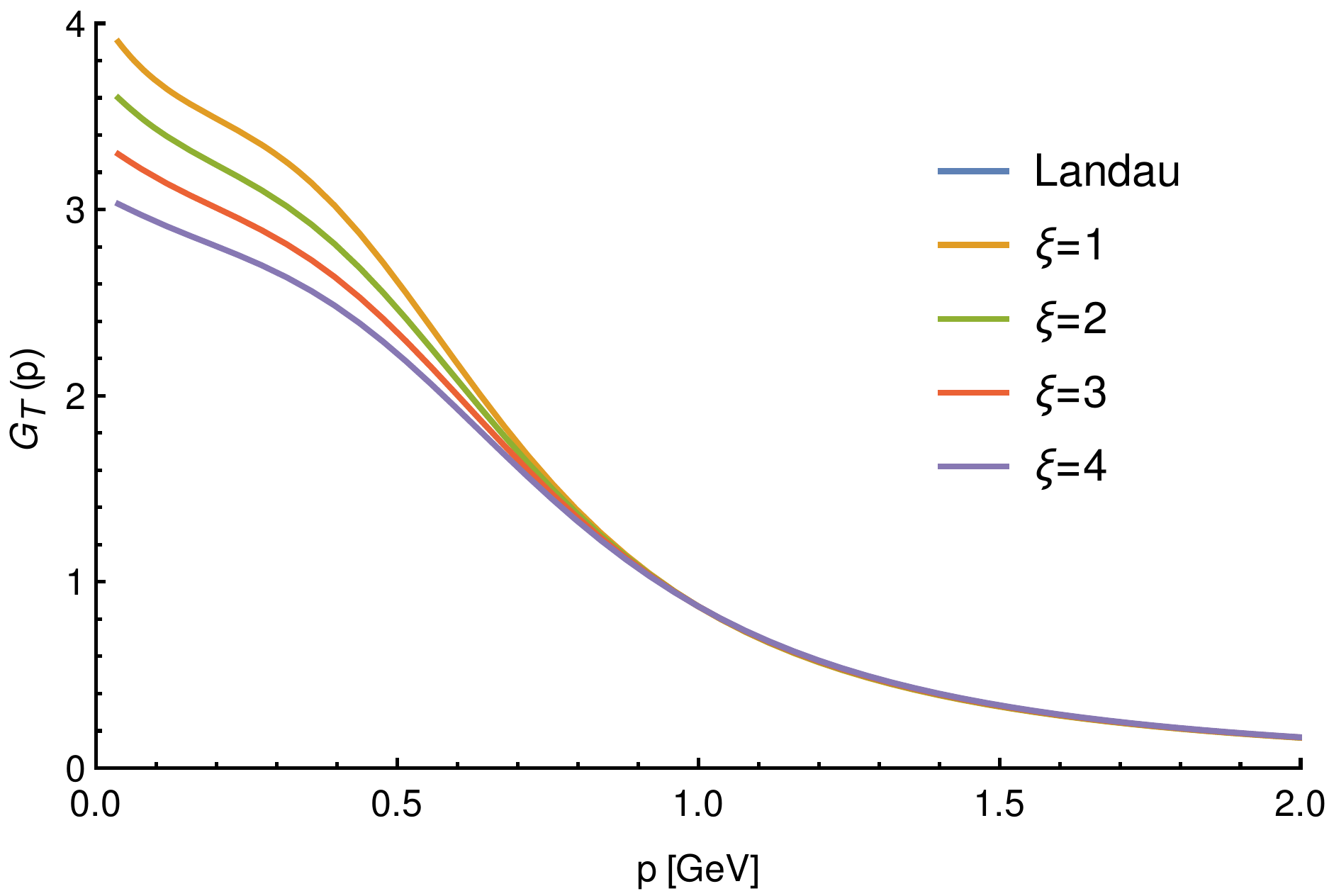}
  \caption{RG-improved ghost (top) and transverse gluon (bottom) propagators as functions of momentum in the infrared-safe scheme with $\mu=p$, for various values of $\xi (\mu_0) \equiv \xi$. }
  \label{fig_gluon_propagator_schemeIRsafe_RG}
\end{figure}

We show in \Fig{fig_gluon_propagator_schemeIRsafe_RG} the RG-improved ghost and (transverse) gluon propagators, where we used, for the RG scale, $\mu=p$. Using \Eqn{eq:relationgammac}, we obtain, for the RG-improved ghost propagator, 

\beq
\label{eq:rgghost}
 G_{\rm gh}(p)=\frac{m^2_{\rm gh}(p)}{m^2_{\rm gh}(\mu_0)}\frac{1}{p^2+m^2_{\rm gh}(p)},
\eeq
where $m_{\rm gh}(\mu)$ is defined in \Eqn{eq:ghmass}. Since the latter saturates to a constant value in the infrared, as discussed above, we check that the value of the ghost propagator at zero momentum does not receive any RG correction:
\beq
 G_{\rm gh}(0)=\frac{1}{m^2_{\rm gh}(\mu_0)},
\eeq
as expected from the nonrenormalization relation \eqn{eq:jbsoin1}.
This is easily checked in \Fig{fig_gluon_propagator_schemeIRsafe_RG}. The fact that the ghost propagator is fixed to its tree-level value both at $p=0$ and $p=\mu_0$ also explains (together with the relative smallness of loop corrections in the ghost sector discussed previously) that the RG-improved ghost propagator is very similar to the one without RG improvement. 

Things are different for the gluon propagator. In that case, the relation \eqn{eq:gammaAirsafe} implies that the RG-improved transverse propagator reads
\beq
\label{eq:gluproirs}
 G_T(p)=\frac{\xi(\mu_0)}{\xi(p)}\frac{1}{p^2+m^2(p)}.
\eeq
We see in \Fig{fig_gluon_propagator_schemeIRsafe_RG} that the RG effects strongly modify the gluon propagator as compared to its strict one-loop expression; see \Fig{fig_gluon_propagator_schemeIRsafe}. In particular, both its value at $p=0$ and the way the latter is approached are dramatically altered, even for small values of $\xi(\mu_0)$. The flattening near $p=0$, observed in strict perturbation theory, is turned into a linear behavior. This can be understood as follows. We first rewrite \Eqn{eq:gluproirs} as
\beq
 G_T(p)=\frac{\xi(\mu_0)}{m_{\rm gh}^2(p)}\frac{\tilde m^2(p)}{1+\tilde m^2(p)},
\eeq
where we have defined the dimensionless mass parameter $\tilde m(\mu)=m(\mu)/\mu$. The fact that the ghost mass parameter \eqn{eq:ghmass} reaches a plateau for sufficiently small $\mu$ and not too small $\xi(\mu_0)$ implies that the $p\to0$ behavior of the gluon propagator is governed by the function $\tilde m^2(p)/[1+\tilde m^2(p)]$, a monotonously

 \begin{figure}[ht]
\begin{center}
\includegraphics[scale=0.4]{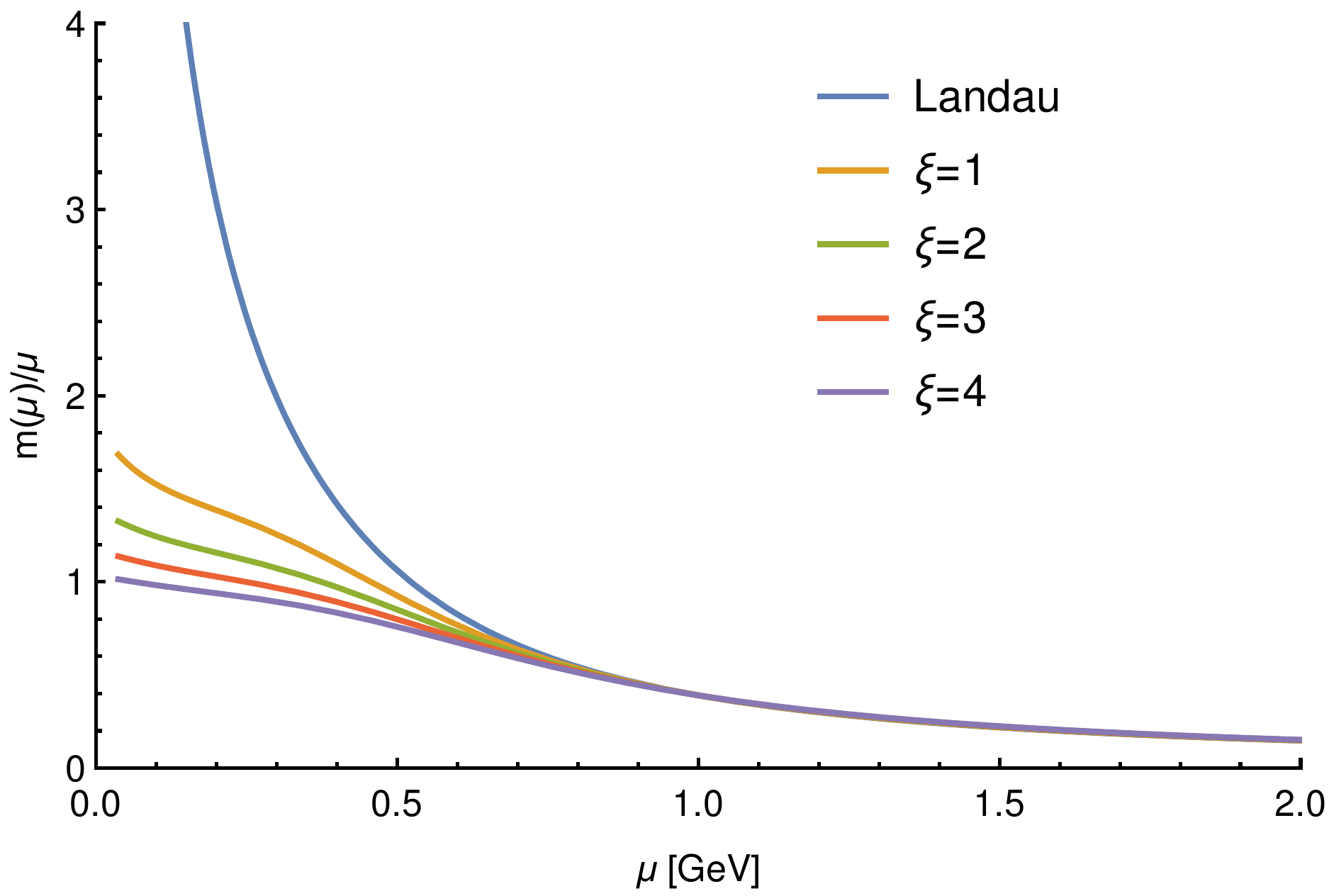}
\end{center}
\caption{Running of the dimensionless mass parameter $\tilde m(\mu)=m(\mu)/\mu$ in the infrared-safe scheme, for various values of $\xi (\mu_0) \equiv \xi$.}
\label{dimlessglu_mass_RG_flow_IRsafe}
\end{figure}
increasing function of $\tilde m^2(p)$. The running of the latter is shown in \Fig{dimlessglu_mass_RG_flow_IRsafe}, where we observe that it is a monotonously decreasing function of $\mu$ and that, for $\xi(\mu_0)$ not too small, it reaches an infrared fixed point, which is approached linearly. We conclude that for $\xi(\mu_0)$ not too small, 
\beq
 G_T(p\to0)=\frac{\xi(\mu_0)}{m_{\rm gh}^2(0)}\frac{\tilde m^2(0)}{1+\tilde m^2(0)}\left[1+{\cal O}(p)\right],
\eeq
where the linear term in $p$ is negative.

\subsection{Comparison with the CF model}

\begin{figure}[ht]
  \centering
  \includegraphics[width=1\linewidth]{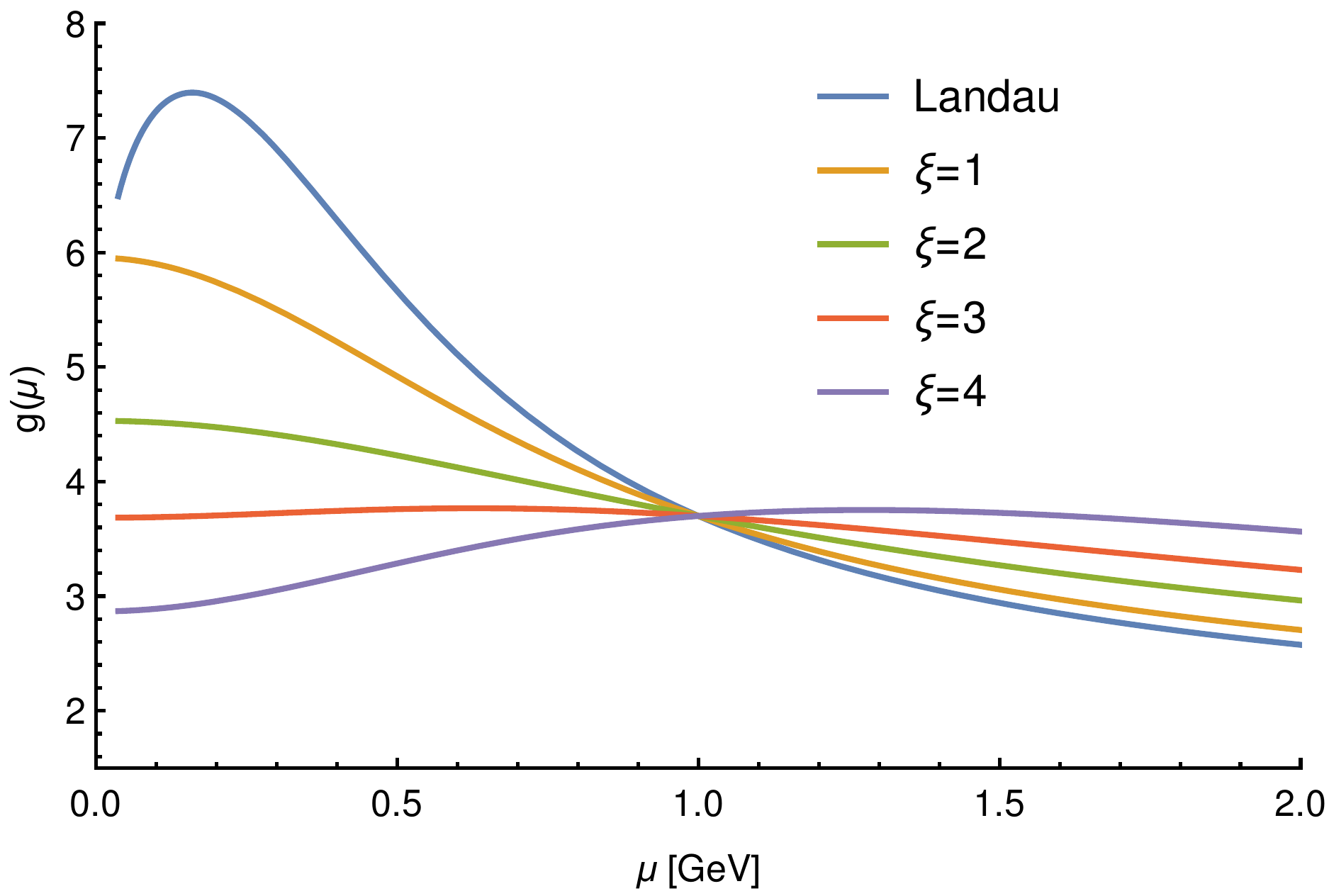}\\
  \includegraphics[width=1\linewidth]{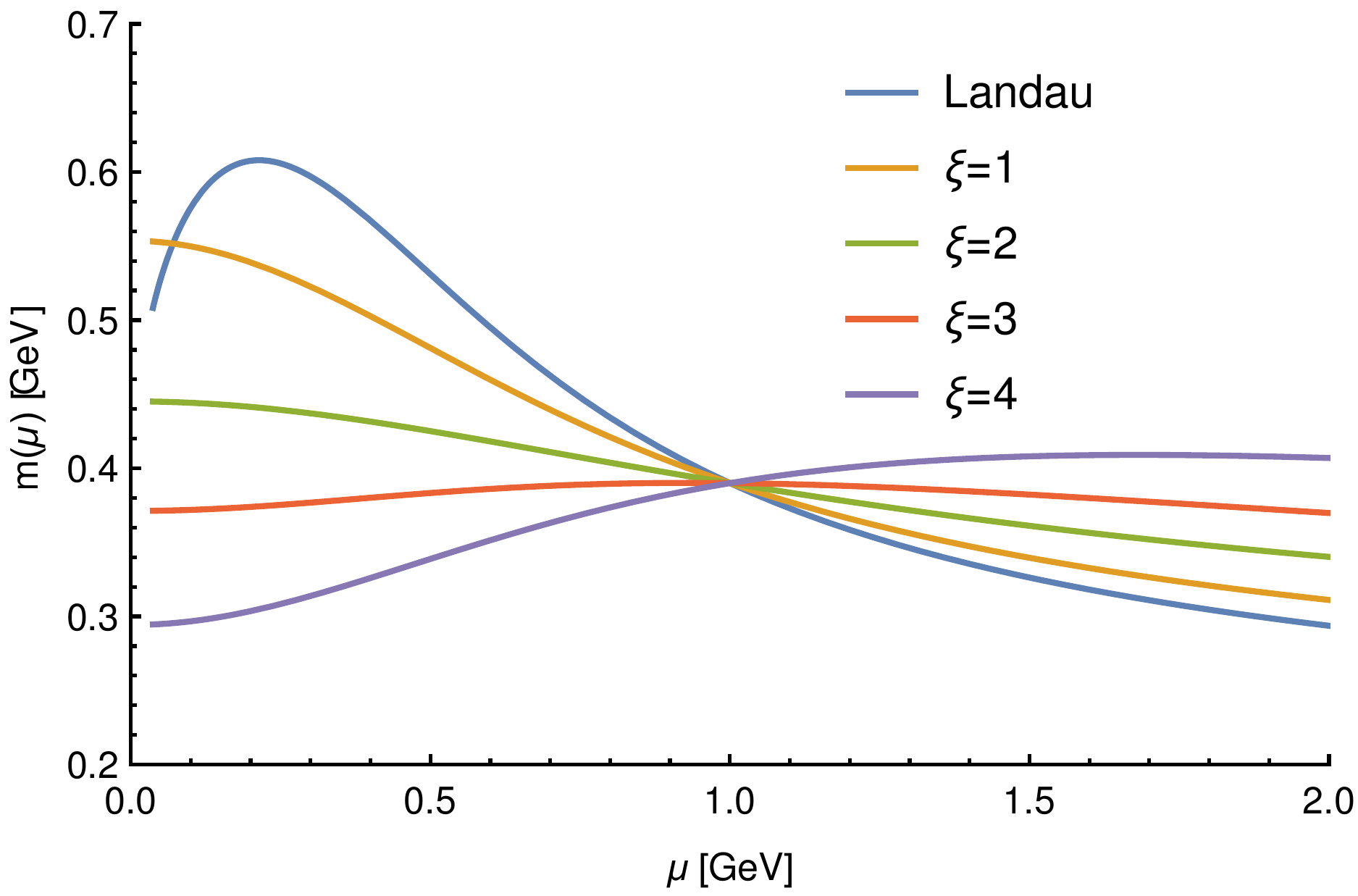}\\
  \includegraphics[width=1\linewidth]{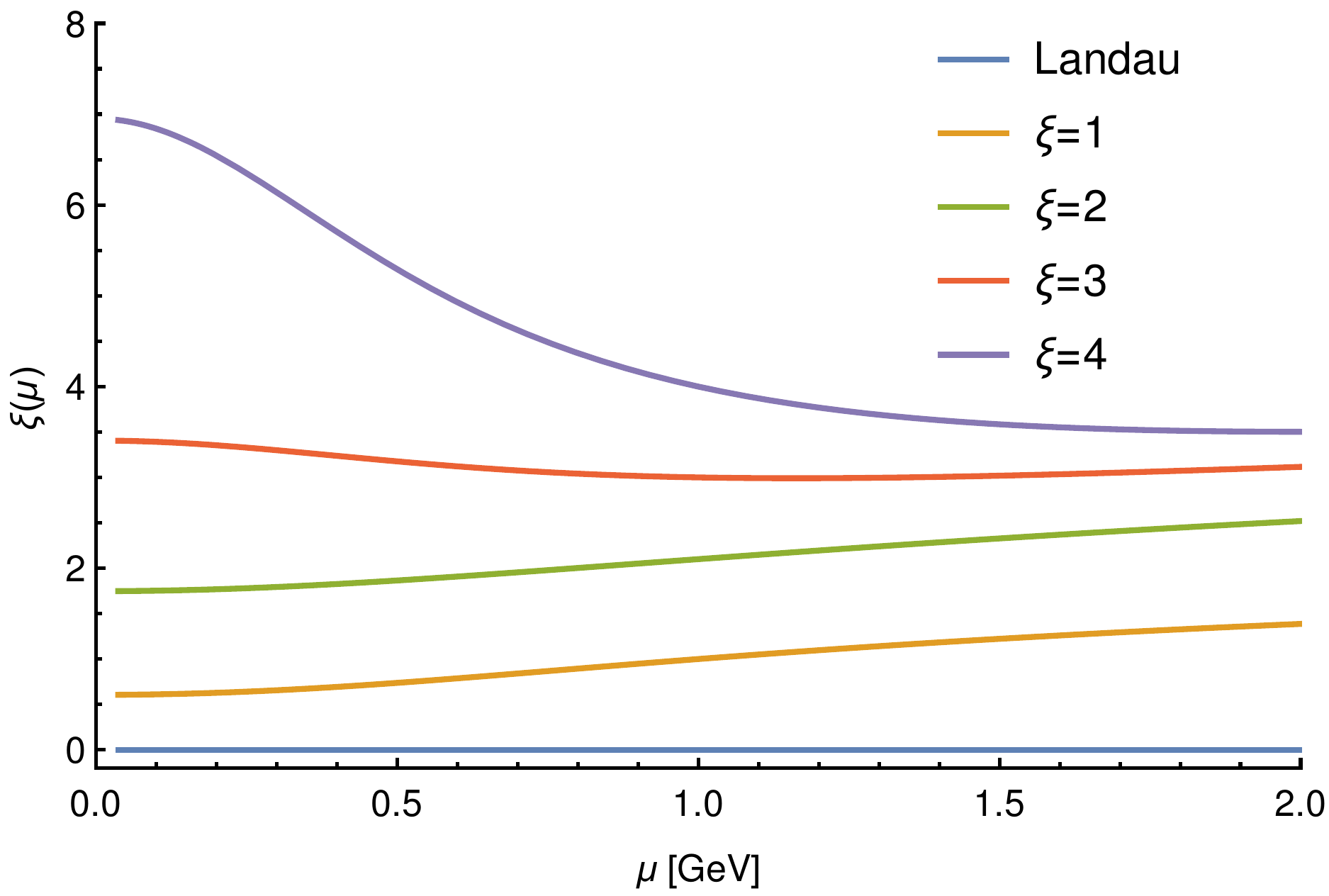}
  \caption{The RG flow of the parameters $g(\mu)$, $m(\mu)$, and $\xi(\mu)$ in the CF model ($n=1$) with the infrared-safe scheme.}
  \label{fig:CFrunning}
\end{figure}

We compare our results for the RG flow of the present theory ($n\to0$) with that of the CF model ($n=1$)\footnote{We recall that, in the CF model, our prescription for the renormalized coupling constant is given by \Eqn{eq_coupling_CF}.}. We only consider the infrared-safe scheme, for which no freezing of the flow in the infrared is put in by hand. The infrared safety of this scheme is also verified in the CF model, as can be seen in \Fig{fig:CFrunning}. 
Clearly, the RG flow of the CF model is dramatically different from that of the case $n\to0$. In particular, we observe that the RG flow freezes below a certain scale. This is to be expected since the CF model only contains massive degrees of freedom which decouple in the deep infrared. The only exception is the case $\xi(\mu_0)=0$, where the ghost becomes massless. In this case, the CF model is equivalent to the case $n\to0$, as already discussed, and we see that the RG flow does not freeze in the infrared. As a consequence of the RG freezing for $\xi(\mu_0)\neq0$, the coupling and mass parameters $g$ and $m$ do not vanish in the infrared and the gauge-fixing parameter $\xi$ does not diverge. All the parameters reach constant values.

\begin{figure}[ht]
  \centering
  \includegraphics[width=1\linewidth]{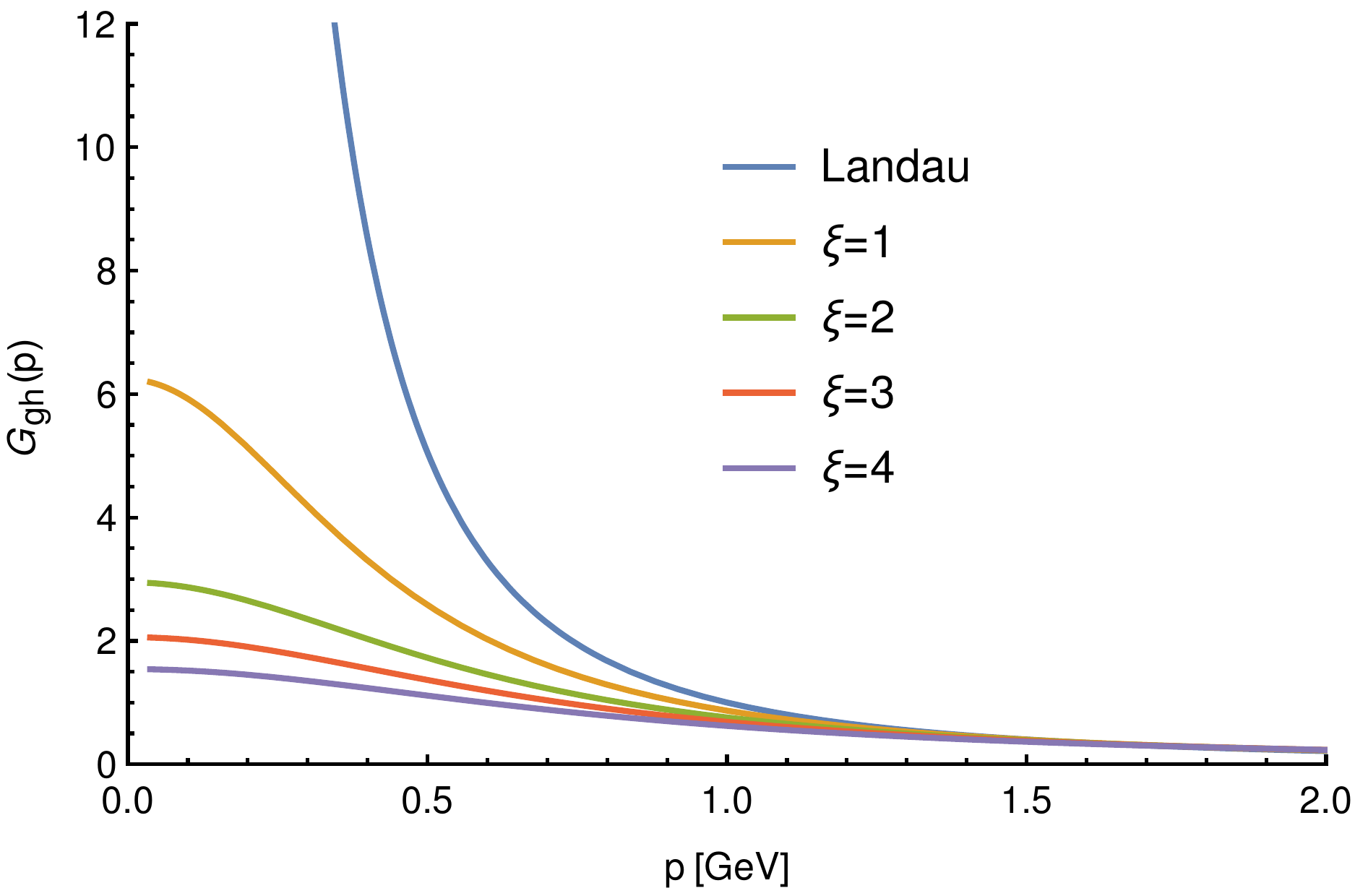}\\
  \includegraphics[width=1\linewidth]{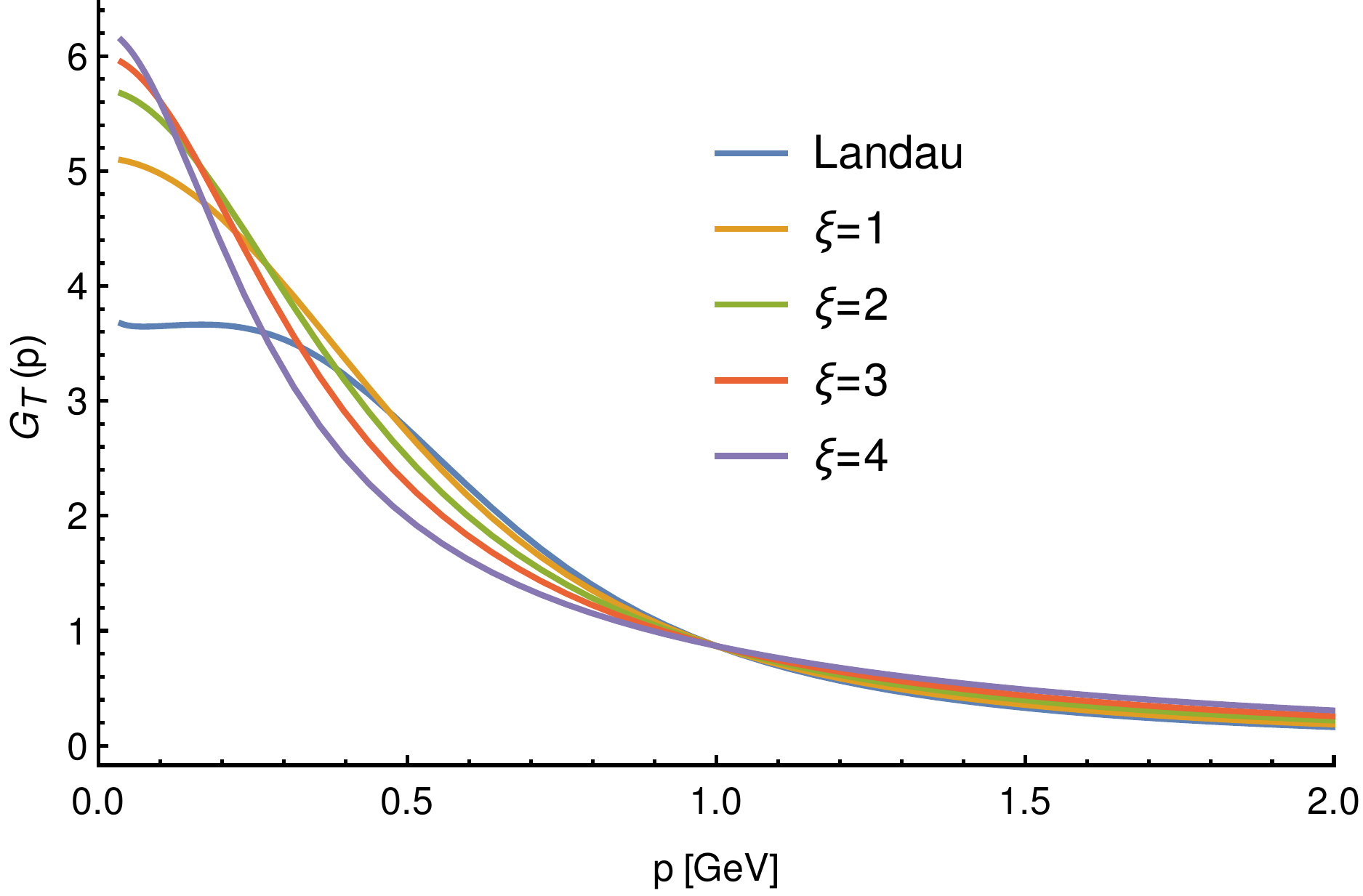}\\
  \includegraphics[width=1\linewidth]{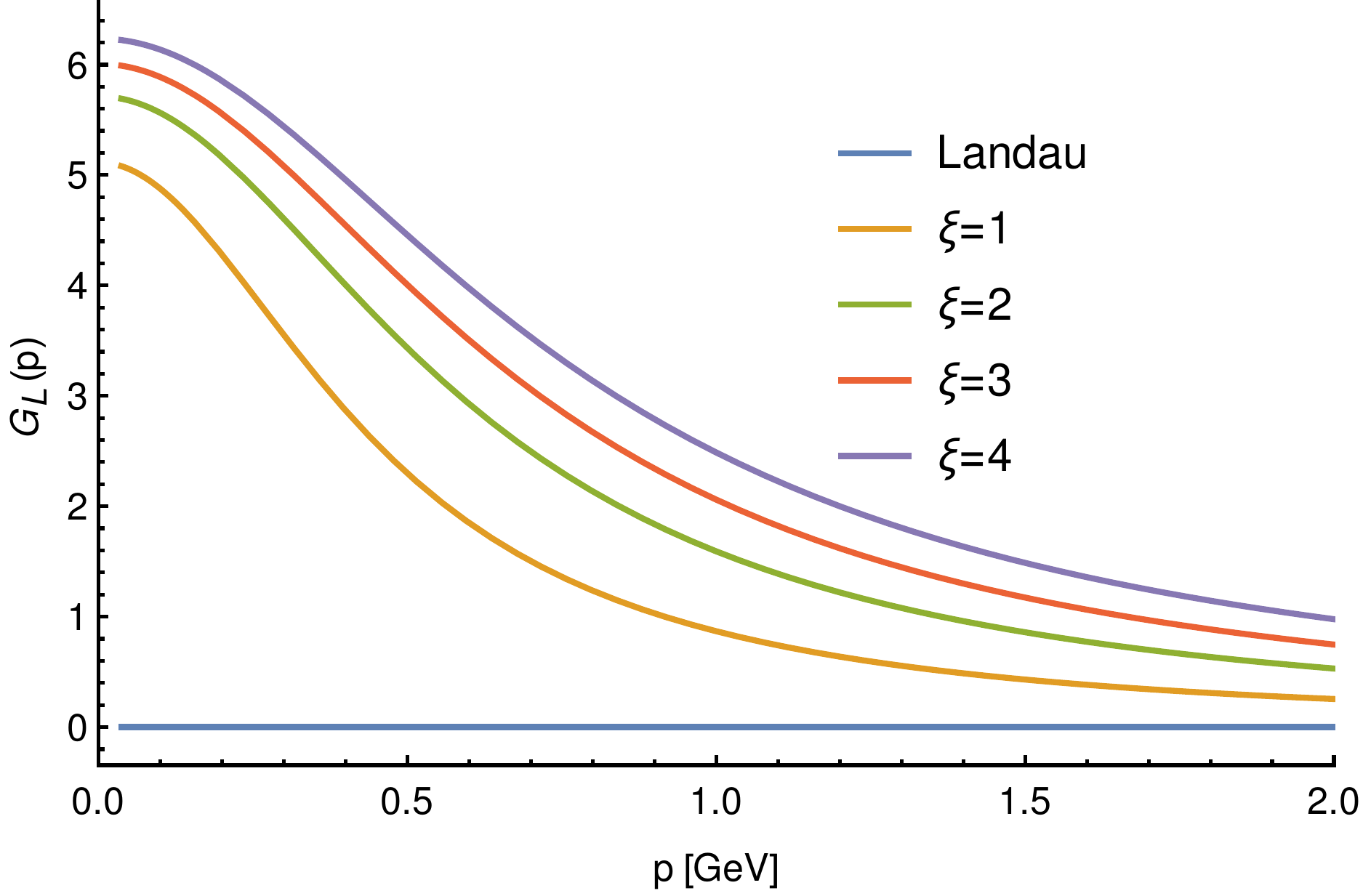}
  \caption{The RG-improved ghost and gluon propagators in the CF model ($n=1$) with the infrared-safe scheme.}
  \label{fig:CFpropagwithRG}
\end{figure} 

The RG-improved ghost and gluon propagators for the CF model are shown in \Fig{fig:CFpropagwithRG} in the infrared-safe scheme. We see little change as compared to the results from strict perturbation theory (Figs.~\ref{fig:CFIRsafe-gh}\,--\,\ref{fig:longCF2}), for $\xi$ not too small, despite the relatively important change of the running parameters in the range of momenta considered here. RG corrections seem more important for small $\xi$. However, the RG-improved results in the CF model are dramatically different from those of the theory considered here ($n\to0$), in particular for the gluon sector. Altogether these qualitative differences illustrate the important role played by the replicated superfield sector of the theory and, in turn, by the Gribov copies, in particular in the infrared.

\section{Conclusions}\label{sec_app_A}

We have investigated Yang-Mills propagators in the class of nonlinear covariant gauges recently proposed in Ref.~\cite{Serreau:2013ila}. These are defined from the minimization of a functional [see \Eqn{eq_func}] that generalizes the one used in the case of the Landau gauge. In particular, it is expected that minimization algorithms routinely employed in the Landau gauge can be applied to the present proposal. This would open the way to lattice simulations in covariant gauges other than the Landau gauge. 

The continuum formulation of these gauges relies on a generalization of the averaging procedure over Gribov copies proposed in Ref. \cite{Serreau:2012cg}. The standard Faddeev-Popov construction---which neglects the Gribov issue---produces the CFDJ action. Instead, our treatment of Gribov copies results in the CF action---a massive generalization of the CFDJ action---coupled to a replicated set of $n-1$ supersymmetric nonlinear sigma models. The dynamics of the replica sector is such that the  theory indeed corresponds to a 
gauge-fixing procedure in the physically relevant limit $n\to0$. Our gauge-fixed action is perturbatively renormalizable in four dimensions, which allows for perturbative calculations. In the Landau gauge limit, this procedure allows one to reproduce, with a one-loop calculation, the lattice results obtained in the minimal Landau gauge, where one selects a unique Gribov copy in the first Gribov region \cite{Tissier_10,Pelaez:2013cpa,Pelaez:2014mxa}. This suggests that for a certain range of values of the averaging parameter $\beta_0$, all copies in the first region are equiprobable and averaging over them is essentially equivalent to picking up a unique one. If this property was to hold away from the Landau gauge limit as well---which can only be assessed by actual lattice calculations--- we expect that the present results provide reasonable predictions for possible lattice calculations in the gauge proposed here.

A key point of the continuum formulation concerns the replica technique and the interplay between the limit \mbox{$n\to0$} and renormalization. Indeed, part of the $n$ dependence of the theory can be absorbed in the definition of the renormalized parameters \cite{Serreau:2012cg,Serreau:2013ila}. In the present work, we have employed a minimal scheme which, first, allows one to reproduce the results of Ref.~\cite{Tissier_10} in the Landau gauge ($\xi=0$) and, second, has a smooth $\xi\to0$ limit. We have studied two renormalization schemes and have presented results for the ghost and gluon propagators at one-loop order with and without RG improvement. Finally, we have compared our results to those of the CF model, obtained by simply setting the number of replicas $n=1$. This allows one to pinpoint the peculiar effects of the superfield sector of our theory, which is related to our particular treatment of the Gribov ambiguities.

The first important aspect of the present treatment of Gribov copies is the fact that the basic fields of the theory acquire effective masses, related to the gauge-fixing parameters $\beta_0$ and $\xi_0$. In contrast to the Landau gauge ($\xi_0=0$) case, not only the transverse gluons, but also the FP ghosts are massive. A striking difference between the CF model ($n=1$) and the gauge-fixed theory ($n\to0$) is the fact that, in the latter case, the gluon propagator remains transverse in momentum space even away from the Landau gauge. This is already visible at tree level. At one-loop order in a strict perturbative expansion---i.e., without RG improvement---we also observe important differences between the $n=1$ and the $n\to0$ cases, mainly in the transverse gluon propagator, which receives direct contributions from the replica sector. It is important to note that these contain massless degrees of freedom, which lead to nonanalyticities at small momentum. These are absent in the case $n=1$.

The role of massless excitations in the gauge-fixed theory is further illustrated by implementing RG improvement. We have devised an infrared-safe renormalization scheme, generalizing the one put forward in Ref. \cite{Tissier_10} for the Landau gauge. We observed that the one-loop flow has no Landau pole and that both the coupling and the mass parameters are attracted towards the Gaussian fixed point, whereas the Landau gauge fixed point appears unstable, both in the UV and in the infrared. This can be seen as a direct effect of the presence of massless modes in the theory. In the Landau gauge limit ($\xi=0$), these are the ghost fields whereas for $\xi\neq0$, these come from the replica sector. Still, we observe that the $\xi\to0$ limit is smooth. This results in strong RG effects on the propagators, mainly the transverse gluon, down to the deep infrared.
These results of the gauge-fixed theory are to be compared to the corresponding ones in the CF model, where all degrees of freedom are massive and the RG flow freezes in the infrared, resulting in quantitative but not qualitative changes as compared to the strict perturbative results. 

The present work provides an explicit example where one can explicitly work out the effect of the Gribov copies in a semianalytical continuum calculation, in a class of gauges that may be amenable to lattice calculations. In the case of the Landau gauge, our treatment of Gribov copies already allowed us to capture some nontrivial infrared physics in perturbation theory, whereas standard techniques based on the FP quantization require the use of nonperturbative tools. However, apart from providing an effective bare gluon mass, the replica sector completely decouples in the Landau gauge. This is not so in the present case and one can thus study explicitly the role of the replica sector---and thus, indirectly, of the Gribov copies---in the dynamics of the theory in the infrared. We believe the present results are of interest for the general question of possible gauge (in)dependences of the infrared sector of the Yang-Mills propagators. We hope that the present work will stimulate lattice studies, e.g., along the lines of Refs.~\cite{Cucchieri:2010ku,Cucchieri:2011pp,Cucchieri:2011aa,Bicudo:2015rma}.

\section*{Acknowledgements}
We are grateful to M.~Pel\'aez, U.~Reinosa and N.~Wschebor for many useful 
discussions.
\appendix

\section{Complete propagators}
\label{sec_complete_propagator}
We detail the inversion of the quadratic part of the action in the sector $(A,ih,\Lambda_k)$. The most general form of the various two-point vertex functions is given in Eqs.~\eqn{eq:decompmunu}--\eqn{eq:decomplast}. These are constrained by the spacetime symmetries, the replica permutation symmetry and the isometries of the Grassmann subspaces for each replica. These can be grouped together in the matrix
\begin{widetext}
\beq
\label{appeq:matrix}
\Gamma^{(2)} =\begin{pmatrix}
\Gamma_T P^T_{\mu \nu}+\Gamma_L P^L_{\mu \nu} &-i p_\mu\Gamma_{ih A} & i p_\mu \Gamma_{4} \\
i p_\nu\Gamma_{ih A}
& \Gamma_{ihih}& 0\\ 
-i p_\nu \Gamma_{4} & 0 & 
\delta_{kl}\!\left[\Gamma_{1} \delta(\underline{\theta}_k,\underline{\theta}_l')+\Gamma_2\square_{\underline{\theta}_k}\delta(\underline{\theta}_k,\underline{\theta}_l') \right]
+(\delta_{kl}-1)\Gamma_3
\end{pmatrix} ,
\eeq
where the scalar functions $\Gamma_T$, $\Gamma_L$, $\Gamma_{ihA}$ and $\Gamma_{1,\ldots,4}$ only depend on $p^2$ and where $\square_{\underline{\theta}}$ is the Laplace operator on the curved Grassmann space, defined as \cite{Tissier_08,Serreau:2013ila}
\beq
\Box_{\underline{\theta}}=\frac{1}{\sqrt{g(\underline{\theta})}}\partial_M \sqrt{g(\underline{\theta})}g^{MN}\partial_N= 2 \beta_0 (\theta \partial_\theta + \bar{\theta}\partial_{\bar{\theta}})+2(1-\beta_0 \bar{\theta}\theta)\partial_\theta \partial_{\bar{\theta}}.
\eeq 
In particular, it satisfies the identity $\square_{\underline{\theta}}\delta(\underline{\theta},\underline{\theta}')=-2+2\beta_0\delta(\underline{\theta},\underline{\theta}')$.

Note that, since there are $n-1$ replicas, the matrix representation \eqn{appeq:matrix} only makes sense for $n>0$ and the limit $n \to 0$ must be done after the inversion. Using the symmetries of the problem , the most general form of the inverse matrix $(\Gamma^{(2)})^{-1}$ reads
\beq
(\Gamma^{(2)})^{-1}= \begin{pmatrix}
\Delta_T P^T_{ \rho \nu}+\Delta_L P^L_{ \rho \nu} &i p_\rho \Delta_{ih A} & -i p_\rho \Delta_{4} \\
- i p_\nu \Delta_{ih A}
& \Delta_{ihih}& \Delta_{5}\\ 
i p_\nu \Delta_{4} & \Delta_{5} & \delta_{lm}\left[\Delta_1 \delta\left(\underline{\theta}_l,\underline{\theta}_m \right)+ \Delta_2\Box_{\underline{\theta}_l}\delta\left(\underline{\theta}_l,\underline{\theta}_m \right) \right]+\left(1-\delta_{lm} \right) \Delta_3
\end{pmatrix}, 
\eeq
where the unknown scalar functions  $ \Delta_T$, $ \Delta_L$, $ \Delta_{ihA}$, and $\Delta_{1,\ldots,5}$ only depend on $p^2$.
The inversion is defined by
\begin{eqnarray}
\Gamma^{(2)} \times \left(\Gamma^{(2)}\right)^{-1}& =& \begin{pmatrix}
\delta_{\mu \rho} &0 &0 \\
0
& 1& 0\\ 
0 & 0& \delta_{km}\delta\left(\underline{\theta}_k,\underline{\theta}_m \right)
\end{pmatrix}, 
\end{eqnarray}
where the product $\times$ involves a sum over Lorentz and replica indices and an integral over the Grassmann variable $\underline{\theta}_l$. The calculation is straightforward. We get, for the gluon propagator,
\beq
\Delta_T= \frac{1}{\Gamma_T}\quad{\rm and} \quad \Delta_L= \frac{\Gamma_{ihih}\left(\Gamma_1+(n-2)\beta_0\Gamma_3\right)}{(\Gamma_L \Gamma_{ihih}-p^2\Gamma_{ihA}^2 )\left(\Gamma_1+(n-2)\beta_0\Gamma_3 \right)-(n-1)\beta_0p^2 \Gamma_{ihih} \Gamma_{4}^2}.
\eeq
The other components in the $(A,ih)$ sector are obtained from these as
\beq
\Delta_{ihA}=\frac{\Gamma_{ihA}}{\Gamma_{ihih}}\Delta_L\quad{\rm and}\quad\Delta_{ih ih}=\frac{1+p^2\Gamma_{ihA} \Delta_{ihA}}{\Gamma_{ihih} }.
\eeq
Finally, the components in the superfield sector read
\begin{align}
\Delta_{ 4}&=-\frac{1}{(n-1)\beta_0 p^2\Gamma_4} \left\{1-
						  \frac{\Delta_{ ihA}(\Gamma_L \Gamma_{ihih}-p^2\Gamma_{ihA}^2 )}{\Gamma_{ihA} }\right\}\\
\Delta_{5} &=-p^2\frac{\Gamma_{ihA}}{\Gamma_{ihih}}\Delta_4\\
\Delta_3&=\frac{1}{\Gamma_1+(n-2)\beta_0 \Gamma_3 }\left\{ p^2 \Gamma_{4} \Delta_{ 4}-\frac{\Gamma_{3}}{(\Gamma_1-\beta_0\Gamma_3  )}\right\}\\
\Delta_1&= \frac{1}{\Gamma_1-\Gamma_3 \beta_0 }+\beta_0 \Delta_3\\
\Delta_2 &=\frac{1}{2\beta_0(\Gamma_1+2\beta_0 \Gamma_2)}-\frac{\Delta_1}{2 \beta_0}.
\end{align}
\end{widetext}
The propagators are obtained by taking the limit $n \to 0$. For instance, the longitudinal gluon propagator is given by
\beq
G_L(p)= \lim_{n \to 0} \Delta_L(p),
\eeq
which is \Eqn{eq:long}.

\section{Tree level}
\label{appsec:tree}

We give here the tree-level expressions of the bare two-point vertex functions. In the transverse gluon and the ghost sectors, 
\begin{align}
 \Gamma_T(p)&=p^2+n\beta_0\\
 \Gamma_{c\bar c}(p)&=p^2+\beta_0\xi_0;
\end{align}
In the longitudinal gluon and $h$ sector,
\begin{align}
 \Gamma_L(p)&=n\beta_0\\
 \Gamma_{ihA}(p)&=1\\
 \Gamma_{ihih}(p)&=-\xi_0;
\end{align}
In the superfield sector
\begin{align}
 \Gamma_1(p)&=p^2\\
 \Gamma_2(p)&=\xi_0/2\\
 \Gamma_3(p)&=0\\
 \Gamma_4(p)&=1.
\end{align}
We check that the Slavnov-Taylor identities of Appendix~\ref{sec:appST} are satisfied at tree level. We also check that the right-hand side of \Eqn{eq:long} yields (for arbitrary $n$)
\beq
 G_L(p)=\frac{\xi_0}{p^2+\beta_0\xi_0}.
\eeq

\section{Slavnov-Taylor identities in the CF model}
\label{sec:appST}
As already discussed, the CF model can be obtained from the theory considered here by setting the number of replicas $n=1$. Its action is given by
\beq
 S[A,c,\bar c,h]=S_{\rm YM}[A]+S_{{\rm CF}}[A,c,\cb,h]
\eeq
with $S_{\rm YM}$ and $S_{\rm CF}$ given in Eqs.~\eqn{eq:SYM} and \eqn{eq_action_CF} respectively. This model possesses a (non-nilpotent) BRST symmetry, whose action on the fields is
\beq
 sA^a_\mu=D_\mu c^a\,,\quad sc^a=-\frac{g}{2}f^{abc}c^bc^c
\eeq
and 
\beq
 s\bar c^a=ih^a\,,\quad sih^a=\beta_0c^a.
\eeq
The Zinn-Justin equation corresponding to this symmetry is obtained as usual, i.e., by introducing external sources for the fields and for all the independent BRST variations, $S\to S-S_1$ with 
\beq
 S_1=\int_x\Big\{J^a_\mu A^a_\mu\!+\!\bar \eta^a c^a\!+\!\bar c^a\eta^a\!+\!M^aih^a\!+\!\bar K^a_\mu sA^a_\mu\!+\!\bar L^a s c^a\Big\},
\eeq
and by performing a Legendre transform with respect to the sources $J_\mu^a$, $\eta^a$, $\bar\eta^a$, and $M^a$. It reads 
\beq
 \int_x\left\{\frac{\delta\Gamma}{\delta\bar K^a_\mu}\frac{\delta\Gamma}{\delta A^a_\mu}+\frac{\delta\Gamma}{\delta\bar L^a}\frac{\delta\Gamma}{\delta c^a}-ih^a\frac{\delta\Gamma}{\delta \bar c^a}-\beta_0c^a\frac{\delta\Gamma}{\delta ih^a}\right\}=0.
\eeq
Taking two derivatives of this equation with respect either to $ih^a$ and $c^a$ or to $A_\mu^a$ and $c^a$ and setting the sources to zero, one obtains the following symmetry identities for the two-point vertex functions in momentum space, as defined in \Eqn{eq:defvertex}:
\beq
 \Gamma_{c\bar K_\mu}(p)\Gamma_{A_\mu ih}(p)-\Gamma_{c\bar c}(p)-\beta_0\Gamma_{ihih}(p)=0
\eeq 
and
\beq
 \Gamma_{c\bar K_\mu}(p)\Gamma_{A_\mu A_\nu}(p)-\beta_0\Gamma_{ihA_\nu}(p)=0.
\eeq
Eliminating $\Gamma_{c\bar K_\mu}(p)$, we obtain, finally,
\beq
p^2\left[\Gamma_{ihA}(p)\right]^2-\Gamma_L(p) \Gamma_{ihih}(p)=\frac{\Gamma_L(p)\Gamma_{c\bar c}(p)}{\beta_0},
\eeq
which is used in obtaining \Eqn{eq:gfgfgf}.

\section{RG flow in the ultraviolet}
\label{appsec:UV}

The flow of the present theory in the ultraviolet is easily obtained from the divergent parts of the renormalization factors; Eqs.~\eqn{eq:okzxfm}--\eqn{eq:oisjdgf}. For instance, the $\beta$ functions for the coupling and the gluon mass parameters read, in the limit $\mu\gg m,m_{\rm gh}$, and taking $n\to0$,
\beq
\frac{\beta_g^{\rm UV}}{g}=-\frac{11}{6}\frac{g^2N}{8\pi^2}\,,\quad{\rm and}\quad \frac{\beta_{m^2}^{\rm UV}}{m^2}=-\frac{35}{12}\frac{g^2N}{8\pi^2}.
\eeq
The UV running of the coupling is the same as in the Landau gauge, that is  
\beq
 \frac{1}{g^2(\mu)}=\frac{1}{g^2(\mu_*)}+\frac{11N}{24\pi^2}\ln\left(\frac{\mu}{\mu_*}\right),
\eeq
where $\mu$ and $\mu_*$ are UV scales. The flow of $m$ is easily integrated. Using Eqs. \eqn{eq:xirunir} and \eqn{eq:ghmass}, we get the following relations
\begin{align}
 \frac{m(\mu)}{m(\mu_*)}&=\left(\frac{g(\mu)}{g(\mu_*)}\right)^{\!\!{35\over44}}\\
 \frac{m_{\rm gh}(\mu)}{m_{\rm gh}(\mu_*)}&=\left(\frac{g(\mu)}{g(\mu_*)}\right)^{\!\!{9\over44}}\\
 \frac{\xi(\mu)}{\xi(\mu_*)}&=\left(\frac{g(\mu)}{g(\mu_*)}\right)^{\!\!-{13\over11}}.
\end{align}
We observe that the Gaussian point $g=0$ and $m=m_{\rm gh}=0$ is an attractive UV fixed point, whereas $\xi=0$ is an unstable UV fixed point.

Using Eqs.~\eqn{eq:rgghost} and \eqn{eq:gluproirs}, we recover the standard one-loop behaviors of the gluon and ghost propagators at large momentum
\begin{align}
 G_{\rm gh}^{\rm UV}(p)&\propto\frac{1}{p^2}\left[1+\frac{11Ng^2(\mu_*)}{48\pi^2}\ln\left(\frac{p^2}{\mu_*^2}\right)\right]^{-{9\over44}}\\
 G_T^{\rm UV}(p)&\propto\frac{1}{p^2}\left[1+\frac{11Ng^2(\mu_*)}{48\pi^2}\ln\left(\frac{p^2}{\mu_*^2}\right)\right]^{-{13\over22}},
\end{align}
where we have used the fact that the contribution from the mass terms are negligible at large momentum: $m_{\rm gh}^2(p),m^2(p)\ll p^2$.


\begin{thebibliography}{}
\bibitem{supp_mat}
This is a Mathematica file where the analytic expressions for all diagrams are given.
\end{thebibliography}
\end{document}